\documentclass[aps,prd,nofootinbib,showpacs,twocolumn,superscriptaddress]{revtex4-1}

\usepackage{graphicx}
\usepackage{float}
\usepackage{amsmath}
\usepackage{cancel}
\usepackage{mathtools}
\usepackage{txfonts}
\usepackage{slashed}

\newcommand{\diffd}{\textrm{d}}

\begin{document}

\title{Description of fully differential Drell-Yan pair production}

\author{Fabian Eichstaedt}\email{fabian.eichstaedt@theo.physik.uni-giessen.de}
\affiliation{Institut f\"ur Theoretische Physik, Universit\"at Giessen, Germany}
\author{Stefan Leupold}
\affiliation{Institutionen f\"or fysik och astronomi, Uppsala Universitet, Sweden}
\author{Kai Gallmeister}
\affiliation{Institut f\"ur Theoretische Physik, Universit\"at Giessen, Germany}
\affiliation{Institut f\"ur Theoretische Physik, Universit\"at Frankfurt/Main, Germany}
\author{Hendrik van Hees}
\affiliation{Institut f\"ur Theoretische Physik, Universit\"at Giessen, Germany}
\affiliation{Institut f\"ur Theoretische Physik, Universit\"at Frankfurt/Main, Germany}
\author{Ulrich Mosel}
\affiliation{Institut f\"ur Theoretische Physik, Universit\"at Giessen, Germany}

\begin{abstract}
We investigate Drell-Yan pair production in a QCD inspired model, which takes into account all relevant
hard processes up to $O(\alpha_s)$. To address the known shortfalls of such a fixed order
calculation we introduce phenomenological parton distributions for initial transverse momentum
and quark mass, and devise a subtraction scheme to avoid double-counting when utilizing the standard 
longitudinal parton distribution functions. We show that we can reproduce Drell-Yan transverse momentum and invariant mass spectra from different proton-proton, proton-nucleus and antiproton-nucleus experiments and at different energies without the need for a $K$ factor. Fixing our parameters at these spectra, we make predictions for Drell-Yan transverse momentum spectra at low hadronic energies, which will be measured for example at $\overline{\text{P}}$ANDA in antiproton-proton collisions.

\end{abstract}

\keywords{drell-yan, parton model, transverse momentum, mass distribution, nlo}

\pacs{12.38.Qk, 12.38.Cy, 13.85.Qk}

\maketitle

\section{Introduction}
\label{sec:intro}
The Drell-Yan (DY) process \cite{Drell:1970wh} has been studied for the last 40 years and provides
an important tool to access the distribution of partons inside the nucleon.
While the primary tool for exploration of the nucleon structure is deep inelastic scattering \cite{Bjorken:1969ja}, DY data give complementary insights, since, for example, it directly probes sea-quark
distributions \cite{Alekhin:2006zm}. A lot of experimental effort
is being devoted to measurements of the DY process: in antiproton-proton ($\overline{\text{p}}$p) collisions at
$\overline{\textrm{P}}$ANDA (FAIR) \cite{Lutz:2009ff} and PAX \cite{Barone:2005pu}, in proton-proton
(pp)
collisions at RHIC \cite{Bunce:2000uv,Bunce:2008aa}, J-PARC 
\cite{Peng:2006aa,Goto:2007aa,Kumano:2008rt}, IHEP \cite{Abramov:2005mk} and JINR
\cite{Sissakian:2008th} and in pion-nucleon collisions at COMPASS \cite{Bradamante:1997mu,Baum:1996yv}.
An overview of the experimental situation can be found in \cite{Reimer:2007iy}.

Studies of this process \cite{Halzen:1978rx,Altarelli:1977kt,Arnold:2008kf}
are generally inspired by perturbative QCD (pQCD). 
The most simple scheme is the parton model description, which is
a leading order (LO) approach ($O(\alpha_s^0$)). However, it 
does not fully describe the interesting observables. While the shape of the invariant
mass ($M$) spectra of the DY pair can be reproduced, the absolute height can only be accounted for by including an additional $K$ factor. Furthermore transverse momentum ($p_T$) spectra are not accessible at all 
\cite{Gavin:1995ch}. The usual approach to handle the latter problem is to fold in a 
phenomenological Gaussian distribution for the parton transverse momentum \cite{D'Alesio:2004up}, the width of which has to be 
fitted to data. But since these distributions are normalized, the absolute size of the
cross sections is still underestimated \cite{D'Alesio:2004up}. The next logical step is
to turn to next-to-leading order (NLO, $O(\alpha_s)$) in the contributing hard subprocesses,
but this brings about additional problems: the calculated $p_T$ spectra are divergent for 
$p_T \rightarrow 0$. In fact, they are divergent in any fixed order of the strong coupling $\alpha_s$,
due to large logarithmic corrections $\ln \left(M/p_T\right)$ \cite{Gavin:1995ch}. It is
possible to remove these divergences by an all-order resummation. However, since
$p_T$ is no longer a hard scale at $p_T \rightarrow 0$, additional non-perturbative (i.e.,
experimental) input 
is needed in these (and all other pQCD) approaches to describe the region of very small $p_T$ 
\cite{Collins:1984kg,Davies:1984sp,Fai:2003zc}. Note that the parton model (i.e.,
LO) description is still a very useful starting point, for example for studying 
spin asymmetries in DY, since there NLO corrections appear to be rather small 
\cite{Shimizu:2005fp,Barone:2005cr,Martin:1997rz}. $\overline{\textrm{P}}$ANDA, however, will allow measurements at hadron c.m.\ energies
of a few GeV, where non-perturbative effects are expected to become more important.
This highlights the need to model these effects in a phenomenological picture.

In \cite{Eichstaedt:2009cn} we revisited a model, which incorporates phenomenological
transverse momentum distributions for quarks and which takes into account the full kinematics
in the hard LO subprocess, i.e., the usual collinear approximation is overcome. It was found
that results differed only slightly from standard parton model calculations, which underestimate the data. In the present paper we improve on our previous work in several ways: First, we include quark mass distributions in the LO process.
We will show that we still underestimate the data. This finding triggered a complete calculation of all hard subprocesses to $O(\alpha_s)$ including the full kinematics, which will also be presented in the current work. As mentioned above, such a calculation would suffer from divergent $p_T$ spectra if the quarks were massless. However, we will show that the phenomenological quark mass distributions we introduced before now effectively smear out the divergent behavior. Such mass distributions or spectral functions are a well known concept in nuclear physics, where they are applied to the strongly coupled system of nucleons in nuclei, see for example \cite{Day:1990mf}. Thus it is worthwhile to test the same concept in the nucleon, which is a strongly coupled system of quarks and gluons. 
For ($p_T$ integrated) $M$ spectra the divergent $O(\alpha_s)$ contributions are commonly absorbed into the parton distribution functions. Therefore, we introduce a subtraction scheme to prevent double counting of
those processes which we consider explicitly.

This paper is structured as follows: We introduce our notation and conventions in Sec.~\ref{subsec:intro-not} and we give an introduction to the general kinematics of DY pair production in Sec.~\ref{subsec:intro-genkin}. In Sec.~\ref{sec:lo} we address DY at LO, in particular we describe the standard parton model and our extensions of it, namely including intrinsic parton transverse momentum and quark mass distributions. We show the results of our extended LO model in Sec.~\ref{subsec:lo-results}. Sec.~\ref{sec:nlo} presents our NLO calculation. The vertex correction is treated in detail in \ref{subsec:nlo-vc} and the calculation of gluon bremsstrahlung and gluon Compton scattering is presented in Secs.~\ref{subsec:nlo-brems} and \ref{subsec:nlo-compton}. In Sec.\ \ref{subsec:ir-div}
the influence of quark mass distributions is discussed. Furthermore, we discuss the treatment
of collinear singularities and describe our subtraction scheme
for the $O(\alpha_s)$ contributions in detail in Sec.\ \ref{subsec:coll_sing}, before we shortly 
comment on the use of initial transverse momentum distributions.
We show the results of our full model in Sec. \ref{sec:results} and compare with
data from different experiments in pp, p-nucleus and $\overline{\text{p}}$-nucleus reactions.
In addition we present our predictions for DY pair production at $\overline{\textrm{P}}$ANDA energies.
Finally we present our conclusions in Sec. \ref{sec:conc}.
The Appendix collects several topics, that are only briefly touched in the main text:
since we assign different masses to the annihilating quarks and antiquarks, we explicitly prove gauge invariance in App.~\ref{app:gauge}. Then we investigate the influence of different quark masses on the form factors $F_1$ and $F_2$ in App.~\ref{app:F1}, and finally we present the details of the phase space evaluation in App.~\ref{app:nlo-kin}.

\subsection{Notation}
\label{subsec:intro-not}

\begin{figure}[H]
	\centering
	\includegraphics[keepaspectratio,width=0.45\textwidth]{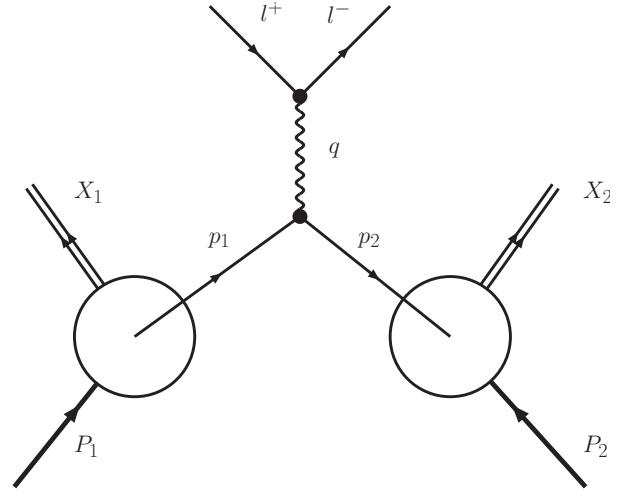}
	\caption{DY production in a nucleon-nucleon collision; $X_1$ and $X_2$ denote the
		 nucleon remnants. See main text for details.}
	\label{fig:DY}	 
\end{figure}

In the following we present the conventions and notations used throughout
this paper:
It will turn out to be useful to write four-momenta using light-cone coordinates. 
We employ the following convention for general four-vectors $a$ and $b$:

\begin{align}
	           a^+ &= a_0 + a_z\ , \\
	           a^- &= a_0 - a_z\ , \\
	\vec a_{\perp} &= \left( a_x, a_y \right)\ , \\
       \Rightarrow a^2 &= a^+ a^- - ({\vec a}_{\perp})^2\ , \\
  \Rightarrow a\cdot b &= \frac{1}{2}\left(a^+ b^- + a^- b^+ - 2\, \vec a_{\perp} \cdot 
		                                                   \vec b_{\perp} \right)\ .
\end{align}       	
Leptons are treated as massless.
We define the target nucleon to carry the four-momentum 
$P_1$ and the beam nucleon to carry the four-momentum $P_2$ (see Fig.~\ref{fig:DY}). In the hadron
center-of-mass (c.m.) frame we choose the $z$-axis as the beam line, and the beam (target) nucleon
moves in the positive (negative) direction.
Therefore, the nucleon four-momenta read
\begin{align}
	P_1 &= \left( \frac{\sqrt{S}}{2} , 0, 0, -\sqrt{\frac{S}{4} - m_N^2 } \right)\ , \label{nucleon1}\\
	P_2 &= \left( \frac{\sqrt{S}}{2} , 0, 0, +\sqrt{\frac{S}{4} - m_N^2 } \right)\ , \label{nucleon2}
\end{align}
which implies
\begin{equation}
	P_1^- = P_2^+ = \frac{\sqrt{S}}{2} + \sqrt{\frac{S}{4} - m_N^2} \xrightarrow{m_N \rightarrow 0}
	\sqrt{S} \label{nucleon_large_comp}
\end{equation}	
for the large momentum components of the nucleons.
Note that in \cite{Eichstaedt:2009cn} we presented a calculation with
vanishing nucleon mass $m_N$. However, we want to study our model also at comparatively low
c.m. energies (e.g.\ $\sqrt{S} \sim 5.5$ GeV at $\overline{\textrm{P}}$ANDA). Therefore, in the present
work we include the nucleon mass since its influence should become significant at these
energies.
We denote the four-momentum of the parton in nucleon 1 (2) as $p_1$ ($p_2$). The on-shell
condition in light-cone coordinates then reads:
\begin{equation}
	m_i^2 = p_i^2 = p_i^+ p_i^- - ({\vec p}_{i_\perp})^2\ .
\end{equation}	
For the virtual photon in Fig.~\ref{fig:DY} the maximal $q_z$ is derived by requiring the invariant
mass of the undetected remnants to vanish and the photon to move collinearly to the nucleons:
\begin{align}
	\left(P_1 + P_2 - q \right)^2 &= X^2 \overset{!}{=} 0 \\
	\Rightarrow S + q^2 - 2 \sqrt{S} \sqrt{q^2 + (q_z)_\textrm{max}^2} &= 0 \\
	\Rightarrow \frac{S - q^2}{2\sqrt{S}} &= (q_z)_\textrm{max} \label{qzmax} \ .
\end{align}
In the literature and in the data presented by many experimental groups the Feynman variable
\cite{Nakamura:2010zzi} is defined as:
\begin{align}
  x_F &=  \frac{2 q_z}{\sqrt{S}}\simeq \frac{q_z}{(q_z)_\textrm{max}} \nonumber \\
  \Rightarrow (q_z)_\textrm{max} &\simeq \frac{\sqrt{S}}{2} \ .
  \label{eq:simple_xF}
\end{align}
Note that this approximation for $x_F$ is obviously only valid for $q^2 \ll S$. Since we perform studies
in the small-$S$ region, our definition of the Feynman variable $x_F'$ is :
\begin{align}
  x_F' = \frac{q_z}{(q_z)_\textrm{max}} = q_z \cdot\frac{2\sqrt{S}}{S-q^2}\ , \label{xFprime}
\end{align}
without any approximations.
Depending on the experiment which we study
we will use $x_F$ or $x_F'$ according to Eq.~(\ref{eq:simple_xF}) and Eq.~(\ref{xFprime}) .

\subsection{General kinematics}
\label{subsec:intro-genkin}

\begin{figure}[htbp]
	\centering
	\includegraphics[keepaspectratio,width=0.4\textwidth]{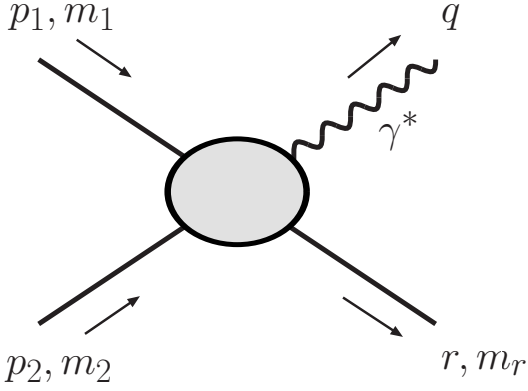}
	\caption{General kinematics of hard subprocesses of DY production.}
       	\label{fig:dy-genkin}	 
\end{figure}	

In this section we shortly present the kinematic scheme,
in which our calculations are performed. Note that we show the most
general form for DY pair production at NLO, of which the LO process
is a special case. Fig.\ \ref{fig:dy-genkin} is a reference for our notation.

For the initial particles we choose four-momenta $p_1$ and $p_2$ and
masses $m_1$ and $m_2$. For the final state we always choose $q$ as
the four-momentum of the virtual photon, i.e., the DY pair, and $M=\sqrt{q^2}$
as its mass. The four-momentum of the remaining final state particle
we define as $r$ and its mass as $m_r$ (both $0$ at LO). The differential partonic
cross section then takes the following general form:
\begin{align}
	\diffd \hat\sigma = F(p_1,p_2,q,r) \cdot \delta^{(4)}(p_1+p_2-q-r)\,
			   \diffd M^2\, \frac{\diffd ^3q}{E_q}\, \frac{\diffd ^3r}{2E_r} \ , 
			   \label{nlo-general-part-xsec}
\end{align}			   
where $F$ contains squared matrix elements, the flux factor and constants, and $E_q$ ($E_r$)
are the energies of the particles with momenta $q$ ($r$). Note that in the LO case there
is only the virtual photon in the final state and thus
\begin{align}
	F(p_1,p_2,q,r) = \tilde F(p_1,p_2,q) \cdot \delta^{(4)}(r)\, 2E_r \diffd E_r\ .
\end{align}
The Mandelstam variables read
\begin{align}
	s &= (p_1 + p_2)^2 \ ,\\
	t &= (p_2 - q)^2 \ ,\\
	u &= (p_1 - q)^2 \ .
\end{align}
In DY measurements the common observables are the invariant mass $M$, the absolute transverse momentum
$p_T$ and the longitudinal momentum $q_z$ of the lepton pair. Thus we can 
integrate Eq.~(\ref{nlo-general-part-xsec}) over the phase space of $r$ as well
as the azimuthal angle $\phi_q$ of $q$. Then we obtain
\begin{align}
	\frac{\diffd \hat\sigma}{\diffd M^2 \diffd t} = \frac{\pi}{2\sqrt{s} 
	      p_{\text{cm}}} \cdot F(s,t,M^2,m_1^2,m_2^2,m_r^2) 
	      \cdot \Theta(\sqrt{s} - E_q)\ ,
	      \label{nlo-double-part-xsec}
\end{align}	
with the center-of-mass momentum of the incoming states \cite{Nakamura:2010zzi}
\begin{align}
	p_{\text{cm}} = \frac{\sqrt{(s - (m_1+m_2)^2) ( s - (m_1-m_2)^2)}}{2\sqrt{s}}\ .
\end{align}	
Now comparing Eqs. (\ref{nlo-general-part-xsec}) and (\ref{nlo-double-part-xsec})
one finds (cf. \cite{Halzen:1978rx})
\begin{align}
	\frac{E_q \diffd \hat\sigma}{\diffd M^2 \diffd ^3q} =
	\frac{2\sqrt{s} p_{\text{cm}}}{\pi} 
	\cdot \frac{\diffd \hat\sigma}{\diffd M^2 \diffd t}
	\cdot \delta\left((p_1+p_2-q)^2 - m_r^2\right) 
\end{align}	
and finally
\begin{align}
	\frac{\diffd \hat\sigma}{\diffd M^2 \diffd p_T^2 \diffd x_F} =&
	\frac{2\sqrt{s} p_\text{cm} (q_z)_\textrm{max}}{E_q}
	\cdot \frac{\diffd \hat\sigma}{\diffd M^2 \diffd t} \nonumber \\
	&\times \delta\left((p_1+p_2-q)^2 - m_r^2\right) \ .
	\label{genkin-master}
\end{align}	
Thus, for all the relevant subprocesses we actually only have to calculate
$\frac{\diffd \hat\sigma}{\diffd M^2 \diffd t}$ and we will use the relation 
(\ref{genkin-master}) especially for the NLO calculation with two
particles (virtual photon + quark/gluon) in the final state.

\section{Leading order Drell-Yan}
\label{sec:lo}

In this section we will present an approach to DY pair production in LO
in the hard subprocess, i.e., $O(\alpha_s^0)$. We will start with
the standard parton model approach and then remedy its shortcomings by introducing
distributions for transverse momentum and mass of the annihilating quarks. Finally
we will present the results of this LO approach.

\subsection{Standard parton model}
\label{subsec:lo-spm}
The LO DY differential cross section in the standard parton
model reads \cite{Drell:1970wh}

\begin{align}
  &\textrm{d}\sigma_\text{LO} = \nonumber \\
			     &\int_0^1 \textrm{d}x_1 \int_0^1 \textrm{d}x_2
		                  \sum_i q_i^2\, f_i(x_1,q^2)\, f_{\, \bar i}(x_2,q^2)\, 
				  \textrm{d}\hat\sigma(x_1,x_2,q^2)\ .  \label{hadr_coll_xsec}
\end{align}
The sum runs over all quark flavors and
antiflavors, $q_i$ denotes the electric charge of quark flavor $i$ and the functions
$f_i$ are parton distribution functions (PDFs).

As usual, $x_1$ and $x_2$ are the momentum fractions carried by the
annihilating partons inside the colliding nucleons,
\begin{align}
	p_1 &= x_1 P_1\ , \label{collinear_p1}  \\
	p_2 &= x_2 P_2\ . \label{collinear_p2}
\end{align}
Note that this implies
\begin{align}
	p_1^- &= x_1 P_1^- \ , \label{x1}\\
	p_2^+ &= x_2 P_2^+     \label{x2}
\end{align}
for the large parton momentum components.
The small components are
\begin{align}
	p_1^+ = \frac{m_1^2}{p_1^-} \ , \\
	p_2^- = \frac{m_2^2}{p_2^+} \ ,
\end{align}
and they are generally neglected in the standard parton model, since
the partons are assumed to be massless. Consistency with the relations (\ref{collinear_p1}, \ref{collinear_p2}) is then achieved by assuming also the nucleon mass to be negligible.

$\textrm{d}\hat\sigma$ is the differential cross section of the partonic subprocess,
\begin{equation}
  \textrm{d}\hat \sigma_\text{LO} = \frac{4 \pi \alpha^2}{9 q^2} \, \delta(M^2 - q^2) \,
				\delta^{(4)}(p_1 + p_2 - q) \, \textrm{d}^4q \, \textrm{d}M^2 \ ,
						\label{part_coll_xsec}
\end{equation}				
and so we find
\begin{align}
  \frac{\diffd \hat\sigma_\text{LO}}{\diffd M^2 \diffd t}
	= \frac{4 \pi \alpha^2}{9 M^2} \, \delta(M^2 - (p_1+p_2)^2) \,
				\delta(t) \ .
\end{align}
Here $q$ is the four-momentum of the virtual photon, $p_1,p_2$ are the four-momenta of the
partons (cf.\ Fig.\ \ref{fig:DY}) and $\alpha \approx 1/137$ is the fine-structure constant.

The factorization into hard (subprocess)
and soft (PDFs) physics is proven in the collinear case at least for leading twist 
(expansion in $1/M$) in \cite{Collins:1989gx}.

Note that it becomes immediately clear from Eqs.~(\ref{collinear_p1}) and (\ref{collinear_p2})
that the incoming partons move collinearly with the nucleons. According to four-momentum conservation
in Eq.\ (\ref{part_coll_xsec}) no transverse momentum can therefore be
generated for the virtual photon (and thus for the DY pair) in the LO process.

The maximal information about the DY pair that can be gained from Eq.\ 
(\ref{hadr_coll_xsec}) is double differential,
\begin{equation}
  \frac{\diffd\sigma_\text{LO}}{\diffd M^2\ \diffd x_F}
	      = \sum_i q_i^2 f_i(x_1,M^2) \, f_{\, \bar i}(x_2,M^2) \,
				  \frac{4 \pi \alpha^2}{9 M^2} \, 
				  \frac{(q_z)_\textrm{max}}
                                       { (P_1^-)^2 E_\textrm{coll} } \ ,
		\label{spm_hadr_xsec}
\end{equation}		
with 

\begin{align}
	x_1 &= \frac{- (q_z)_\textrm{max}\, x_F + E_\textrm{coll}}{P_1^-} \ , 
        \label{x1_spm}	\\
	x_2 &= \frac{(q_z)_\textrm{max}\, x_F + E_\textrm{coll}}{P_1^-} 
	\label{x2_spm}
\end{align}	
and the energy of the collinear DY pair,
\begin{equation}
E_\textrm{coll} = \sqrt{M^2 + \left((q_z)_\textrm{max}\, x_F\right)^2}\ .
\end{equation}

In this section we have presented the standard parton model solution for the LO DY
cross section. The only quantities in this approach not determined by pQCD are the PDFs.
These have to be obtained by fitting parametrizations to experimental data, mainly on deep inelastic
scattering (DIS), but also on measurements of DY production itself \cite{Stirling:1900sj}.

\subsection{Intrinsic transverse momentum}
\label{subsec:lo-intrkt}
As already mentioned above, no DY pair transverse momentum ($p_T$) is generated in the simple 
parton model approach. Nevertheless, measurements indicate a Gaussian form of the $p_T$ spectra
at not too large $p_T$. This has been studied in approaches including initial quark transverse
momentum distributions, e.g.~\cite{D'Alesio:2004up,Anselmino:2005sh}. In \cite{Eichstaedt:2009cn} we also presented
an approach
incorporating primordial quark transverse momentum to address this issue. However, we found
that additional unphysical solutions for the momentum fractions $x_i$ appear, which have to be removed
properly. These unphysical solutions are an artifact of rewriting the momentum variables using light-cone
coordinates and they reveal themselves as one of two possible solutions of a quadratic equation.
The correct solutions can always be identified by putting all transverse momenta to
zero and then by checking whether the well known parton model solutions for the $x_i$ as given
in Eqs.~(\ref{x1_spm},\ref{x2_spm}) are recovered. In \cite{Eichstaedt:2009cn} we found that
in the transverse momentum dependent approach the LO DY differential
cross section reads
\begin{align}
  &\textrm{d}\sigma_\text{LO} = \fint_0^1 \textrm{d}x_1 \fint_0^1 \textrm{d}x_2
       		           \int \textrm{d}{\vec p}_{1_\perp} \int \textrm{d}{\vec p}_{2_\perp}
				\nonumber \\
		          &\times \sum_i q_i^2 {\tilde f}_i(x_1,{\vec p}_{1_\perp},q^2)
				{\tilde f}_{\, \bar i}(x_2,{\vec p}_{2_\perp},q^2) 
		           \textrm{d}\hat\sigma(x_1,{\vec p}_{1_\perp},x_2,{\vec p}_{2_\perp},q^2)
			   \ .        \label{hadr_full_xsec_intrkt}
\end{align}
The functions ${\tilde f}_i(x,{\vec p}_{\perp},q^2)$ are now extensions of the standard
longitudinal PDFs, since they also describe the distribution of quark transverse momentum.
We will show our ansatz for these functions in Sec.~\ref{subsec:lo-distr}. Note that we
take into account the full kinematics in the partonic cross section, i.e.
$\textrm{d}\hat\sigma_\text{LO}=\textrm{d}\hat\sigma_\text{LO}(x_1,{\vec p}_{1_\perp},x_2,{\vec p}_{2_\perp},q^2)$,
however, the quark masses are neglected as in the previous section.
In this approach the transverse momentum ($p_T=|{\vec q}_\perp|$) of the DY pair is accessible,
since the annihilating quark and antiquark can have finite initial transverse momenta.

The symbol $\fint$ represents the requirement that the unphysical solutions for the $x_i$ have to be removed,
as mentioned above. This is discussed in detail in \cite{Eichstaedt:2009cn}. Note that in
the current paper we take into account a finite nucleon mass. Thus the following formulas for the
momentum fractions $x_i$ are recovered
from the corresponding formulas in \cite{Eichstaedt:2009cn} by simply replacing
$\sqrt{S}$ by $P_1^-$, cf. Eq.~(\ref{nucleon_large_comp}).
Then we find for the triple-differential hadronic cross section:
\begin{align}
  &\frac{\textrm{d} \sigma_\text{LO}}{\diffd M^2\ \diffd x_F \diffd p_T^2} 
	                   = \int_0^{2\pi} \diffd \phi_\perp 
			   \int_0^{({\vec k}_{\perp})^2_\textrm{max}} \frac{1}{2}
				   \diffd {\vec k}_{\perp}^{\,2}
			       \frac{\pi\ (q_z)_\textrm{max}}{E} \nonumber \\
			   \times& \left| (P_1^-)^2 - \frac{(\hat {\vec p}_{1_\perp})^2
			           (\hat {\vec p}_{2_\perp})^2 }
				  {(x_1)_-^2 (x_2)_+^2 (P_1^-)^2}\right|^{-1}
				  F_\text{LO}((x_1)_-,\hat {\vec p}_{1_\perp},(x_2)_+,\hat {\vec p}_{2_\perp},M^2)
			  \label{master_lo_intrkt}
\end{align}
with
\begin{align}
  &F_\text{LO}((x_1)_-,\hat {\vec p}_{1_\perp},(x_2)_+,\hat {\vec p}_{2_\perp},M^2)
	= \nonumber \\
	  &\sum_i q_i^2\, {\tilde f}_i\left((x_1)_-,\hat {\vec p}_{1_\perp},M^2\right)\,
		       {\tilde f}_{\, \bar i}\left((x_2)_+,\hat {\vec p}_{2_\perp},M^2\right)\,
		       \frac{4 \pi \alpha^2}{9 M^2} \ ,
\end{align}	
\begin{align}
	&(x_1)_- = \nonumber \\
	&\frac{1}{P^-_1}\left(\frac{q^-}{2} 
			 - \frac{\vec k_{\perp} \cdot \vec q_{\perp}}{q^+} + 
	      		\sqrt{ \left( \frac{\vec k_{\perp} \cdot \vec q_{\perp}}{q^+} \right)^2 
			+ \frac{q^-}{q^+} \left(\frac{1}{4} M^2 - \vec k_{\perp}^{\,2} \right) }\right)
											\ , \\
	&(x_2)_+ = \nonumber \\
	&\frac{1}{P_1^-} \left(\frac{q^+}{2}
		         + \frac{\vec k_{\perp} \cdot \vec q_{\perp}}{q^-} 
			 + \sqrt{ \left( \frac{\vec k_{\perp} \cdot \vec q_{\perp}}{q^-} \right)^2 
			 + \frac{q^+}{q^-} \left(\frac{1}{4} M^2 - \vec k_{\perp}^{\,2} \right) }\right)
											\ 
\end{align}	
and
\begin{align}
	\hat {\vec p}_{1_\perp}	&= \frac{1}{2}{\vec q}_{\perp} - {\vec k}_{\perp} \ , \\
	\hat {\vec p}_{2_\perp}	&= \frac{1}{2}{\vec q}_{\perp} + {\vec k}_{\perp} \ , \\
	E &= \sqrt{M^2 + p_T^2 + x_F^2(q_z)_\textrm{max}^2} \\
	q^+ &= E + x_F (q_z)_\textrm{max} \ , \\
	q^- &= E - x_F (q_z)_\textrm{max} \ , \\
	\left|{\vec q}_{\perp}\right| &= p_T \ ,\\
	\vec k_{\perp} \cdot \vec q_{\perp} &= |\vec k_{\perp}| 
	       p_T \cos(\phi_\perp) \ , \\ 
       ({\vec k}_{\perp}^{\,2})_\textrm{max} &= \frac{(M^2 + p_T^2) \frac{M^2}{4}}
                                                  {M^2 + p_T^2(1-\cos^2(\phi_\perp))} \ .
\end{align}

In Sec.~\ref{subsec:lo-results} we will show and compare the results of this approach to the standard 
parton model and the approach with mass distributions as described in Sec.~\ref{subsec:lo-quarkmass}.

\subsection{Quark masses}
\label{subsec:lo-quarkmass}
In Sec.\ \ref{subsec:lo-intrkt} we have extended the standard collinear PDFs towards quark distributions which
also include quark transverse momentum. However, we have kept the masses of the quarks fixed at zero.
Since in light cone coordinates the onshell condition reads 
$ m^2 = p^2 = p^+ p^- - ({\vec p}_{\perp})^2 $, this is equivalent to varying only two of the
three, in principal independent quark momentum components $p^+$, $p^-$ and $p_\perp$. A fully 
unintegrated parton distribution should depend on all three of these components. Therefore, we once
more extend the parton distributions of Sec.\ \ref{subsec:lo-intrkt} by
\begin{equation}
     {\tilde f}_i(x,{\vec p}_{\perp},q^2) \rightarrow {\hat f}_i(x,{\vec p}_{\perp},m^2,q^2) \ .
\end{equation}	
We will present our ansatz for $\hat f$ in Sec.\ \ref{subsec:lo-distr}.
The differential hadronic cross section now becomes
\begin{align}
  \textrm{d}\sigma_\text{LO} =& \fint_0^1 \textrm{d}x_1 \fint_0^1 \textrm{d}x_2
       		           \int \textrm{d}{\vec p}_{1_\perp} \int \textrm{d}{\vec p}_{2_\perp}
			   \int \textrm{d}{m_1^2} \int \textrm{d}{m_2^2} \nonumber \\
		           &\times \sum_i q_i^2 {\hat f}_i(x_1,{\vec p}_{1_\perp},m_1^2,q^2)
				{\hat f}_{\, \bar i}(x_2,{\vec p}_{2_\perp},m_2^2,q^2) \nonumber \\
			   &\times \textrm{d}\hat\sigma(x_1,{\vec p}_{1_\perp},x_2,{\vec p}_{2_\perp},
					        m_1^2,m_2^2,q^2)
			   \ .        \label{hadr_full_xsec_offshell}
\end{align}
The partonic cross section is now given by
\begin{align}
  \textrm{d}\hat \sigma_\text{LO} =& \frac{4 \pi \alpha^2}{9 q^4} 
  \frac{2 q^4 - q^2 (m_1^2 - 6 m_1 m_2 + m_2^2) - (m_1^2 - m_2^2)^2}
  {2\sqrt{(q^2-m_1^2-m_2^2)^2 - m_1^2 m_2^2}} \nonumber \\
	                        \times&\, \delta(M^2 - q^2) \,
				\delta^{(4)}(p_1 + p_2 - q) \, \textrm{d}^4q \, \textrm{d}M^2 \ .
						\label{part_offshell_xsec}
\end{align}				
Note that by assigning different masses $m_1$ and $m_2$ to the quarks, gauge invariance is
{\it not} preserved at the quark-photon vertex. However, the unphysical polarization states of the virtual photon produced in this manner are projected out at the lepton-photon vertex
and thus the entire amplitude is indeed gauge invariant. See Appendix \ref{app:gauge} for
details.

Now we have to perform the same procedure as described in \cite{Eichstaedt:2009cn} to remove the
unphysical solutions for the longitudinal momentum fractions $x_i$. In complete analogy we find
(the details of this calculation can be found in Appendix \ref{appsub:lo-kin})

\begin{widetext}
\begin{align}
  \frac{\textrm{d} \sigma_\text{LO}}{\diffd M^2\ \diffd x_F \diffd p_T^2} 
	                   =& \int_0^{2\pi} \diffd \phi_\perp 
			     \int_0^{({\vec k}_{\perp})^2_\textrm{max}} \frac{1}{2}
			     \diffd ({\vec k}_{\perp})^2
			     \int_0^{(m_1)^2_\textrm{max}} \diffd m_1^2
			     \int_0^{(m_2)^2_\textrm{max}} \diffd m_2^2 \
			     \frac{\pi\ (q_z)_\textrm{max}}{E}  \nonumber \\
			&\times   \left| \left(P_1^-\right)^2 - \frac{ \left[
	           \left(\frac{1}{2}\vec q_{\perp} - {\vec k}_{\perp}\right)^2 + m_1^2 \right]
	           \left[\left(\frac{1}{2}\vec q_{\perp} + {\vec k}_{\perp}\right)^2 + m_2^2\right]}{(x_1)_-^2 (x_2)_+^2 \left( P_1^- \right)^2}\right|^{-1} 
		   F_\text{LO}((x_1)_-,\hat {\vec p}_{1_\perp},m_1^2,
		          (x_2)_+,\hat {\vec p}_{2_\perp},m_2^2,M^2) \nonumber \\
	           &\times
		   \Theta\left(1-(x_1)_-\right)\,\Theta\left((x_1)_-\right)\,
                   \Theta\left(1-(x_2)_+\right)\,\Theta\left((x_2)_+\right) \ .
			  \label{master_lo_offshell}
\end{align}
\end{widetext}
with
\begin{align}
  &F_\text{LO}((x_1)_-,\hat {\vec p}_{1_\perp},m_1^2,(x_2)_+,\hat {\vec p}_{2_\perp},m_2^2,M^2) \nonumber \\
	&= \sum_i q_i^2\, {\hat f}_i\left((x_1)_-,\hat {\vec p}_{1_\perp},m_1^2,M^2\right)\,
		       {\hat f}_{\, \bar i}\left((x_2)_+,\hat {\vec p}_{2_\perp},m_2^2,M^2\right) \nonumber \\
	&\times \frac{4 \pi \alpha^2}{9 M^4} 
	\frac{2 M^4 - M^2 (m_1^2 - 6 m_1 m_2 + m_2^2) - (m_1^2 - m_2^2)^2}
	{2\sqrt{(q^2-m_1^2-m_2^2)^2 - m_1^2 m_2^2}}
\end{align}	

\begin{align}
	&(x_1)_- = \frac{1}{P^-_1} \left(\frac{q^-}{2} 
			 - \frac{\vec k_{\perp} \cdot \vec q_{\perp}}{q^+} 
			 - \frac{m_2^2-m_1^2}{2 q^+} \right. \nonumber \\
			  &+ \left.
	      		\sqrt{ \left( \frac{\vec k_{\perp} \cdot \vec q_{\perp}}{q^+} 
				     +\frac{m_2^2-m_1^2}{2 q^+} \right)^2 
	       	    + \frac{q^-}{q^+} \left(\frac{1}{4} M^2 - \vec k_{\perp}^2 
			                    -\frac{m_1^2+m_2^2}{2} \right) }\right) \ ,
\end{align}					    
\begin{align}
	&(x_2)_+ = \frac{1}{P^-_1}\left(\frac{q^+}{2} 
			 + \frac{\vec k_{\perp} \cdot \vec q_{\perp}}{q^-} 
			 + \frac{m_2^2-m_1^2}{2 q^-} \right. \nonumber \\
			 &+ \left.
	      		\sqrt{ \left( \frac{\vec k_{\perp} \cdot \vec q_{\perp}}{q^-} 
				     +\frac{m_2^2-m_1^2}{2 q^-} \right)^2 
	       	    + \frac{q^+}{q^-} \left(\frac{1}{4} M^2 - \vec k_{\perp}^2 
			                    -\frac{m_1^2+m_2^2}{2} \right) }\right) \ 
\end{align}
and
\begin{align}
	\hat {\vec p}_{1_\perp}	&= \frac{1}{2}{\vec q}_{\perp} - {\vec k}_{\perp} \ , \\
	\hat {\vec p}_{2_\perp}	&= \frac{1}{2}{\vec q}_{\perp} + {\vec k}_{\perp} \ , \\
	E &= \sqrt{M^2 + p_T^2 + x_F^2(q_z)_\textrm{max}^2} \\
	q^+ &= E + x_F (q_z)_\textrm{max} \ , \\
	q^- &= E - x_F (q_z)_\textrm{max} \ , 
\end{align}
\begin{align}      
	\left|{\vec q}_{\perp}\right| &= p_T \ ,\\
	\vec k_{\perp} \cdot \vec q_{\perp} &= |\vec k_{\perp}| 
	       p_T \cos(\phi_\perp) \ ,\\ 
	({\vec k}_{\perp})^2_\textrm{max} &= \frac{(M^2 + p_T^2) \frac{M^2}{4}}
		{M^2 + p_T^2(1-\cos^2(\phi_\perp))} \ , \\
	(m_1)^2_\textrm{max} &= 2\vec k_{\perp} \cdot \vec q_{\perp} + q^+q^-
				-\sqrt{4 q^+ q^-\left(\vec k_{\perp} 
				       +\frac{1}{2}\vec q_{\perp}\right)^2 } \ , \\
	(m_2)^2_\textrm{max} &= -2\vec k_{\perp} \cdot \vec q_{\perp} + m_1^2 + q^+q^- \nonumber \\
				&\ -\sqrt{4 q^+ q^- m_1^2 
			       + 4 q^+ q^-\left(\vec k_{\perp} -\frac{1}{2}\vec q_{\perp}\right)^2 } \ .
\end{align}

In Sec.~\ref{subsec:lo-results} we show the results of this calculation as well as of the approaches in Secs.~\ref{subsec:lo-spm} and \ref{subsec:lo-intrkt}.

\subsection{Distributions}
\label{subsec:lo-distr}

The Bjorken limit and the corresponding infinite momentum frame, in which the standard parton model
is well defined and derived from LO pQCD, is an
idealization of real experiments. There the nucleons will always move with some finite
momentum and thus the partons inside the nucleons will interact before the collision.
These interactions will generate momentum components, which are neglected in the (purely
collinear) standard parton model, namely momentum components perpendicular to
the beam line, $\vec p_{1_\perp},\vec p_{2_\perp}$, as well as the small light-cone components
$p_1^+,p_2^-$. The latter translate to non-vanishing quark masses. 

Note that the factorization into hard (subprocess) and soft (PDFs) physics 
is proven in the transverse case at least for partons with low transverse 
momentum in \cite{Ji:2004xq}. For the case of mass distributions for the quarks we {\it assume}
this factorization.

\subsubsection{Transverse momentum distributions}
\label{subsubsec:tmd}

In Sec. \ref{subsec:lo-intrkt} we introduced transverse momentum dependent parton distribution
functions ${\tilde f}_i$. They are functions of the light-cone momentum fraction $x_i$, 
the transverse 
momentum ${\vec p}_{i_\perp}$ and the hard scale of the subprocess $q^2$. However, the general form of these
functions is unknown. Known rather well are the longitudinal PDFs.
Since data of DY pair production are compatible with a Gaussian form of the $p_T$ spectrum up 
to a certain $p_T$ \cite{Webb:2003bj,Webb:2003ps}, we assume factorization of the longitudinal 
and the transverse part of ${\tilde f}_i$ and make the common ansatz 
\cite{Wang:1998ww,Raufeisen:2002zp,D'Alesio:2004up}
\begin{equation}
   {\tilde f}_i(x,{\vec p}_{\perp},q^2) = f_i(x,q^2) \cdot f_{\perp}({\vec p}_{\perp})\ .
   \label{ftilde}
\end{equation}	
Here $f_i$ are the usual longitudinal PDFs and for $f_{\perp}$ we choose a Gaussian form,
\begin{equation}
	f_{\perp}({\vec p}_{\perp}) = \frac{1}{4\pi D^2} 
				\exp\left( -\frac{({\vec p}_{\perp})^2} {4 D^2}\right) \ .
	\label{fperp}
\end{equation}	
The width parameter $D$ is connected to the average squared transverse momentum via
\begin{equation}
	\left< ({\vec p}_{\perp})^2 \right> = 
	\int \diffd {\vec p}_{\perp} ({\vec p}_{\perp})^2 f_{\perp}({\vec p}_{\perp})
	= 4 D^2 \ 
\end{equation}
and it has to be fitted to the available data.

\subsubsection{Mass distributions}
\label{subsubsec:spectralfunc}
In Sec.\ \ref{subsec:lo-quarkmass} we have extended our model by also distributing quark masses.
This approach is motivated by studies of quark correlations and quark spectral functions,
see for example \cite{Froemel:2001iy,Fromel:2007rc}. Note that in such a mass distribution approach 
one effectively parametrizes higher twist effects, i.e.\ effects
which are suppressed by inverse powers of the hard scale $M$. These higher twist contributions
should become particurlaly important in the description of DY pair production in the region of small energies 
(and thus small $M$), which is one aim of our studies. 

For the fully unintegrated parton distributions ${\hat f}_i$ we make the ansatz,
\begin{equation}
	{\hat f}_i(x,{\vec p}_{\perp},m^2,q^2) = f_i(x,q^2) \cdot f_{\perp}({\vec p}_{\perp})
						 \cdot A(p) \ .
\end{equation}
Again $f_i$ are standard PDFs and  $f_{\perp}$ are the transverse momentum distributions of Sec.~\ref{subsubsec:tmd}. Since the distribution of longitudinal parton momenta is determined by the argument of 
the PDFs $x \sim p^+$, we now allow for a distribution of the remaining degree of freedom, i.e. the small
component $p^-$, by writing
\begin{equation}
	A(p)\, \diffd p^- = \frac{1}{N} \frac{\hat \Gamma(m^2)}
                                 {\left(p^- - \frac{p_T^2}{p^+}\right)^2 + \hat \Gamma^2(m^2)}\, 
				 \diffd p^- \ ,
	\label{spec-func-pminus}
\end{equation}
with $m^2 = p^2$. Rewriting in terms of $m^2$ yields
\begin{equation}
	A(p)\, \diffd m^2 = \frac{1}{N} \frac{\hat\Gamma(m^2) p^+}{m^4 + (p^+)^2 \hat\Gamma^2(m^2)}\, \diffd m^2 \ . \label{spec-func}
\end{equation}
We choose a non-constant width such that the quark can never become heavier than its parent
nucleon,
\begin{equation}
	\hat\Gamma(m^2) = \frac{m_N^2 - m^2}{m_N^2} \, \Gamma \ ,
	\label{spec-func-mod-width}
\end{equation}
for $0 < m^2 < m_N^2$ and $\hat\Gamma(m^2) = 0$ otherwise, where
$\Gamma$ is a free parameter.
The factor $\frac{1}{N}$ normalizes the spectral function such that
\begin{equation}
	\int_0^\infty \diffd m^2 A(p) = \int_0^{m_N^2} \diffd m^2 A(p) = 1\ .
	\label{spec-func-norm}
\end{equation}	
\begin{figure}[H]
	\centering
	\includegraphics[keepaspectratio,angle=-90,width=0.45\textwidth]{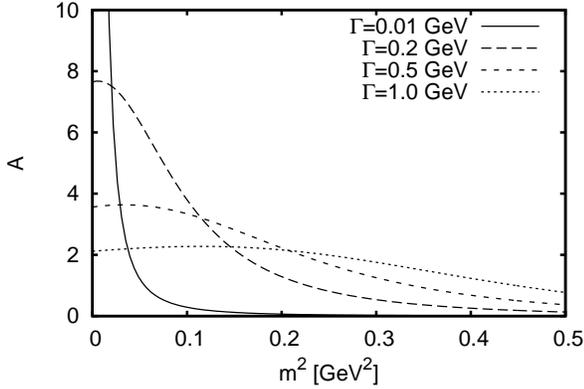}
	\caption{Spectral function $A$ plotted for different values of the width $\Gamma$. Everywhere $p^+=0.5$ GeV.}
	\label{fig:specfunc}	 
\end{figure}	
In Fig.~\ref{fig:specfunc} we plot $A(p)$ as a function of $m^2$ for fixed $p^+=0.5$ GeV for different values of the width $\Gamma$. Note that for not too small $\Gamma$ the region near $m^2=0$ is heavily suppressed compared to, for example, $\Gamma=0.01$ GeV.

\subsection{Results of the LO calculation}
\label{subsec:lo-results}

In this section we present and compare our results for the different LO 
approaches of Secs.~\ref{subsec:lo-spm} - \ref{subsec:lo-quarkmass}. The data are from the NuSea Collaboration (E866) 
\cite{Webb:2003bj,Webb:2003ps} and from FNAL-E439 \cite{Smith:1981gv}. 
For the collinear PDFs we used the \emph{leading order} MSTW2008LO68cl parametrization
\cite{Martin:2009iq} available through the LHAPDF library, version 5.8.4 \cite{Whalley:2005nh}.

\subsubsection{E866 -- $p_T$ spectra}
\label{subsubsec:E866_triple_lo}

Experiment E866 measured continuum dimuon production in pp collisions at $S \approx 1500$ GeV$^2$.
The triple-differential cross section as given by the E866 collaboration is
\begin{equation}
	E\frac{\diffd \sigma}{d^3p} \equiv \frac{2E}{\pi \sqrt{S}} \frac{\diffd \sigma}
							     	   {\diffd x_F \diffd p_T^2} \ ,
        \label{E866_triple_xsec}								   
\end{equation}	
where an average over the azimuthal angle has been taken and where $E$ is the energy of the DY pair,
cf.\  Eq.\ (\ref{DYenergy}). The data are given in several
bins of $M$, $x_F$ and $p_T$ and for every datapoint the average values $\left< M \right>$,
$\left< x_F \right>$ and $\left< p_T \right>$ are given. Since our schemes provide 
Eqs.~(\ref{master_lo_intrkt}) and (\ref{master_lo_offshell}) we calculate the quantity of 
Eq.~(\ref{E866_triple_xsec}) for every datapoint using these averaged values and 
then perform a simple average in each $M^2$ bin:
\begin{align}
	\frac{2E}{\pi \sqrt{S}} \frac{\diffd \sigma}{\diffd x_F \diffd p_T^2}
       &\rightarrow\frac{2E}{\pi \sqrt{S}} \int_{M^2\textrm{-bin}}
         \frac{\diffd \sigma}{\diffd M^2 \diffd x_F \diffd p_T^2} \diffd M^2 \nonumber \\
       &\approx 	\frac{2E}{\pi \sqrt{S}} \Delta M^2
                \frac{\diffd \sigma}{\diffd M^2 \diffd x_F \diffd p_T^2}\left(\left< M \right>,
									      \left< x_F \right>,
									      \left< p_T \right>
									\right)
									  \ ,
       \label{eq:exp_triple_xsec}
\end{align}
where
\begin{equation}
      E = \sqrt{\vrule height 10pt width 0pt 
	         \left< M \right>^2 + \left< p_T \right>^2 + 
	         \left< x_F \right>^2 \left<(q_z)_\textrm{max}\right>^2 }
		 \label{DYenergy}
\end{equation}
and $\Delta M^2 = M_\textrm{max}^2 - M_\textrm{min}^2$ with $M_\textrm{max}$ ($M_\textrm{min}$)
the upper (lower) limit of the bin. 	

\begin{figure}[H]
	\centering
	\includegraphics[angle=-90,keepaspectratio,width=0.45\textwidth]
                        {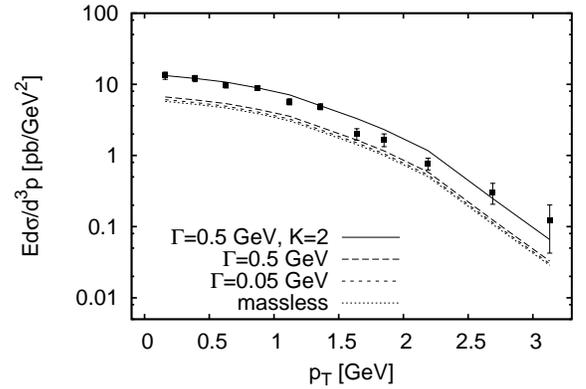}
	\caption{$p_T$ spectrum obtained at LO from the massless and the mass distribution approach with different values of $\Gamma$. Everywhere $D=0.5$ GeV. The solid line is the mass distribution approach for $\Gamma=0.5$ GeV multiplied by a factor $K=2$. Data are from E866 binned with $4.2$ GeV $< M < 5.2$ GeV, $-0.05 < x_F < 0.15$.  Only statistical errors are shown.}
	\label{fig:E866_triple_4.2_lo}	 
\end{figure}

We plot the results for the two different approaches of Secs.~\ref{subsec:lo-intrkt}
and \ref{subsec:lo-quarkmass} in Fig.~\ref{fig:E866_triple_4.2_lo}.
The different dashed lines represent the massless and the mass distribution approach for different values of $\Gamma$ and they all agree within $\approx20\%$. Note, however, that
with increasing $\Gamma$ the calculated cross section is slightly enhanced. Everywhere a value of $D=0.5$ GeV for the transverse momentum dispersion is chosen. With this choice of the parameter $D$ the shape of the spectra is described rather well. However, in both approaches the absolute size of the cross section is underestimated:
we have to multiply the result of the mass distribution approach for $\Gamma=0.5$ GeV by $K=2$ to fit the data (solid line).

\subsubsection{E439 - $M$ spectrum}
\label{subsubsec:E439_lo}

Experiment E439 measured dimuon production in pW collisions at $S \approx 750$ GeV$^2$.
The double differential cross section,
\begin{equation}
	\frac{\diffd \sigma}{\diffd M \diffd x_F'}          \label{E439_double_xsec}								   \ ,
\end{equation}	
has been given at a fixed $x_F' = 0.1$.

\begin{figure}[H]
	\centering
	\includegraphics[angle=-90,keepaspectratio,width=0.45\textwidth]
                        {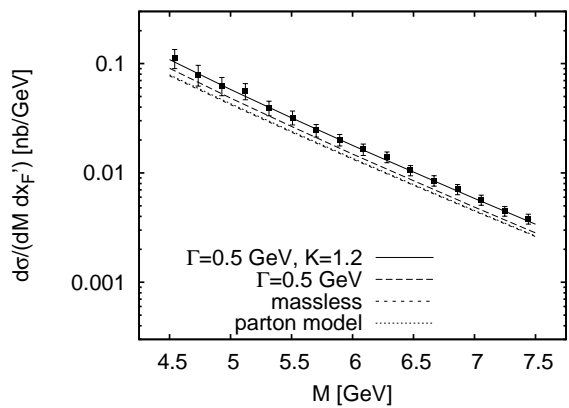}
	\caption{$M$ spectrum obtained from the standard parton model, $k_T$ approach (massless) and mass distribution approach ($D=0.5$ GeV for the latter two). The solid line is the result of the mass distribution approach multiplied by a factor $K=1.2$. Data are from E439 with $x_F' = 0.1$. Only statistical errors are shown.}
	\label{fig:E439_lo_double}	 
\end{figure}

As before we begin with Eqs.~(\ref{master_lo_intrkt}) and (\ref{master_lo_offshell}) and calculate
the quantity 
Eq.~(\ref{E439_double_xsec}) by integrating over $p_T^2$ and performing a simple transformation
from $x_F$ to $x_F'$:

\begin{align}
	\frac{\diffd \sigma}{\diffd M \diffd x_F'}
 	=&\int_0^{(p_T)^2_\textrm{max}} \diffd p_T^2
	      \frac{\diffd \sigma}{\diffd M \diffd x_F' \diffd p_T^2} \nonumber \\
	=&\int_0^{(p_T)^2_\textrm{max}} \diffd p_T^2\ 
	 2 M \left(1-\frac{M^2}{S}\right) \nonumber \\
        &\times \frac{\diffd \sigma}{\diffd M^2 \diffd x_F \diffd p_T^2} 
           \left(M, x_F=x_F'\left(1-\frac{M^2}{S}\right) \right) \ .
\end{align}
Since the experiment was done on tungsten we calculate the cross section for pp and pn and average accordingly (74 protons and 110 neutrons).
We compare the results in Fig.~\ref{fig:E439_lo_double}. Everywhere $D=0.5$ GeV
except for the simple parton model, which has no $k_T$ distribution. The lowest
curve represents the indistinguishable results of the standard parton model (Sec. \ref{subsec:lo-spm}) and of the (massless) initial $k_T$ approach (Sec. \ref{subsec:lo-intrkt}). The result of the mass distribution approach (Sec. \ref{subsec:lo-quarkmass}) for 
$\Gamma=0.5$ GeV (long dashed) is somewhat larger but still underestimates the data: The solid line is the result of this mass distribution approach multiplied by a factor
$K=1.2$ and it fits the data very well.

\subsection{Conclusion for the LO calculation}
\label{subsec:lo-concl}
In this section we have presented and compared three different approaches
to DY pair production at LO in the partonic subprocess. The
standard parton model approach describes invariant mass spectra only
up to a $K$ factor and it cannot describe transverse momentum ($p_T$) spectra.
The latter issue was addressed in the initial $k_T$ approach.
We found that with a suitable choice of an initial $k_T$
distribution the DY $p_T$ spectra can be described very well, however, still only
up to a $K$ factor. The mass distribution approach can improve the picture somewhat,
but the enhancement of the calculated cross sections is too small to describe
the data. Still an a priori undetermined multiplicative factor $K$ is needed
to reproduce the measured cross sections. This finding has triggered the NLO calculations,
which will be presented in Sec.~\ref{sec:nlo}.

\section{Next-to-leading order Drell-Yan}
\label{sec:nlo}
Building on the LO ($O(\alpha_s^0)$) calculations of Sec. \ref{sec:lo} we here present
an approach, which incorporates all relevant DY pair production processes up to $O(\alpha_s)$.
We will show that by introducing initial $k_T$ as well as quark mass distributions we
can soften the divergences at low $p_T$ of the NLO processes and describe $p_T$ and
$M$ spectra without the need for a $K$ factor.

\begin{figure}[htbp]
	\centering
	\includegraphics[keepaspectratio,width=0.22\textwidth]{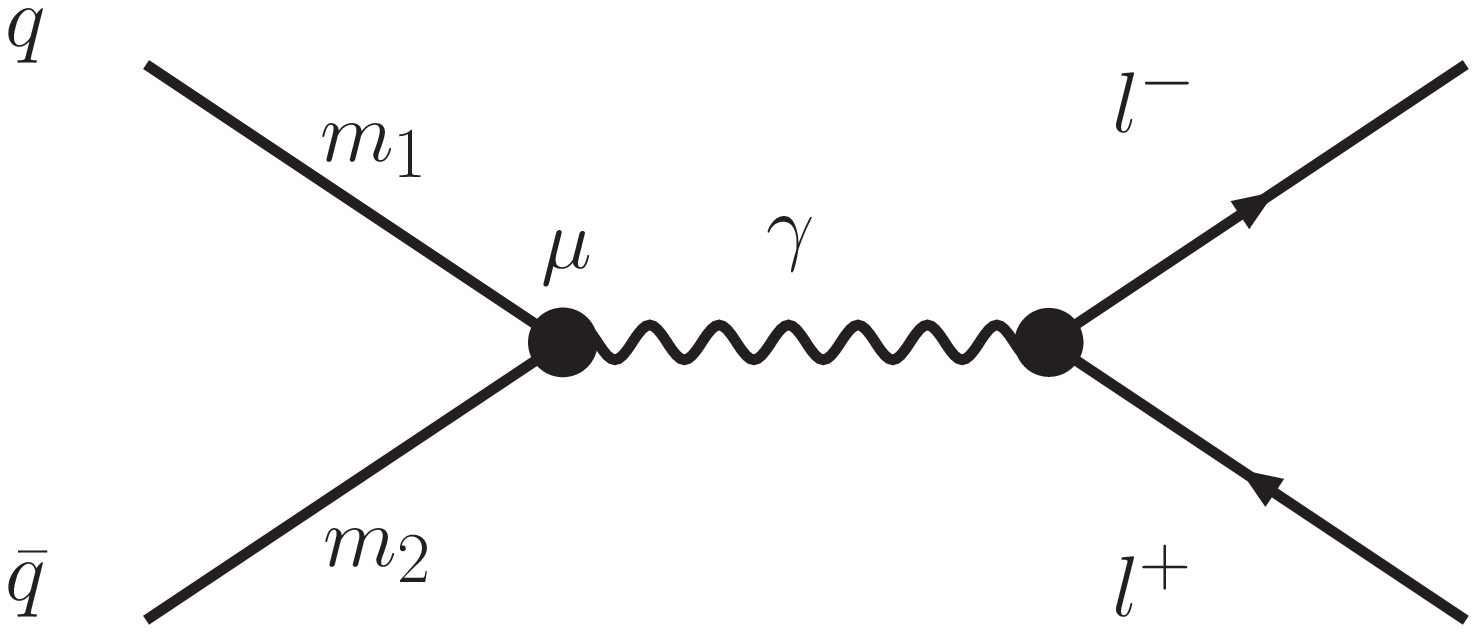}
	\includegraphics[keepaspectratio,width=0.22\textwidth]{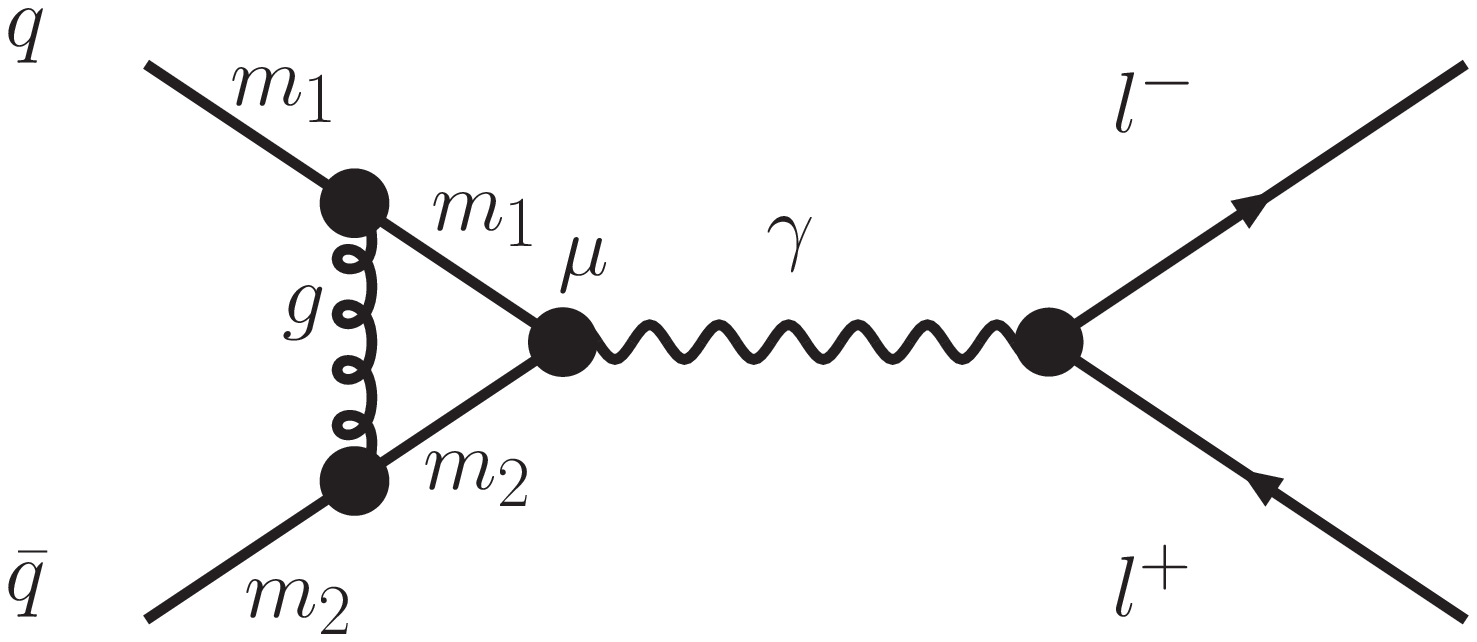}
	\caption{Leading order and vertex correction processes to DY production. Note that only the interference of the two processes contributes at NLO.}
	\label{fig:vertex-correction}	 
\end{figure}

\begin{figure}[htbp]
	\centering
	\includegraphics[keepaspectratio,width=0.22\textwidth]{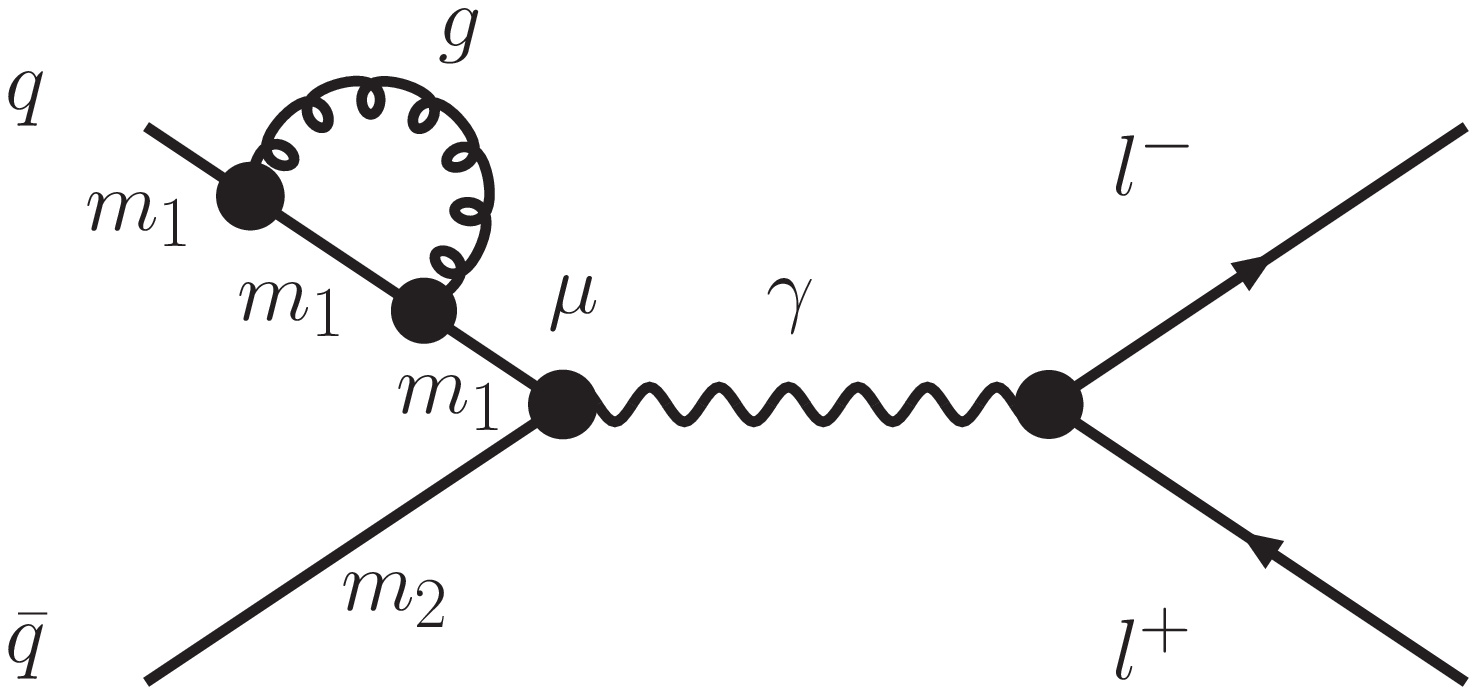}
	\includegraphics[keepaspectratio,width=0.22\textwidth]{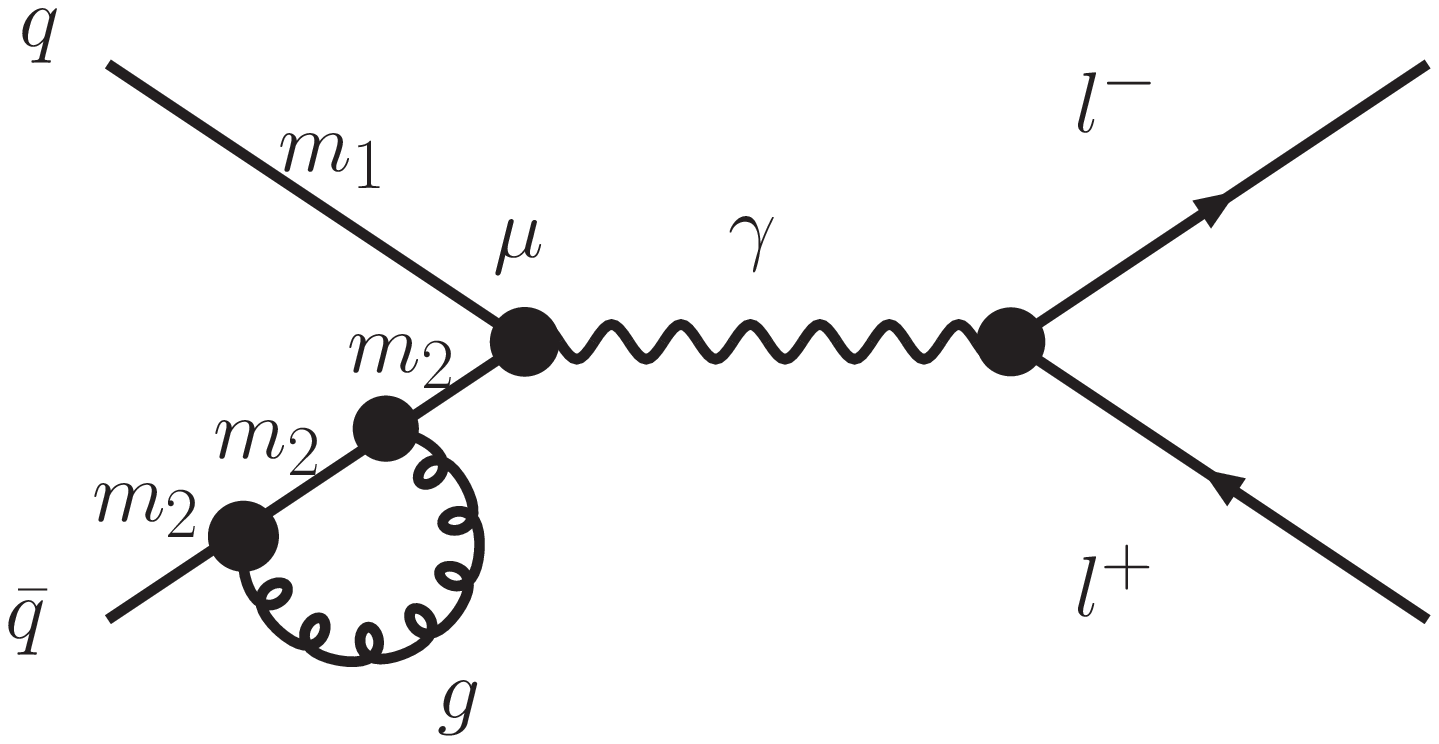}
	\caption{Wave function renormalization processes for DY production at NLO.}
	\label{fig:wave-renorm}	 
\end{figure}

\begin{figure}[htbp]
	\centering
	\includegraphics[keepaspectratio,width=0.23\textwidth]{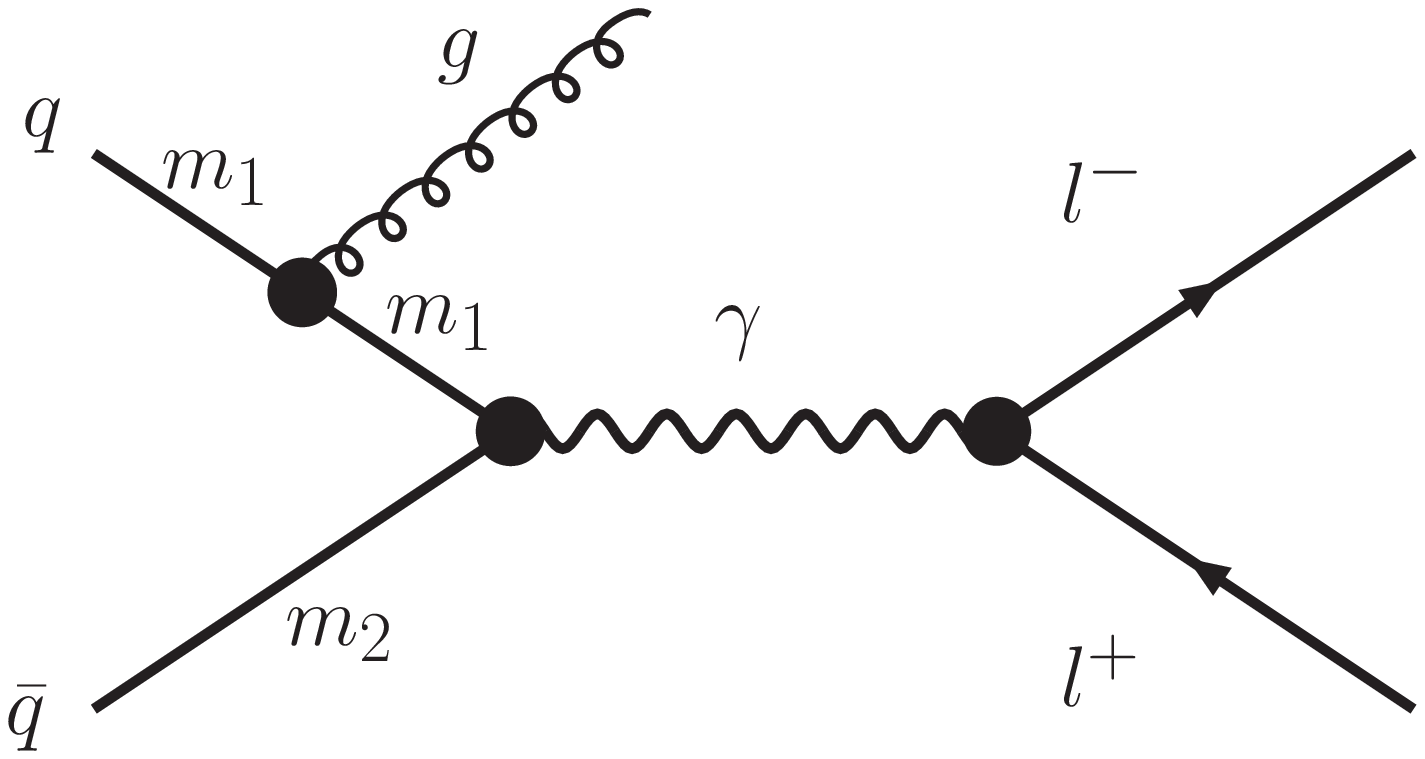}
	\includegraphics[keepaspectratio,width=0.19\textwidth]{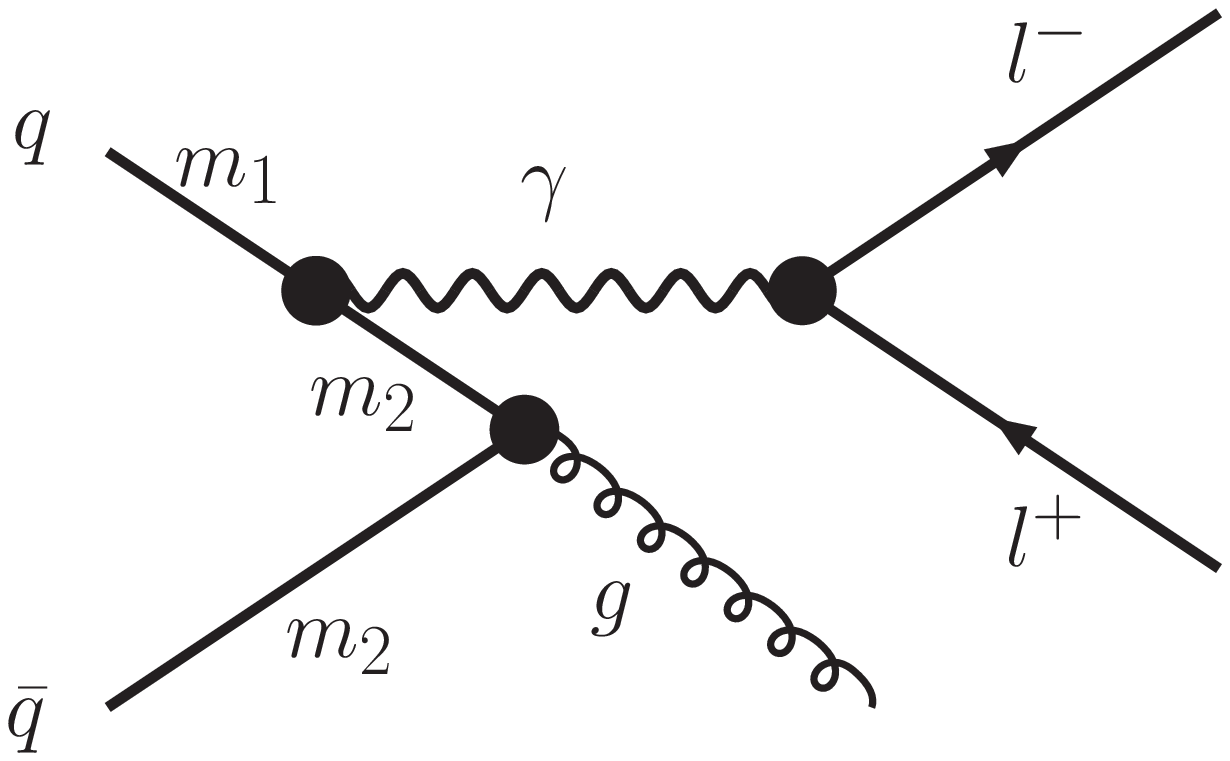}
	\caption{Gluon bremsstrahlung processes for DY production at NLO.}
	\label{fig:gluon-brems}	 
\end{figure}

\begin{figure}[htbp]
	\centering
	\includegraphics[keepaspectratio,width=0.24\textwidth]{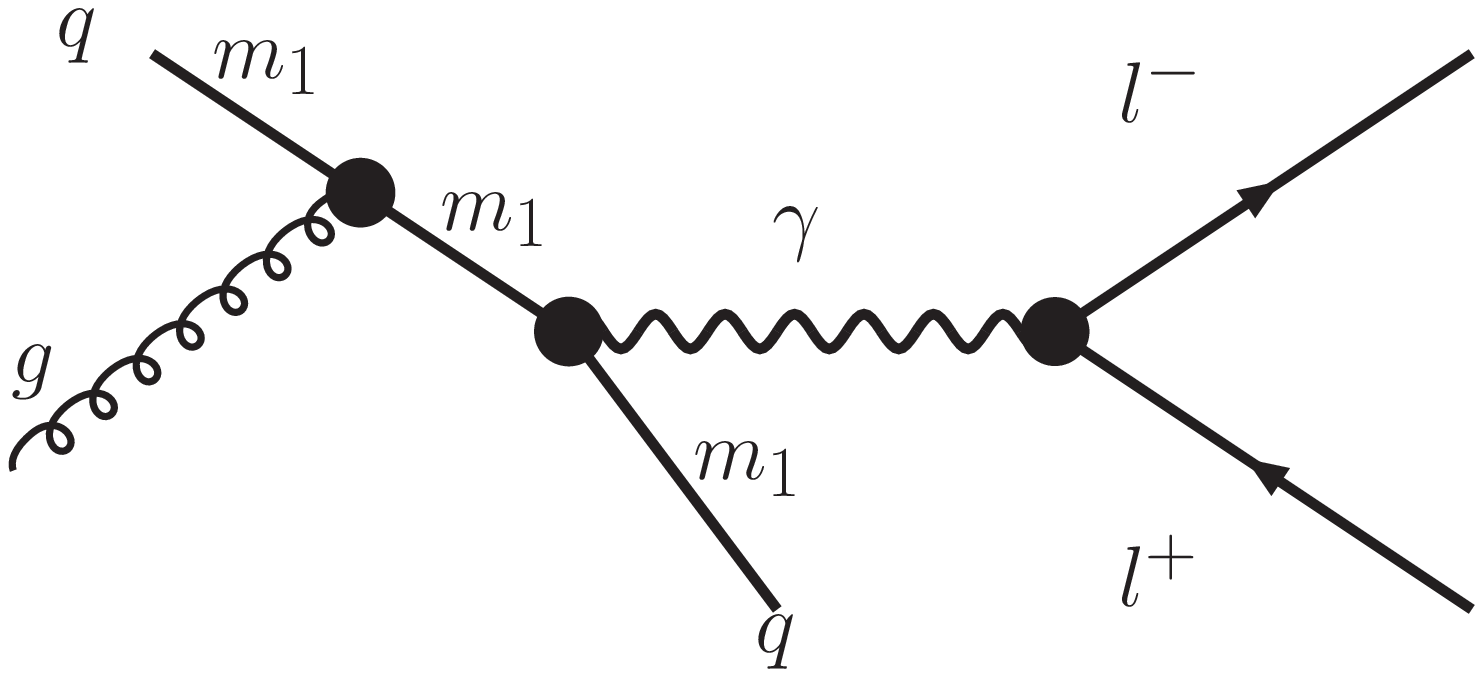}
	\includegraphics[keepaspectratio,width=0.18\textwidth]{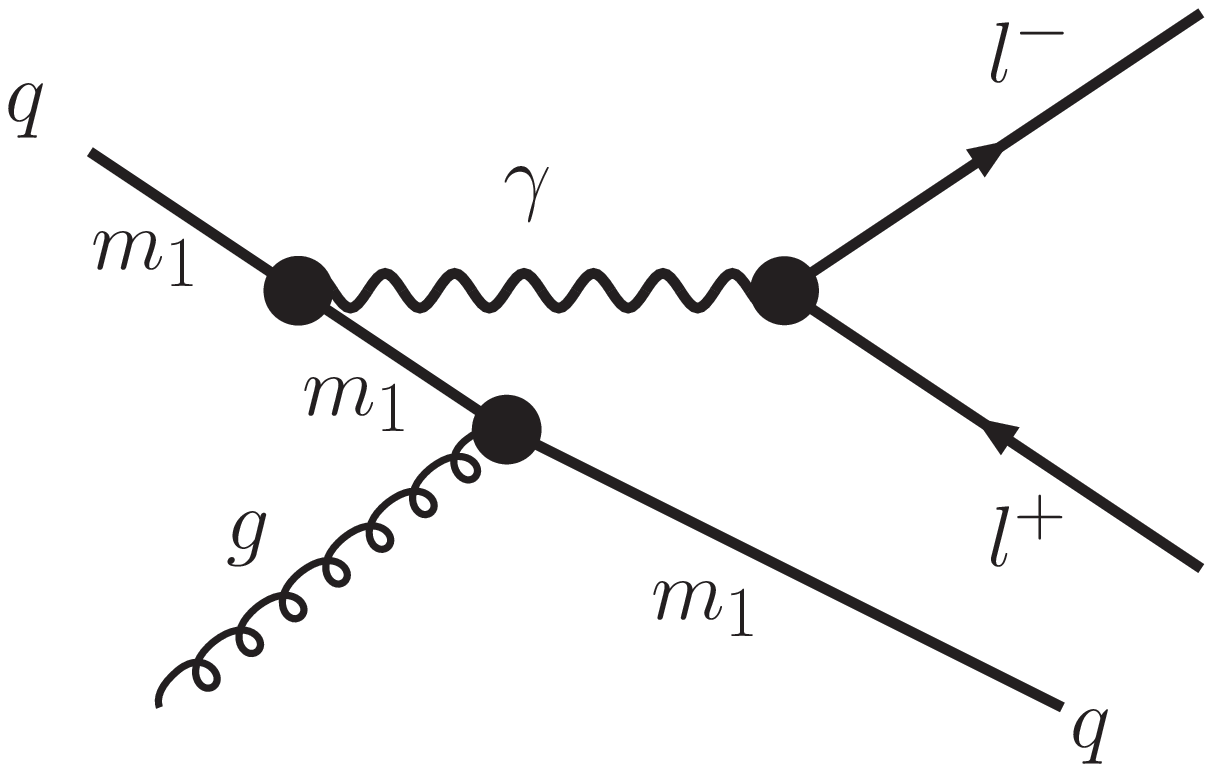}
	\caption{Gluon Compton scattering processes for DY production at NLO.}
	\label{fig:gluon-compton}	 
\end{figure}

In addition to the LO the following processes contribute to DY pair production to $O(\alpha_s)$. First we have the vertex correction diagram of Fig.~\ref{fig:vertex-correction} (right). This process alone does contribute at order $\alpha_s^2$, however,
due to identical initial and final states it interferes with the LO process of 
Fig.~\ref{fig:vertex-correction} (left) and the interference is of order $\alpha_s$. The
same is true for the wave function renormalization processes of Fig.~\ref{fig:wave-renorm}.
Then there is gluon bremsstrahlung, where either the quark or the antiquark emits a real gluon
before annihilating, see Fig.~\ref{fig:gluon-brems}. Somewhat different is gluon Compton 
scattering
since there a gluon and a quark/antiquark fuse before or after emitting the virtual photon,
see Fig.~\ref{fig:gluon-compton}.

It is important to note at this point that our initial state quarks have different masses
$m_1$ and $m_2$ when we evaluate the matrix elements of the loop correction and
the bremsstrahlung diagrams, just as in the LO calculation in Sec.~\ref{subsec:lo-quarkmass}.
However, we keep the quark mass fixed at the quark-gluon vertex, see Figs.~\ref{fig:vertex-correction}, \ref{fig:wave-renorm} and \ref{fig:gluon-brems}. This guarantees that also in the strong sector gauge invariance is preserved, as we show in Appendix \ref{app:gauge}. For the same reason all gluons are treated as massless. 

In the case of gluon Compton scattering we keep the quark mass fixed at every vertex, see Fig.~\ref{fig:gluon-compton}, for the following reasons: In principle the final state quark is supposed to be ``free'' and thus one would assign to it a mass $m_1=0$. To preserve gauge invariance the quark mass at the gluon
vertex must not change and thus the exchange quark in the right diagram in Fig.~\ref{fig:gluon-compton} 
would have to be also massless. This, however, immediately generates an infrared (IR) divergence, as will be illustrated in Sec.~\ref{subsec:ir-div}. Therefore, we assign a mass $m_1$ to the entire quark line.

\subsection{Vertex correction}
\label{subsec:nlo-vc}

\subsubsection{Form factors}
\label{subsubsec:nlo-vc-formf}

The vertex correction process of Fig.~\ref{fig:vertex-correction} together with the wave function
renormalization of Fig.~\ref{fig:wave-renorm} modifies the bare quark-photon vertex:
\begin{equation}
\gamma^\mu \rightarrow \Gamma^\mu = \gamma^\mu + \delta\Gamma^\mu(\alpha_s) + O(\alpha_s^2) \ .
\end{equation}
From general principles one now can decompose the correction into
\begin{align}
	\delta\Gamma^\mu = A \cdot \gamma^\mu + B \cdot (p_1-p_2)^\mu \ + C \cdot (p_1+p_2)^\mu,
\end{align}	
where $A$, $B$ and $C$ are functions of $q^2$, $m_1$ and $m_2$. However, in DY pair-production 
the term $(p_1+p_2)^\mu$ does not contribute, as we show in detail in Appendix \ref{app:gauge-em}.
Therefore, from now on we neglect this term.

The Gordon identity \cite{Peskin:1995ev} for the case of different masses $m_1$ and $m_2$ reads:
\begin{align}
	&\bar v (p_2,m_2) \gamma^\mu u(p_1,m_1) = \nonumber \\
        &\bar v (p_2,m_2) \left( \frac{(p_1-p_2)^\mu}{m_1+m_2} 
			      + \frac{i \sigma^{\mu\nu}q_\nu}{m_1+m_2} \right) u(p_1,m_1)
\end{align}	
and thus we find:
\begin{align}
	\Gamma^\mu = \gamma^\mu \cdot \left( 1 + A + (m_1+m_2)B \right)
		   + \frac{i \sigma^{\mu\nu}q_\nu}{m_1+m_2} \cdot (-(m_1+m_2)B) \ .
\end{align}
We can now identify the well known form factors $F_1$ and $F_2$ \cite{Peskin:1995ev}:
\begin{align}
	F_1 &= 1 + A + (m_1+m_2)B \ , \\
	F_2 &= -(m_1+m_2)B\ .
\end{align}
The calculation of $A$ and $B$ is tedious, but straightforward. We will only need the real parts of $A$ and
$B$, see Eq. (\ref{T_VC}):
\begin{widetext}
\begin{align}
  &\text{Re}(A)=  \nonumber \\ &\frac{\alpha_s}{4\pi} \cdot \text{Re}
      	   \left(-3 + \log(1-v^2) 
		 - \frac{1}{2}\left(\log(1-\alpha^2) + 
				    \log(1-(\alpha+\phi)^2) \right)
		 - \frac{\alpha}{2} \left(\log\frac{\alpha+1}{\alpha-1}
			 		 +\log\frac{\alpha+\phi+1}{\alpha+\phi-1} \right)
		 - \frac{\phi}{2} \log\frac{\alpha+\phi+1}{\alpha+\phi-1} + I_1 \right)
   	\label{vc-A}	 \ , 
\end{align}	
with
\begin{align}
	I_1 =& -1 + \frac{\phi}{2}\log\frac{1-v^2}{1-v^2-2\phi}
              -\frac{4v^2+3\phi+\tau+\frac{\phi^2}{2}}{2\alpha+\phi}
               \left(\log\frac{\alpha-1}{\alpha+1}
	            +\log\frac{\alpha+\phi-1}{\alpha+\phi+1}\right)
              + 2(1+v^2+\phi) \cdot I_3  
	      - 2\log(\kappa) + \frac{3}{2}\log\left( \frac{m_1^2}{m_2^2} \right)
	      \ ,
\end{align}	      
\begin{align}
        I_3 =& \frac{1}{2r}\left[\log(\kappa) \log\frac{r+1-\frac{\phi}{2}}{r-1-\frac{\phi}{2}}
                       +\log\frac{v^2-1}{2r}\log\frac{r+1-\frac{\phi}{2}}{r-1-\frac{\phi}{2}}
	               -\frac{1}{2}\log^2(r+1-\frac{\phi}{2})
	               +\frac{1}{2}\log^2(r-1-\frac{\phi}{2}) \nonumber \right. \\
		       &\left. +\text{Li}_2\left(\frac{r+1-\frac{\phi}{2}}{2r}\right)
		      -\text{Li}_2\left(\frac{r-1-\frac{\phi}{2}}{2r}\right)
		       +\log^2\left(\sqrt{\frac{1-v^2}{1-v^2-2\phi}}\right)
		       +\log\left(\sqrt{\frac{1-v^2}{1-v^2-2\phi}}\right)\log(\kappa)\right] \ ,
\end{align}
\begin{align}
  &\text{Re}(B)=  \frac{\alpha_s}{4\pi} \frac{1}{m_1} \cdot \text{Re}
      	    \left[\frac{\frac{\tau}{4}(1-\alpha) + \frac{v^2-1}{2}}{2\alpha+\phi}
	          \left(\log\frac{\alpha-1}{\alpha+1}
		              +\log\frac{\alpha+\phi-1}{\alpha+\phi+1}\right)
		  - \frac{\tau}{4}\log\frac{\alpha+\phi+1}{\alpha+\phi-1} \right]
	\label{vc-B}
\end{align}

\end{widetext}
and
\begin{align}
        v =& \sqrt{1-\frac{4 m_1^2}{q^2}} \ , \\
	\tau =& (1-v^2) \cdot \left(1-\frac{m_2}{m_1}\right) \ ,
\end{align}
\begin{align}
	\phi =& \frac{1}{2} \cdot \left(1-\frac{m_2^2}{m_1^2}\right) \cdot (1-v^2) \ ,\\
	\alpha =& -\frac{\phi}{2} + \sqrt{\phi + v^2 + \frac{\phi^2}{4}} \ ,
\end{align}
\begin{align}
	r =& \sqrt{\phi + v^2 + \frac{\phi^2}{4}} \ ,\\
	\kappa =& \frac{\lambda^2}{m_1^2} \label{vc-kappa}\ . 
\end{align}	
$\text{Li}_2$ is the Dilogarithm or Spence function.

We have checked and confirmed that in the limit of equal masses $m_2 \rightarrow m_1$ the well known formula for $F_1(q^2,m^2)$ \cite{Muta:1998vi,Berends:1973tz} is
recovered. We note that also for unequal masses the ultraviolet (UV) divergences of the loops, displayed in Figs.~\ref{fig:vertex-correction}, \ref{fig:wave-renorm}, cancel. This is by virtue of the Ward-Takahashi identities,
which are fulfilled at the quark-gluon vertices, since there by construction the mass of the quark does not
change. However, for $m_1 \neq m_2$ one finds that gauge invariance
is broken at the quark-photon vertex (the full amplitude is gauge invariant, see Appendix \ref{app:gauge-em}).
The gauge dependence of the quark-photon vertex results in a finite renormalization
of the charge, which cannot be canceled by gauge invariant counterterms. Therefore one finds:
\begin{align}
	\underset{q^2 \rightarrow 0}{\lim} F_1(q^2,m_1^2,m_2^2) \neq 1 \ .
\end{align}	
On the other hand, $q^2=M^2$ sets the hard scale
in DY pair production and thus our model should only be applied at reasonably
large $q^2$. A sensible lower limit would be $q^2>1 \text{ GeV}^2$. Thus we probe $F_1$ far away from $q^2=0$ and we show in Appendix \ref{app:F1} that for these physically interesting $q^2$
the influence of the renormalized charge is negligible.

\subsubsection{Soft gluon divergence}
\label{subsubsec:nlo-vc-soft}

To obtain Eqs.\ (\ref{vc-A}-\ref{vc-kappa}) we have assigned to the gluon 
a mass $\lambda$ which serves as an IR regulator
in the loop integral. According to the theorems by Kinoshita-Poggio-Quinn 
\cite{Kinoshita:1962ur,Kinoshita:1975ie,Poggio:1976qr,Sterman:1976jh} and
Kinoshita-Lee-Nauenberg \cite{Kinoshita:1962ur,Lee:1964is} the (IR) divergence
of the loop integral cancels against a similar divergence of the bremsstrahlung
processes in Fig.~\ref{fig:gluon-brems}. Therefore we will also introduce
the same gluon mass $\lambda$ in the calculation of the bremsstrahlung in
Sec.\ \ref{subsec:nlo-brems} and we will show in Sec.\ \ref{subsub:nlo-res-E866-pT},
that the two divergences actually cancel numerically, as they should.

\subsubsection{Interference cross section}
\label{subsubsec:nlo-vc-interf}
The LO partonic cross section can be written as ($M^2=q^2$)
\begin{align}
	\frac{\diffd \hat\sigma_\text{LO}}{\diffd M^2 \diffd t} =
	\frac{1}{3} \frac{\pi\alpha^2}{6 M^3 p_\textrm{cm}}
	\cdot T_\text{LO} \cdot
	\delta(s-M^2)\, \delta(t-m_1^2) \ ,
\end{align}
where $\frac{1}{3}$ is the color factor and 
\begin{align}
	T_\text{LO} = \left(-g_{\mu\nu} + \frac{q_\mu q_\nu}{q^2}\right)
	\cdot \text{Tr}\left[(\slashed{p}_2 - m_2) \gamma^\mu
	                    (\slashed{p}_1 + m_1) \gamma^\nu\right] \ .
\end{align}			    
From this one easily finds that the cross section of the interference of the LO process and the vertex correction can be written as
\begin{align}
	\frac{\diffd \hat\sigma_\text{VC}}{\diffd M^2 \diffd t} =
	\frac{1}{3} \frac{\pi\alpha^2}{6 M^3 p_\textrm{cm}}
	\cdot T_\text{VC} \cdot
	\delta(s-M^2)\, \delta(t-m_1^2) \ ,
\end{align}
with
\begin{align}
	T_\text{VC} = &\left(-g_{\mu\nu} + \frac{q_\mu q_\nu}{q^2}\right) \nonumber \\
	\times&\left( 2\,\text{Re}(A)\cdot \text{Tr}\left[(\slashed{p}_2 - m_2) \gamma^\mu
	                                      (\slashed{p}_1 + m_1) \gamma^\nu\right] \right.
					      			\nonumber \\
			&\left.\ 2\,\text{Re}(B) \cdot \text{Tr}\left[(\slashed{p}_2 - m_2) \gamma^\mu
	                                      (\slashed{p}_1 + m_1) \right]
      				      \cdot (p_1 - p_2)^\nu \right) \ .
		\label{T_VC}
\end{align}

Note that $\diffd \hat\sigma_\text{VC}$ depends on the gluon mass $\lambda$ of 
Eq.~(\ref{vc-kappa}) and so does the cross section for gluon bremsstrahlung of Sec.~\ref{subsec:nlo-brems}. Only the sum of the two cross sections is a physically meaningful quantity
and thus we will only plot the sum of the two in our results in Sec.~\ref{sec:results}.

\subsubsection{Kinematics}
\label{subsubsec:nlo-vc-kinematics}
Since the vertex correction shares initial and final states with the LO 
process, the hadronic cross section of the vertex correction is calculated exactly as 
described for the LO mass distribution case in Sec.~\ref{subsec:lo-quarkmass}.

\subsection{Gluon bremsstrahlung}
\label{subsec:nlo-brems}
In the case of gluon bremsstrahlung we assign the same fictitious mass $\lambda$ to the gluon
as for the vertex correction process. This ensures the cancellation of the soft gluon divergences,
as we will show in Sec.\ \ref{sec:results}.
Then the partonic cross section becomes ($E_g \geq \lambda$)
\begin{align}
	\frac{\diffd \hat\sigma_\text{B}}{\diffd M^2 \diffd t} =
	\frac{4}{9} \frac{\alpha^2 \alpha_s}{48 M^2}
	\cdot \frac{T_\text{B}}{s p_\textrm{cm}^2} \cdot \Theta(E_g)\ .
	\label{gluon-brems-part-xsec}
\end{align}
Here $\frac{4}{9}$ is the color factor and $T_B$ is given by
\begin{align}
	T_\text{B} = \left(g_{\mu\nu} - \frac{q_\mu q_\nu}{q^2}\right)
	\cdot \text{Tr}\left[(\slashed{p}_2-m_2)S^{\alpha\mu}(\slashed{p}_1+m_1)
	                  S^\nu_{\ \alpha}\right]
\end{align}			  
with 
\begin{align}
	S^{\alpha\beta} = \gamma^\alpha \, 
	                \frac{\slashed{p}_1 - \slashed{q} + m_2}
			     {(p_1 - q)^2 - m_2^2}
			\gamma^\beta
			+	
			\gamma^\beta \, 
	                \frac{\slashed{q} - \slashed{p}_2 + m_1}
			     {(p_2 - q)^2 - m_1^2}
			\gamma^\alpha  \ .
	\label{eq:S_Brems}
\end{align}		

The calculation of the hadronic cross section basically follows along the same
lines as for the LO case in Sec.~\ref{subsec:lo-quarkmass}. Once
again one has to remove unphysical solutions for the momentum fractions $x_i$,
however, the calculation of the phase space is more subtle. The details of this
calculation are given in appendix \ref{appsub:nlo-kin-brems}.

\subsection{Gluon Compton scattering}
\label{subsec:nlo-compton}

For gluon Compton scattering we choose for the initial quark/antiquark to have
four-momentum $p_1$ and for the gluon to have four-momentum $p_2$, thus $m_2=0$ since
the gluon is real. For the outgoing quark/antiquark we then have $m_r=m_1$.
The partonic cross section then reads ($E_r \geq m_r$):
\begin{align}
	\frac{\diffd \hat\sigma_\text{C}}{\diffd M^2 \diffd t} =
	\frac{1}{6} \frac{\alpha^2 \alpha_s}{12 M^2 (s-m_1^2)^2}
	\cdot T_\text{C} \cdot \Theta(E_r)\ ,
	\label{gluon-compton-part-xsec}
\end{align}
where $\frac{1}{6}$ is the color factor and $T_C$ is given by
\begin{align}
	T_\text {C} = \left(g_{\mu\nu} - \frac{q_\mu q_\nu}{q^2}\right)
	\cdot \text{Tr}\left[(\slashed{p}_1 + \slashed{p}_2 - \slashed{q} +m_1)
                          S^{\mu\alpha}(\slashed{p}_1+m_1)
	                  S_\alpha^{\ \nu}\right]
\end{align}			  
with 
\begin{align}
	S^{\alpha\beta} = \gamma^\alpha \, 
	                \frac{\slashed{p}_1 + \slashed{p}_2 + m_1}
			     {(p_1 + p_2)^2 - m_1^2}
			\gamma^\beta
			+	
			\gamma^\beta \, 
	                \frac{\slashed{p}_1 - \slashed{q} + m_1}
			     {(p_1 - q)^2 - m_1^2}
			\gamma^\alpha  \ .
	\label{eq:S_Compton}
\end{align}		

The calculation of the hadronic cross section is similar to the case of gluon bremsstrahlung, however, the inherent asymmetry of the initial state (massive quark hits massless gluon) requires additional care. The details can be found in appendix \ref{appsub:nlo_kin_compton}.

\subsection{Influence of quark mass distributions on DY $p_T$ spectra}
\label{subsec:ir-div}
As already mentioned in the introduction, massless pQCD calculations of DY pair production at NLO produce
divergent $p_T$ spectra \cite{Gavin:1995ch}. The origin of these IR divergences are twofold:
first there is soft gluon emission (bremsstrahlung). This type of soft divergence, however, is not problematic,
since it exactly cancels against a divergence in the virtual processes (vertex correction), cf.\ Sec.\ 
\ref{subsubsec:nlo-vc-soft}. Second there is emission (bremsstrahlung) or capture (Compton scattering) of a 
gluon by a massless participant quark (also called mass or collinear singularity): the $u$-channel exchange quarks in Figs.\ \ref{fig:gluon-brems} and \ref{fig:gluon-compton} can become onshell at $p_T=0$ and thus for $m_1=m_2=0$ the propagators in Eqs.~(\ref{eq:S_Brems},\ref{eq:S_Compton}) produce a non-integrable (in $p_T^2$) singularity at $p_T=0$. To address this problem was one reason for
introducing mass distributions for the participating quarks. This procedure aims at smearing out the divergence and so making the $p_T$ spectra integrable. 

\begin{figure}[H]
	\centering
	\includegraphics[angle=-90,keepaspectratio,width=0.45\textwidth]
                        {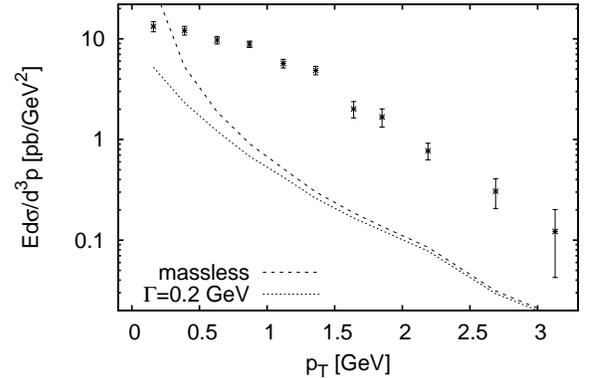}
	\caption{$p_T$ spectrum of gluon Compton scattering obtained from massless and mass distribution approach with initially collinear quarks. The PDFs are the MSTW2008LO68cl set. Data are from E866 binned with $4.2$ GeV $< M < 5.2$ GeV, $-0.05 < x_F < 0.15$. Only statistical errors are shown.}
	\label{fig:compton_ir}	
\end{figure}

We will illustrate this procedure on the example of 
the gluon Compton scattering process. In Fig.\ \ref{fig:compton_ir} we compare $p_T$ spectra produced by gluon Compton scattering in two different schemes: one calculation with massless quarks and a calculation which includes quark mass distributions. In both cases the quark's initial transverse momentum is set to zero.
It is seen that now indeed a transverse momentum of the dileptons is generated. However, its magnitude is
significantly underestimated. One can also see clearly that the rise for $p_T \rightarrow 0$ of the calculation with mass distributions is slower than for the calculation with massless quarks. This is a consequence of the effective cut-off, which is introduced by distributing the quark masses. We find that the divergence in $p_T$ is softened enough to make the $p_T$ spectra integrable.

\subsection{Collinear (mass) singularities and parton distribution functions}
\label{subsec:coll_sing}

As we have just shown, we have regularized the collinear singularities of the NLO processes by introducing
quark mass distributions. However, for the calculation of the cross sections we would like to use PDFs
as supplied in the literature. But exactly those collinear singularities, that we have just regularized,
are commonly absorbed into the definition of the standard PDFs. Thus, to avoid double-counting, 
in this section we present a subtraction scheme, that leaves us with a consistent cross section
to the order of the hard processes we are considering.

To set the stage we first review briefly the introduction of the renormalized PDFs into pQCD. For the DY process this concerns the calculation of $M$ spectra, because the spectra differential in $p_T$ are not accessible by pQCD. Since we are interested in the description of fully differential DY spectra by our model, we have to modify the way towards the standard renormalized PDFs. This is outlined in a second subsection. 

\subsubsection{Collinear singularities in pQCD}
\label{subsubsec:pQCD_coll_sing}
If Bjorken-scaling were not violated the PDFs found in deep inelastic scattering were functions of the 
momentum fraction $x$ only.
However, it is well known that the interactions among the quarks and gluons induce scaling violations
via processes like gluon bremsstrahlung and gluon quark-antiquark production. To calculate
the contributions of these processes to the longitudinal PDFs one has to integrate over
the transverse momentum of the emitted particle (quark or gluon) and finds, that they suffer from collinear 
(or mass) singularities, i.e.\ they are singular because the quarks are treated as massless.  
The divergences appear at the boundaries of the transverse momentum integrals
(which is why they are called collinear divergences). Thus,
one can regulate these divergences by introducing a regulating cut-off $\eta^2$ in the transverse
momentum integral.
In this scheme one can define renormalized longitudinal quark PDFs by absorbing the collinear
singularities and the (non-measurable, scaling) bare quark and gluon PDFs,
$f_i^0(x)$ and $g^0(x)$, into one function \cite{Ellis:1991qj}:
\begin{align}
  &f_i(x,\mu^2) \nonumber \\
  = &f_i^0(x) + \frac{\alpha_s}{2\pi} \int_x^1 \diffd y\ \frac{1}{y}
      \left\{ f_i^0(y) \left[ P_{qq}\left( \frac{x}{y} \right) \log\left( \frac{\mu^2}{\eta^2} \right)
         + C^S_q\left( \frac{x}{y} \right) \right] \right. \nonumber \\
	 &\phantom{f_i^0(x) + \frac{\alpha_s}{2\pi} \int_x^1 \diffd y\ }
         +\left. g^0(y) \left[ P_{qg}\left( \frac{x}{y} \right) \log\left( \frac{\mu^2}{\eta^2} \right)
      + C^S_g\left( \frac{x}{y} \right) \right]  \right\} \ ,
  \label{eq:quark-pdf}
\end{align}
with the hard scale $\mu^2$. The coefficient functions of the divergent logarithms are the splitting 
functions $P_{qq}$ and $P_{qg}$, which are given below. The functions $C^\text{S}_q$ and $C^\text{S}_g$ contain
possible finite contributions of the scaling violating processes and the superscript S reminds us of
the fact, that these finite contributions depend on the chosen renormalization scheme, since only the divergent contributions actually have to be absorbed into the renormalized PDFs. The splitting functions found
in deep inelastic scattering to order $\alpha_s$ are given by \cite{Ellis:1991qj}
\begin{align}
  P_{qq}(x) &= \frac{4}{3} \left[ \frac{1+x^2}{\left( 1-x \right)_+} + \frac{3}{2} \delta\left( 1-x \right) \right] \ , \\
  P_{qg}(x) &= \frac{1}{2} \left[ x^2 + (1-x)^2 \right] \ ,
  \label{eq:splitting}
\end{align}
where $\frac{4}{3}$ and $\frac{1}{2}$ are color factors. The plus prescription reads:
\begin{align}
  \int_0^1 \diffd x \frac{f(x)}{\left( 1-x \right)_+} = 
  \int_0^1 \diffd x \frac{f(x) - f(1)}{1-x}  \ ,
  \label{eq:pluspres}
\end{align}
where $f$ is a smooth function on $[0,1]$. 

Remarkably one finds the same splitting functions when
calculating order $\alpha_s$ corrections to DY pair production \cite{Ellis:1991qj}: $P_{qq}$
collects all the contributions from the processes with $q\bar q$ in the initial state, i.e.\ 
the vertex correction and gluon bremsstrahlung processes of Figs.\ \ref{fig:vertex-correction},
\ref{fig:wave-renorm} and \ref{fig:gluon-brems}. $P_{qg}$ contains the contributions from
the gluon Compton scattering processes in Fig.\ \ref{fig:gluon-compton}.
Then the partonic cross section for DY pair production to order $\alpha_s$ integrated
over the transverse momentum of the DY pair can be written schematically as \cite{Ellis:1991qj}
\begin{align}
  \frac{ \diffd \hat \sigma}{\diffd M^2}
  = \frac{4\pi\alpha^2}{9M^4} e_i^2 z \left[
  \delta\left( 1-z \right) 
                     + \frac{\alpha_s}{2\pi}\left( \mathcal{F}_{q\bar q}(z) 
		                                  +\mathcal{F}_{qg}(z)  \right) \right]\ ,
  \label{eq:dy-nlo-part-xsec}
\end{align}
for a quark of flavor $i$ and with
\begin{align}
  z = \frac{M^2}{\hat s} = \frac{M^2}{x_1 x_2 S} = \frac{\tau}{x_1 x_2} \ ,
  \label{eq:dy-nlo-z}
\end{align}  
where $\sqrt{\hat s}$ ($\sqrt{S}$) is the partonic (hadronic) c.m.\ energy and $x_1,x_2$ the momentum
fractions of the quarks.
The $\delta$-function in (\ref{eq:dy-nlo-part-xsec}) gives just the leading-order contribution and the functions $\mathcal{F}_{q\bar q}$
and $\mathcal{F}_{qg}$ give the contributions with initial states consisting of quark-antiquark 
(vertex correction and gluon bremsstrahlung) and quark-gluon (gluon Compton scattering), respectively:
\begin{align}
  \mathcal{F}_{q\bar q}(z) &=
  2 P_{qq}(z) \log\left( \frac{M^2}{\eta^2} \right) + \hat C^\text{S}_q(z) \ , 
  \label{eq:dy-nlo-functions}
  \\
  \mathcal{F}_{qg}(z) &=
  P_{qg}(z) \log\left( \frac{M^2}{\eta^2} \right) + \hat C^\text{S}_g(z) \ ,
  \label{eq:dy-nlo-functions2}
\end{align}
where again the cut-off $\eta^2$ was introduced to regulate the transverse momentum integration and
where the functions $\hat C^\text{S}_q$ and $\hat C^\text{S}_g$ are again renormalization scheme dependent, 
finite contributions. In principle one could obtain the hadronic cross section by folding the 
partonic cross section (\ref{eq:dy-nlo-part-xsec}) with the bare parton distributions and summing
over all quark flavors:
\begin{align}
   &\frac{ \diffd \sigma}{\diffd M^2} \nonumber \\
   = &\frac{4\pi\alpha^2}{9M^4} \sum_i e_i^2 \int_0^1 \diffd x_1 \diffd x_2 
   \frac{\tau}{x_1 x_2} \Theta\left( x_1 x_2 - \tau \right)\nonumber \\
     &\times \left\{ \left(f_i^0(x_1) f_{\bar i}^0 (x_2) + (f_i^0 \leftrightarrow f^0_{\bar i}) \right)
   \left[ \delta\left( 1-z \right) 
                     + \frac{\alpha_s}{2\pi} \mathcal{F}_{q\bar q}(z) \right] \right. \nonumber \\
         &+\left. \left( g^0(x_1) \left(f_i^0(x_2) + f_{\bar i}^0 (x_2)\right) 
	 + (g^0 \leftrightarrow f^0_i,f^0_{\bar i}) \right) \frac{\alpha_s}{2\pi} \mathcal{F}_{qg}(z)
						  \right\} \ .
  \label{eq:dy-nlo-hadr-xsec}
\end{align}
This cross section cannot be evaluated straigtforwardly,
since neither the bare parton distributions are available, nor
are $\mathcal{F}_{q\bar q}$ and $\mathcal{F}_{qg}$ well defined, since they depend on the arbitrary
cut-off $\eta^2$. However, we note the following relation for a general function $P$:
\begin{align}
  &\int_0^1 \diffd x_1 \diffd x_2 \frac{\tau}{x_1 x_2} \Theta\left( x_1 x_2 - \tau \right)
  f_i^0(x_1) f_{\bar i}^0(x_2) P\left( \frac{\tau}{x_1 x_2} \right) \nonumber \\
  = &\tau \int_\tau^1\frac{\diffd x_1}{x_1} f_i^0(x_1) 
   \int_{\frac{\tau}{x_1}}^1 \frac{\diffd x_2}{x_2} P\left( \frac{\tau}{x_1 x_2} \right) f_{\bar i}^0(x_2)
   \nonumber \\
   = &\tau \int_\tau^1\frac{\diffd x_1}{x_1} f_i^0(x_1) \int_\tau^1 \diffd x_2 
           \delta\left( x_2 - \frac{\tau}{x_1} \right)
	   \int_{x_2}^1 \frac{\diffd y}{y} P\left( \frac{x_2}{y} \right) f_{\bar i}^0(y)
   \nonumber \\	   
   = &\tau \int_0^1 \diffd x_1 \diffd x_2 \diffd z \delta\left( 1-z \right)
      \delta\left(x_1 x_2 z - \tau\right) f_i^0(x_1) \nonumber \\
     &\times
      \int_{x_2}^1 \frac{\diffd y}{y} P\left( \frac{x_2}{y} \right) f_{\bar i}^0(y) \ .
  \label{eq:dy-nlo-renorm-pdf-rel}
\end{align}
In addition one finds for the product of two renormalized PDFs of the type of Eq.\ (\ref{eq:quark-pdf})
\begin{align}
  &f_i(x_1,M^2) f_{\bar i}(x_2,M^2) \nonumber \\
  = &f_i^0(x_1) f_{\bar i}^0(x_2)  \nonumber \\
     + &f_i^0(x_1)
     \frac{\alpha_s}{2\pi} \int_{x_2}^1 \diffd y\ \frac{1}{y}
     \left\{ f_{\bar i}^0(y) \left[ P_{qq}\left( \frac{x_2}{y} \right) \log\left( \frac{M^2}{\eta^2} \right)
         + C^S_q\left( \frac{x_2}{y} \right) \right] \right. \nonumber \\
	 &\phantom{f_i^0(x_2) + \frac{\alpha_s}{2\pi} \int_x^1 \ \ }
         +\left. g^0(y) \left[ P_{qg}\left( \frac{x_2}{y} \right) \log\left( \frac{M^2}{\eta^2} \right)
      + C^S_g\left( \frac{x_2}{y} \right) \right]  \right\} \nonumber \\
      + &f_{\bar i}^0(x_2)
     \frac{\alpha_s}{2\pi} \int_{x_1}^1 \diffd y\ \frac{1}{y}
      \left\{ f_i^0(y) \left[ P_{qq}\left( \frac{x_1}{y} \right) \log\left( \frac{M^2}{\eta^2} \right)
         + C^S_q\left( \frac{x_1}{y} \right) \right] \right. \nonumber \\
	 &\phantom{f_i^0(x_2) + \frac{\alpha_s}{2\pi} \int_x^1 \ \ }
         +\left. g^0(y) \left[ P_{qg}\left( \frac{x_1}{y} \right) \log\left( \frac{M^2}{\eta^2} \right)
      + C^S_g\left( \frac{x_1}{y} \right) \right]  \right\} \nonumber \\
      + & O\left( \alpha_s^2 \right)\ .
  \label{eq:dy-nlo-renorm-pdf-prod}
\end{align}
Comparing the last two equations, one finds, that 
one can express the hadronic cross section (\ref{eq:dy-nlo-hadr-xsec}) in terms of
the renormalized PDFs (\ref{eq:quark-pdf}). One obtains \cite{Ellis:1991qj}
\begin{align}
   &\frac{ \diffd \sigma}{\diffd M^2} \nonumber \\
   = &\frac{4\pi\alpha^2}{9M^4} \tau \sum_i e_i^2 
      \int_0^1 \diffd x_1 \diffd x_2 \diffd z 
      \delta\left(x_1 x_2 z - \tau\right) \nonumber \\
     &\times \left\{ \left(f_i(x_1,M^2) f_{\bar i} (x_2,M^2) + (f_i \leftrightarrow f_{\bar i}) \right)
   \left[ \delta\left( 1-z \right) 
            + \frac{\alpha_s}{2\pi} \tilde C^\text{S}_{q}(z) \right] \right. \nonumber \\
   &+ \left[ g(x_1,M^2) \left(f_i(x_2,M^2) + f_{\bar i} (x_2,M^2)\right) \right. \nonumber \\
   &\phantom{+\ \ } +\left. \left. (g \leftrightarrow f_i,f_{\bar i}) \right] \frac{\alpha_s}{2\pi} 
     \tilde C^\text{S}_g(z)
						  \right\} \ ,
  \label{eq:dy-nlo-hadr-xsec2}
\end{align}
which is correct to $O(\alpha_s)$. The collinear divergences and the bare PDFs have been absorbed
into the renormalized PDFs. Again there remain scheme-dependent finite contributions $\tilde C_q$,
$\tilde C_g$.

\subsubsection{Collinear singularities in our model}
\label{subsubsec:model_coll_sing}
The introduction of the renormalized PDFs gives rise to the famous DGLAP evolution equation which describe successfully the scaling violations \cite{Ellis:1991qj}. Clearly for our model we want to inherit this fundamental QCD property. Also from a pragmatic point of view we want to use the standard (renormalized) PDFs from the literature. On the other hand, in the pQCD approach to the DY process the reshuffling of the collinear singularities into the renormalized PDFs is only possible for the $p_T$ integrated $M$ spectrum. In our model we also want to describe the DY spectra differential in $p_T$. Therefore, we cannot follow the steps outlined in the previous subsection. However, we have an explicit regularization of the collinear singularities. This allows to make contact between the bare and the renormalized PDFs by a kind of backward engineering, which we will describe next. 

Since we explicitly take into account all $O(\alpha_s)$ processes, 
cf.\ Secs.\ \ref{subsec:nlo-vc}-\ref{subsec:nlo-compton}, we now have to ensure that all the
$O(\alpha_s)$ contributions to the cross section, that were absorbed into the renormalized PDFs
are subtracted. Otherwise we would double-count the $O(\alpha_s)$ contributions. Schematically
the hadronic cross section for DY pair production to next-to-leading order in $\alpha_s$ can be written as
\begin{align}
  \diffd \sigma = \int \sum_i e_i^2 &\left[
                 \underbrace{\diffd \hat \sigma_\text{LO}}_{O(\alpha_s^0)=O(1)} f_i^0 \cdot f_{\bar i}^0
		 +\underbrace{\diffd \hat \sigma_\text{VC+B}}_{O(\alpha_s)} 
		 \ f_i^0 \cdot f_{\bar i}^0 \right. \nonumber \\
		 &+\left. \underbrace{\diffd \hat \sigma_\text{C}}_{O(\alpha_s)} 
		 \ g^0 \cdot (f_i^0 + f_{\bar i}^0) \right] \ ,
  \label{eq:dy-model-schema}
\end{align}
with the bare PDFs $f_i^0$, $g^0$. We note that $f_i \cdot f_{\bar i} = f_i^0 f_{\bar i}^0 + O(\alpha_s)$
and $g \cdot ( f_i + f_{\bar i}) = g^0 \cdot ( f_i^0 +  f_{\bar i}^0 ) + O(\alpha_s)$, and so to
$O(\alpha_s)$ nothing changes if we replace the bare PDFs multiplying the NLO partonic cross sections:
\begin{align}
  \diffd \sigma = \int \sum_i e_i^2 &\left[
                 \underbrace{\diffd \hat \sigma_\text{LO}}_{O(\alpha_s^0)=O(1)} f_i^0 \cdot f_{\bar i}^0
		 +\underbrace{\diffd \hat \sigma_\text{VC+B}}_{O(\alpha_s)} 
		 \ f_i \cdot f_{\bar i} \right. \nonumber \\
		 &+\left. \underbrace{\diffd \hat \sigma_\text{C}}_{O(\alpha_s)} 
		 \ g \cdot (f_i + f_{\bar i}) \right]  + O(\alpha_s^2)\ .
  \label{eq:dy-model-schema2}
\end{align}
However, if we were to do the same with the LO term, we would get additional $O(\alpha_s)$ contributions,
cf.\ Eq.\ (\ref{eq:dy-nlo-renorm-pdf-prod}). How do we subtract these contributions? After all, we cannot
calculate the integrals in Eq.\ (\ref{eq:dy-nlo-renorm-pdf-prod}), since we do not know $\eta$ or the
bare PDFs.

In our model we set out to calculate exactly those transverse momentum spectra
that were integrated in the derivation of the renormalized PDFs in the pQCD case above.
To accomplish this,
we have introduced quark mass distributions to handle the collinear divergences that 
enter the quark PDF in Eq.\ (\ref{eq:quark-pdf}). The important difference between the pQCD approach
and our model is, therefore, the following: in pQCD the regulating cut-off $\eta^2$ is completely
arbitrary and physical results can only be obtained by ``absorbing'' this arbitrariness into
the renormalized PDFs. All the finite and scheme-depedent contributions can be calculated analytically.
In our model we know the regulator $\eta$ in (\ref{eq:quark-pdf}),
it is nothing but our quark mass $m$ (or better $m^2$). However,
we do not know the finite contributions $C^\text{S}_{q,g}$ in our model. To estimate them, we introduce two new parameters
$\kappa_q$ and $\kappa_g$, so that the functions $\mathcal{F}_{q\bar q}$ and $\mathcal{F}_{qg}$ 
become:
\begin{align}
  \mathcal{F}^m_{q\bar q}(z) &=
  2 P_{qq}(z) \log\left( \frac{M^2}{\kappa_q^2 m^2} \right)  \ , 
  \label{eq:dy-model-nlo-functions}
  \\
  \mathcal{F}^m_{qg}(z) &=
  P_{qg}(z) \log\left( \frac{M^2}{\kappa_g^2 m^2} \right)  \ .
  \label{eq:dy-model-nlo-functions2}
\end{align}
Then we can rewrite the renormalized PDFs in Eq.\ (\ref{eq:dy-nlo-renorm-pdf-prod}) as
\begin{align}
  &f_i(x_1,M^2) f_{\bar i}(x_2,M^2) \nonumber \\
  = &f_i^0(x_1) f_{\bar i}^0(x_2)  \nonumber \\
     + &f_i^0(x_1)
     \frac{\alpha_s}{2\pi} \int_{x_2}^1 \diffd y\ \frac{1}{y}
     \left\{ f_{\bar i}^0(y) \left[ P_{qq}\left( \frac{x_2}{y} \right) \log\left( \frac{M^2}{\kappa_q^2 m^2} \right)
         \right] \right. \nonumber \\
	 &\phantom{f_i^0(x_2) + \frac{\alpha_s}{2\pi} \int_x^1 \ \ }
         +\left. g^0(y) \left[ P_{qg}\left( \frac{x_2}{y} \right) \log\left( \frac{M^2}{\kappa_g^2 m^2} \right)
      \right]  \right\} \nonumber \\
      + &f_{\bar i}^0(x_2)
     \frac{\alpha_s}{2\pi} \int_{x_1}^1 \diffd y\ \frac{1}{y}
      \left\{ f_i^0(y) \left[ P_{qq}\left( \frac{x_1}{y} \right) \log\left( \frac{M^2}{\kappa_q^2 m^2} \right)
          \right] \right. \nonumber \\
	 &\phantom{f_i^0(x_2) + \frac{\alpha_s}{2\pi} \int_x^1 \ \ }
         +\left. g^0(y) \left[ P_{qg}\left( \frac{x_1}{y} \right) \log\left( \frac{M^2}{\kappa_g^2 m^2} \right)
      \right]  \right\} \nonumber \\
      + & O\left( \alpha_s^2 \right)\ .
  \label{eq:dy-nlo-model-renorm-pdf-prod}
\end{align}
Solving for the product of the bare PDFs gives
\begin{align}
    &f_i^0(x_1) f_{\bar i}^0(x_2) \nonumber \\
    = &f_i(x_1,M^2) f_{\bar i}(x_2,M^2) \nonumber \\
     - &f_i^0(x_1)
     \frac{\alpha_s}{2\pi} \int_{x_2}^1 \diffd y\ \frac{1}{y}
     \left\{ f_{\bar i}^0(y) \left[ P_{qq}\left( \frac{x_2}{y} \right) \log\left( \frac{M^2}{\kappa_q^2 m^2} \right)
         \right] \right. \nonumber \\
	 &\phantom{f_i^0(x_2) + \frac{\alpha_s}{2\pi} \int_x^1 \ \ }
         +\left. g^0(y) \left[ P_{qg}\left( \frac{x_2}{y} \right) \log\left( \frac{M^2}{\kappa_g^2 m^2} \right)
      \right]  \right\} \nonumber \\
      - &f_{\bar i}^0(x_2)
     \frac{\alpha_s}{2\pi} \int_{x_1}^1 \diffd y\ \frac{1}{y}
      \left\{ f_i^0(y) \left[ P_{qq}\left( \frac{x_1}{y} \right) \log\left( \frac{M^2}{\kappa_q^2 m^2} \right)
          \right] \right. \nonumber \\
	 &\phantom{f_i^0(x_2) + \frac{\alpha_s}{2\pi} \int_x^1 \ \ }
         +\left. g^0(y) \left[ P_{qg}\left( \frac{x_1}{y} \right) \log\left( \frac{M^2}{\kappa_g^2 m^2} \right)
      \right]  \right\} \nonumber \\
      + & O\left( \alpha_s^2 \right)\ .
  \label{eq:dy-nlo-renorm-pdf-prod2}
\end{align}
On the right hand side we can replace all the bare PDFs by their renormalized version, since all additional
corrections introduced by this procedure are $O(\alpha_s^2)$:
\begin{align}
    &f_i^0(x_1) f_{\bar i}^0(x_2) \nonumber \\
    = &f_i(x_1,M^2) f_{\bar i}(x_2,M^2) \nonumber \\
     - &f_i(x_1,M^2)
     \frac{\alpha_s}{2\pi} \int_{x_2}^1 \diffd y\ \frac{1}{y}
     \left\{ f_{\bar i}(y,M^2) 
             \left[ P_{qq}\left( \frac{x_2}{y} \right) \log\left( \frac{M^2}{\kappa_q^2 m^2} \right)
         \right] \right. \nonumber \\
	 &\phantom{f_i^0(x_2) + \frac{\alpha_s}{2\pi} \int_x^1 \ \ }
         +\left. g(y,M^2) \left[ P_{qg}\left( \frac{x_2}{y} \right) \log\left( \frac{M^2}{\kappa_g^2 m^2} \right)
      \right]  \right\} \nonumber \\
      - &f_{\bar i}(x_2,M^2)
      \frac{\alpha_s}{2\pi} \int_{x_1}^1 \diffd y\ \frac{1}{y}
      \left\{ f_i(y,M^2)\left[ P_{qq}\left( \frac{x_1}{y} \right) \log\left( \frac{M^2}{\kappa_q^2 m^2} \right)
          \right] \right. \nonumber \\
	 &\phantom{f_i^0(x_2) + \frac{\alpha_s}{2\pi} \int_x^1 \ \ }
         +\left. g(y,M^2) \left[ P_{qg}\left( \frac{x_1}{y} \right) \log\left( \frac{M^2}{\kappa_g^2 m^2} \right)
      \right]  \right\} \nonumber \\
      + & O\left( \alpha_s^2 \right)\ .
  \label{eq:dy-nlo-renorm-pdf-prod3}
\end{align}
We define
\begin{align}
  f_i^\text{sub}(x,M^2) &=  \frac{\alpha_s}{2\pi} \int_{x}^1 \diffd y\ \frac{1}{y}
                           f_i(y,M^2) 
                           P_{qq}\left( \frac{x}{y} \right) \log\left( \frac{M^2}{\kappa_q^2 m^2} \right)
	\ ,			   \nonumber \\
  g^\text{sub}(x,M^2)   &= \frac{\alpha_s}{2\pi} \int_{x}^1 \diffd y\ \frac{1}{y}
      g(y,M^2) P_{qg}\left( \frac{x}{y} \right) \log\left( \frac{M^2}{\kappa_g^2 m^2} \right) \ ,
  \label{eq:dy-nlo-sub-pdfs}
\end{align}
and so we find
\begin{align}
    &f_i^0(x_1) f_{\bar i}^0(x_2) \nonumber \\
    = &f_i(x_1,M^2) f_{\bar i}(x_2,M^2) \nonumber \\
    - &f_i(x_1,M^2) f_{\bar i}^\text{sub}(x_2,M^2) - f_i(x_1,M^2) g^\text{sub}(x_2,M^2)  \nonumber \\
    - &f_{\bar i}(x_2,M^2) f_i^\text{sub}(x_1,M^2) - f_{\bar i}(x_2,M^2) g^\text{sub}(x_1,M^2) \nonumber \\
      + & O\left( \alpha_s^2 \right)\ .
  \label{eq:dy-nlo-renorm-pdf-prod4}
\end{align}
In this scheme the hadronic cross section (\ref{eq:dy-model-schema2}) becomes
\begin{align}
  \diffd \sigma
   = &\int \sum_i e_i^2 \left[
                 \diffd \hat \sigma_\text{LO} f_i \cdot f_{\bar i} \right. \nonumber \\
		 & -\diffd \hat \sigma_\text{LO} \left(f_i \cdot f^\text{sub}_{\bar i}
		   +f^\text{sub}_i \cdot f_{\bar i} \right) 
		 +\diffd \hat \sigma_\text{VC+B} 
		 \ f_i \cdot f_{\bar i} \nonumber \\
		 &\left. -\diffd \hat \sigma_\text{LO} \left(f_i \cdot g^\text{sub}
		 +f_{\bar i} \cdot g^\text{sub} \right) 
                 + \diffd \hat \sigma_\text{C}
		 \ g \cdot (f_i + f_{\bar i}) \right]  \nonumber \\
     &+ O(\alpha_s^2)\ .
  \label{eq:dy-model-schema3}
\end{align}
From now on, we label the different contributions to the cross section as sketched in the
following: 
\begin{align}
  \diffd \sigma_\text{LO} = \int \sum_i e_i^2  &\diffd \hat \sigma_\text{LO} f_i \cdot f_{\bar i} \ ,
  \label{eq:dy-model-lo}
  \\
  \diffd \sigma_{q\bar q} = \int \sum_i e_i^2
         		      &\diffd \hat \sigma_\text{VC+B} \ f_i \cdot f_{\bar i} 
			     -\diffd \hat \sigma_\text{LO} \left(f_i \cdot f^\text{sub}_{\bar i}
			     +f^\text{sub}_i \cdot f_{\bar i} \right) \ , \label{eq:dy-model-qqbar}
			      \\
  \diffd \sigma_{qg} = \int \sum_i e_i^2
                                &\diffd \hat \sigma_\text{C}
                               \ g \cdot (f_i + f_{\bar i})  \nonumber \\
			       -&\diffd \hat \sigma_\text{LO} \left(f_i \cdot g^\text{sub}
				+f_{\bar i} \cdot g^\text{sub} \right) 
				\label{eq:dy-model-qg}
			     \ .
\end{align}
In Eqs.\ (\ref{eq:dy-model-qqbar}) and (\ref{eq:dy-model-qg}) we subtract now precisely those 
$O(\alpha_s)$ contributions which were absorbed before into the
renormalized PDFs. Thus, by this procedure we have produced in our model a consistent cross section 
to $O(\alpha_s)$.
Indeed, we have checked explicitly that the collinear divergences which appear in 
$\diffd \sigma_\text{VC+B}$, if the quark masses are sent to zero, cancel the corresponding divergence 
of $f_i^\text{sub}$ as given in Eq.\ (\ref{eq:dy-nlo-sub-pdfs}). The same is true for 
$\diffd \sigma_\text{C}$ and $g^\text{sub}$. Thus the expressions (\ref{eq:dy-model-qqbar}) 
and (\ref{eq:dy-model-qg}) remain finite for vanishing quark masses. 

For the calculation of our results in Sec.\ \ref{sec:results} we will use
Eqs.\ (\ref{eq:dy-model-lo})-(\ref{eq:dy-model-qg}). The quantities which enter are, on the one hand,
the standard PDFs which we can take from the literature, and, on the other hand, the parameters 
$\kappa_q$ and $\kappa_g$ which appear in Eq.\ (\ref{eq:dy-nlo-sub-pdfs}).
We recall that these parameters introduced in Eqs.\ (\ref{eq:dy-model-nlo-functions}) and
(\ref{eq:dy-model-nlo-functions2}) correspond to finite, i.e. infrared safe, contributions which appear,
e.g., as $\hat C^S_{q,g}$ in Eqs.\ (\ref{eq:dy-nlo-functions}) and (\ref{eq:dy-nlo-functions2}).
We will vary the parameters $\kappa_q$ and $\kappa_g$ around natural values (i.e.\ around 1),
to estimate these finite contributions.

\subsection{Initial transverse momentum distributions}
\label{subsec:ir-div-initialkt}
Even in the mass distribution approach at NLO we find that the $p_T$ data are heavily underestimated. Therefore, we have also introduced parton initial transverse momentum distributions for the NLO processes, just as we did at LO in Sec.\ \ref{subsec:lo-intrkt}. Taking into account these distributions we obtain a good description of measured cross sections without $K$ factors, as we will show in Sec.\ \ref{sec:results}.

\section{Results}
\label{sec:results}
In this section we present our combined results of the LO and NLO calculations of Secs.~\ref{sec:lo} and \ref{sec:nlo}. Since one of our main goals is to gain access to the
DY $p_T$ spectra, we compare our model with data on $p_T$ spectra from different experiments. Most of these experiments were done in proton-nucleus collisions. However, the recent E866 experiment \cite{Webb:2003ps,Webb:2003bj} measured pp collisions and we will begin to fix the parameters of our model at the data for this more elementary reaction in Sec.~\ref{subsub:nlo-res-E866-pT}. Using the fixed parameters we compare our results to data for different pp and pA experiments in Secs.~\ref{subsubsec:nlo-res-E866-M}-\ref{subsub:nlo-res-E439}. In Sec.~\ref{subsub:nlo-res-E537} we study our results for an antiproton-nucleus experiment 
($\overline{\text{p}}$N) (E537), which is useful, since one of our aims is to make predictions for DY pair production at $\overline{\text{P}}$ANDA, which will measure antiproton-proton ($\overline{\text{p}}$p) collisions. These predictions can be found in Sec.\ \ref{subsub:nlo-res-panda}.

Our model has effectively four parameters: The width of the initial parton transverse momentum distribution, represented by $D$, the width $\Gamma$ of the quark mass distribution (spectral function) and
the two subtraction constants $\kappa_q$ and $\kappa_g$ which estimate the finite contributions
which were absorbed into the PDFs. We will vary $\kappa_q$ and $\kappa_g$ in a natural range $\frac{1}{2}
\dots 2$.
The gluon mass $\lambda$ is not really a parameter but a numerical necessity and we will explore its influence on our calculations in Sec.~\ref{subsub:nlo-res-E866-pT}. In addition our model utilizes standard parton distribution functions. These are calculated by several different groups, for example \cite{Martin:2009iq,Gluck:2007ck}, and they are not unique. We will show results obtained with different \emph{leading order} PDFs to get an
estimate for the 
theoretical uncertainty. Once again we employ PDFs available through the LHAPDF library, version 5.8.4 
\cite{Whalley:2005nh}.

At this point we note two things: first, the divergence of the NLO processes near $p_T=0$ can also be cured
by choosing a finite and fixed quark mass. However it is much more natural to assume a broad mass distribution,
since the nucleon is composed of strongly interacting partons; we will show below, that we can describe the
experimental data quantitatively well assuming broad quark mass distributions. Second, in principle we could have chosen different transverse momentum and mass distributions for valence quarks, sea quarks and gluons (only transverse momentum in this case). We have not done so for two reasons: on the one hand this would have forced us to introduce additional parameters, while one is of course always inclined to keep the number of parameters as small as possible. On the other hand, as one will see in the results, the data do not require additional modelling. Thus we always use the same distributions from Sec.\ \ref{subsec:lo-distr} for all quarks and gluons (again no mass distributions for the latter).

\subsection{E866 - $p_T$ spectrum}
\label{subsub:nlo-res-E866-pT}
In this section we present the results of our full model. The data are from E866 \cite{Webb:2003bj,Webb:2003ps} for pp collisions at $S=1500 \text{ GeV}^2$.

\begin{figure}[H]
	\centering
	\includegraphics[angle=-90,keepaspectratio,width=0.45\textwidth]
                        {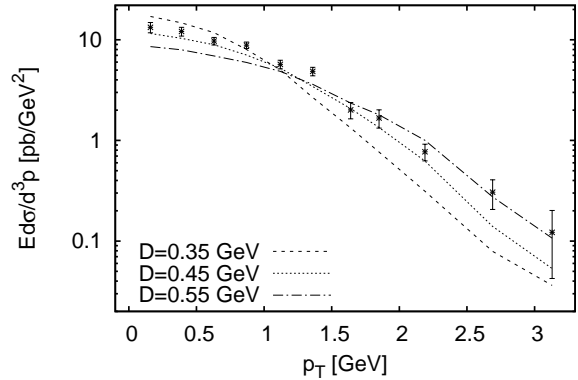}
	\caption{$p_T$ spectrum obtained from our full model with different values of $D$. Everywhere $\Gamma=0.2$ GeV, $\lambda=5$ MeV, $\kappa_q=1$ and $\kappa_g=2$. The PDFs are the MSTW2008LO68cl set. Data are from E866 binned with $4.2$ GeV $< M < 5.2$ GeV, $-0.05 < x_F < 0.15$. Only statistical errors are shown.}
	\label{fig:E866_triple_4.2_nlo_varD}	 
\end{figure}	

\begin{figure}[H]
	\centering
	\includegraphics[angle=-90,keepaspectratio,width=0.45\textwidth]
                        {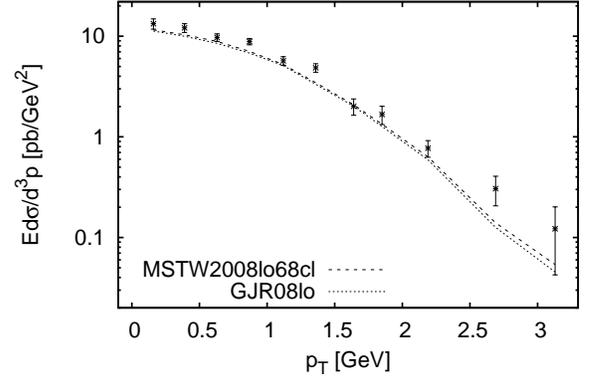}
	\caption{$p_T$ spectrum obtained from our full model with different PDF sets. Everywhere $D=0.45$ GeV, $\Gamma=0.2$ GeV, $\lambda=5$ MeV, $\kappa_q=1$ and $\kappa_g=2$. Data are from E866 binned with $4.2$ GeV $< M < 5.2$ GeV, $-0.05 < x_F < 0.15$. Only statistical errors are shown.}
	\label{fig:E866_triple_4.2_nlo_pdfcomp}	 
\end{figure}

First we want to stress, that the calculation in our full model reproduces measured DY $p_T$ spectra without the need for a $K$ factor, see for example Fig.\ \ref{fig:E866_triple_4.2_nlo_pdfcomp}. In order to better understand
the parameter dependence of our model, in the following we explore the parameter space. The details of how we obtain the presented cross sections were given in Sec.\ \ref{subsubsec:E866_triple_lo}. First we show several plots, in which we vary only one parameter and keep the others fixed.

In Fig.\ \ref{fig:E866_triple_4.2_nlo_varD} we plot the results of our full NLO model for different $D$. As one can see for $D$ around $0.45$ GeV, which corresponds to an average squared initial quark transverse momentum $\langle k_T^2 \rangle = (0.9)^2 \text{ GeV}^2$, the data are reproduced quite well. Obviously, $D$ determines the shape of the $p_T$ spectra.

\begin{figure}[H]
	\centering
	\includegraphics[angle=-90,keepaspectratio,width=0.45\textwidth]
                        {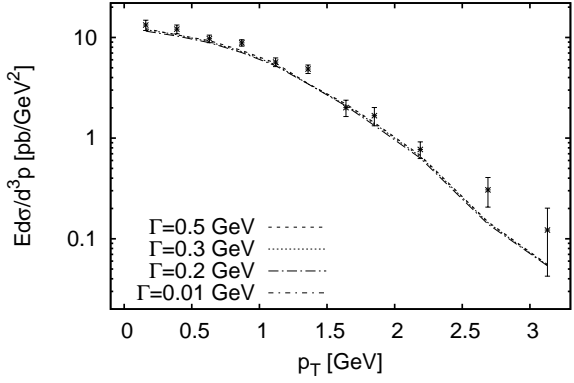}
			\caption{$p_T$ spectrum obtained from our full model with different values of $\Gamma$. Everywhere $D=0.45$ GeV, $\lambda=5$ MeV, $\kappa_q=1$ and $\kappa_g=2$. The PDFs are the MSTW2008LO68cl set. Note that the curves for $\Gamma \geq 0.2$ GeV are on top of each other. Data are from E866 binned with $4.2 \text{ GeV } < M < 5.2$ GeV, $-0.05 < x_F < 0.15$. Only statistical errors are shown.}
	\label{fig:E866_triple_4.2_nlo_varGamma}	 
\end{figure}

In Fig.\ \ref{fig:E866_triple_4.2_nlo_varGamma} we show results for different $\Gamma$. The results for several values of $\Gamma$ all agree very well with each other and at the same time reproduce the data quite well. Thus at E866 energies our model appears to be rather insensitive to changes of $\Gamma$ over a wide range.

\begin{figure}[H]
	\centering
	\includegraphics[angle=-90,keepaspectratio,width=0.45\textwidth]
                        {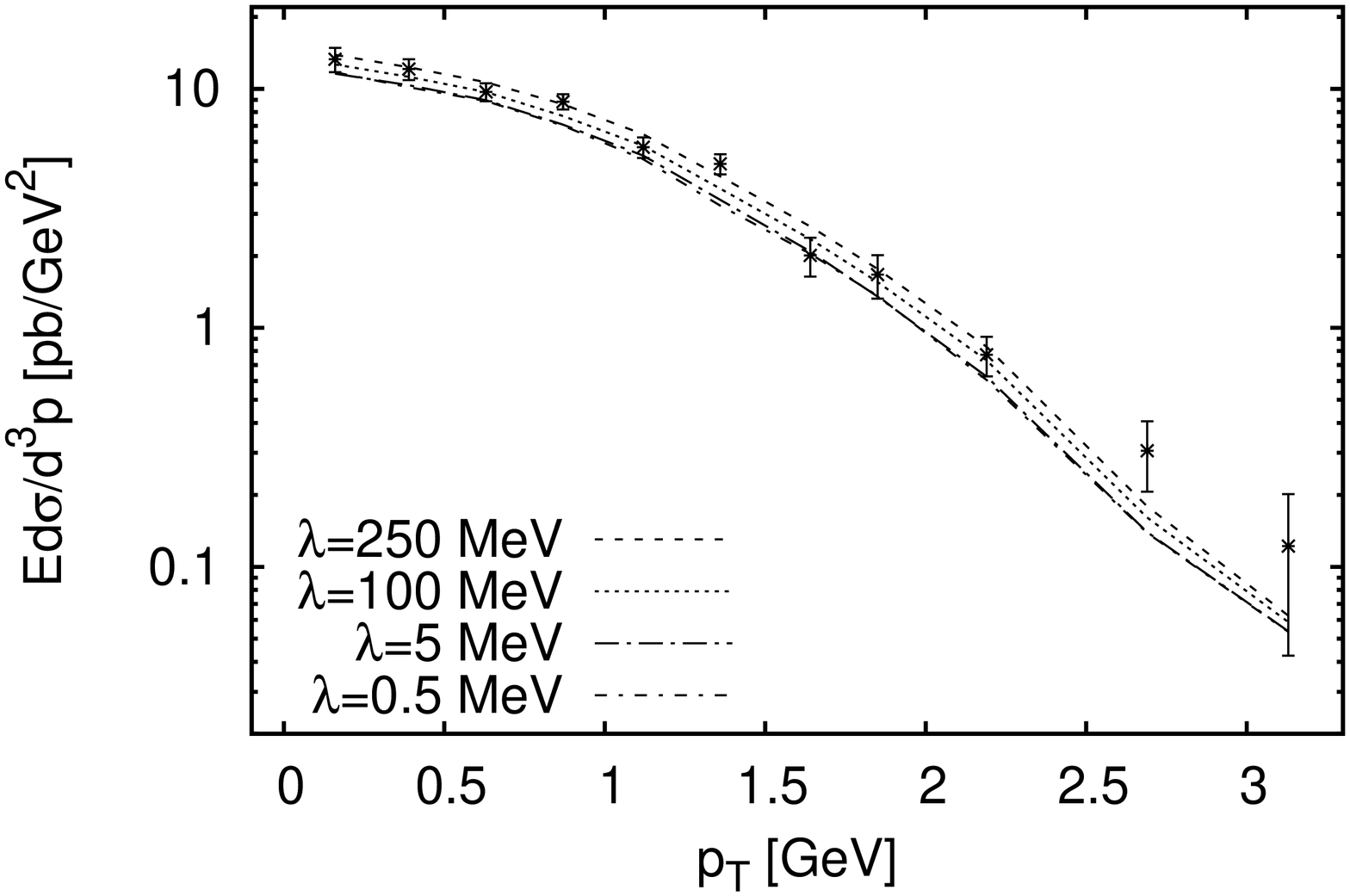}
	\caption{$p_T$ spectrum obtained from our full model with different values of $\lambda$. Everywhere $D=0.45$ GeV, $\Gamma=0.2$ GeV, $\kappa_q=1$ and $\kappa_g=2$. The PDFs are the MSTW2008LO68cl set. Note that the curves for $\lambda \leq 5$ MeV are on top of each other. Data are from E866 binned with $4.2$ GeV $< M < 5.2$ GeV, $-0.05 < x_F < 0.15$. Only statistical errors are shown.}
	\label{fig:E866_triple_4.2_nlo_varomega}	 
\end{figure}

Remember that the gluon mass $\lambda$ is introduced to regulate divergences that occur when the gluons become
very soft. In the limit of $\lambda \rightarrow 0$ these divergences of the bremsstrahlung and the vertex correction
processes should exactly cancel. Therefore, it is sensible to choose $\lambda$ as small as numerically feasible. 
Results for different choices of the gluon mass $\lambda$ are shown in Fig.\ \ref{fig:E866_triple_4.2_nlo_varomega}. While the results for $\lambda = 100,250$ MeV are still visibly larger than the results for $5$ and $0.5$ MeV, the latter two agree very well with each other. Thus the calculated cross section appears to converge in the $\lambda \rightarrow 0$ limit. Therefore, we chose $\lambda=5$ MeV for the calculation of all the following results.

We show an example of the influence of different parton distribution functions in Fig.\ \ref{fig:E866_triple_4.2_nlo_pdfcomp}. The results with the MSTW2008LO68cl \cite{Martin:2009iq} and GJR08lo \cite{Gluck:2007ck} sets agree quite well with each other, with only small deviations. This is an illustration of the uncertainties induced by the different integrated parton distributions.

\begin{figure}[H]
	\centering
	\includegraphics[angle=-90,keepaspectratio,width=0.45\textwidth]
                        {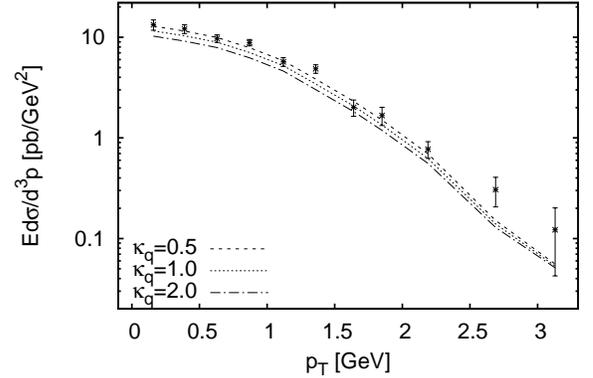}
	\caption{$p_T$ spectrum obtained from our full model with different values of the subtraction
	parameter $\kappa_q$. Everywhere $D=0.45$ GeV, $\Gamma=0.2$ GeV, $\lambda=5$ MeV and $\kappa_g=2$. The PDFs are the MSTW2008LO68cl set. Data are from E866 binned with $4.2$ GeV $< M < 5.2$ GeV, $-0.05 < x_F < 0.15$. Only statistical errors are shown.}
	\label{fig:E866_triple_4.2_nlo_varkappaq}	 
\end{figure}	

\begin{figure}[H]
	\centering
	\includegraphics[angle=-90,keepaspectratio,width=0.45\textwidth]
                        {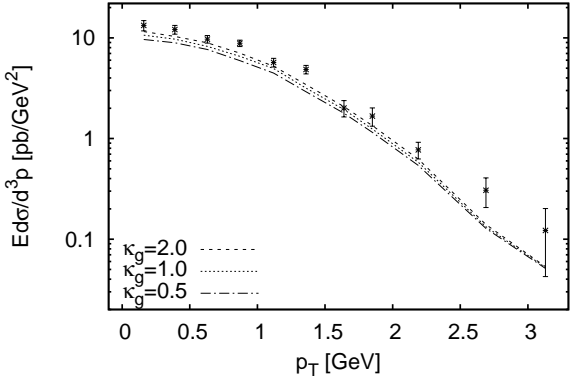}
	\caption{$p_T$ spectrum obtained from our full model with different values of the subtraction
	parameter $\kappa_g$.  Everywhere $D=0.45$ GeV, $\Gamma=0.2$ GeV, $\lambda=5$ MeV and $\kappa_q=1$. The PDFs are the MSTW2008LO68cl set. Data are from E866 binned with $4.2$ GeV $< M < 5.2$ GeV, $-0.05 < x_F < 0.15$. Only statistical errors are shown.}
	\label{fig:E866_triple_4.2_nlo_varkappag}	 
\end{figure}

To determine the subtraction parameters $\kappa_q$ and $\kappa_g$ we explore the dependence of the
cross section on these two parameters in Figs.\ \ref{fig:E866_triple_4.2_nlo_varkappaq} and 
\ref{fig:E866_triple_4.2_nlo_varkappag}. In the range of natural choices ($\kappa_q,\kappa_g = \frac{1}{2} \dots 2$) we find, that with $\kappa_q=1$ and $\kappa_g=2$ the data are desribed rather well. Although
Fig.\ \ref{fig:E866_triple_4.2_nlo_varkappaq} would indicate a better fit for smaller $\kappa_q$, we stick
to $\kappa_q=1$, since for this value the slope of the $M$ spectra fits also very well, as we show
for example in Sec.\ \ref{subsub:nlo-res-E605}.

\begin{figure}[H]
	\centering
	\includegraphics[angle=-90,keepaspectratio,width=0.45\textwidth]
                        {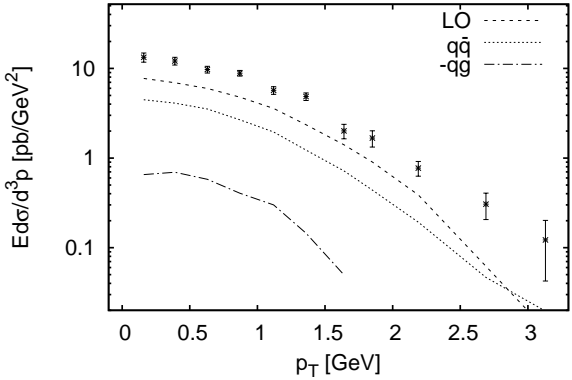}
			\caption{$p_T$ spectrum obtained from our model decomposed into the different contributions as described at the end of Sec.\ \ref{subsec:coll_sing}. Note that we plot the \emph{negative} quark-gluon contribution. Everywhere $D=0.45$ GeV, $\Gamma=0.2$ GeV, $\lambda=5$ MeV, $\kappa_q=1$ and $\kappa_g=2$. The PDFs are the MSTW2008LO68cl set. Data are from E866 binned with $4.2$ GeV $< M < 5.2$ GeV, $-0.05 < x_F < 0.15$. Only statistical errors are shown.}
	\label{fig:E866_triple_4.2_nlo_decompose}	 
\end{figure}	

In Fig.\ \ref{fig:E866_triple_4.2_nlo_decompose} the cross sections of the different contributions to the full result are plotted individually. The definition of the LO, quark-antiquark ($q\bar q$) and quark-gluon (qg) contributions
is given at the end of Sec.\ \ref{subsec:coll_sing}. Note that the sum of the LO contribution and the $q\bar q$ corrections make up most of the cross section. The contribution of the $qg$ correction is relatively small and negative for not too large $p_T$. 

\begin{figure}[H]
	\centering
	\includegraphics[angle=-90,keepaspectratio,width=0.45\textwidth]
                        {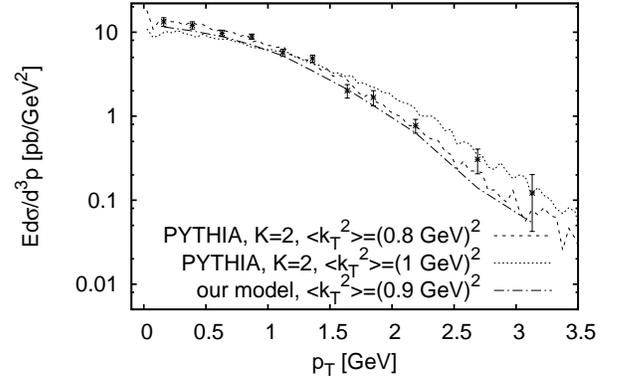}
	\caption{$p_T$ spectrum comparison of PYTHIA results with our full model. 
	         For the PYTHIA calculations version 6.225
	         with CTEQ5L PDFs were used. The PYTHIA result is plotted for two different values of the
		 average initial $k_T^2$ and multiplied by a factor $K=2$. Our model was calculated
		 with $D=0.45$ GeV, $\Gamma=0.2$ GeV, $\lambda=5$ MeV, $\kappa_q=1$, $\kappa_g=2$
		 and MSTW2008LO68cl PDF set. 
		 Data are from E866 binned with $4.2$ GeV $< M < 5.2$ GeV, $-0.05 < x_F < 0.15$. 
		 Only statistical errors are shown.}
	\label{fig:E866_Pythia}	 
\end{figure}

Fig.\ \ref{fig:E866_Pythia} shows a comparison of the results of our model with the results of a PYTHIA
event generator calculation. For this specific plot PYTHIA version 6.225 \cite{Sjostrand:2000wi} with
CTEQ5L PDFs \cite{Lai:1999wy} is used and the results are scaled up by a factor $K=2$. 
The comparison with the experimental data suggests a good
fit of the shape of the spectrum for a value of $\langle k_T^2\rangle =(0.8 \text{ GeV})^2$ in PYTHIA 
(internal
parameter $\text{PARP(91)}=0.8$). For our value of $D$ we get
$\langle k_T^2\rangle = (2 D)^2 = (0.9 \text{ GeV})^2$.
Although the PYTHIA parameter is obviously intended to have the same meaning as 
our definition for the average initial transverse momentum, the complex internal treatment of the interaction in the PYTHIA code leads to some numerical mismatch with our implementation.

\begin{figure}[H]
	\centering
	\includegraphics[angle=-90,keepaspectratio,width=0.45\textwidth]
                        {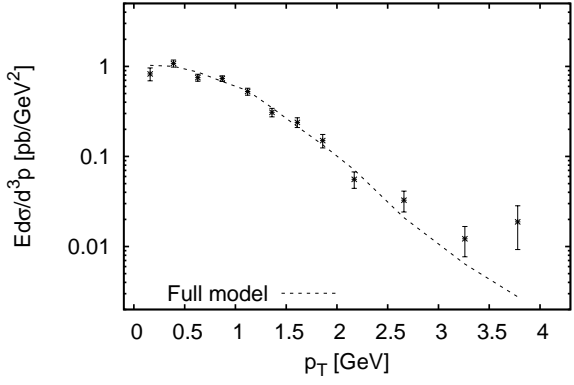}
	\caption{$p_T$ spectrum obtained from our full model. Everywhere $D=0.45$ GeV, $\Gamma=0.2$ GeV, $\lambda=5$ MeV, $\kappa_q=1$ and $\kappa_g=2$. The PDFs are the MSTW2008LO68cl set. Data are from E866 binned with $7.2$ GeV $< M < 8.7$ GeV, $-0.05 < x_F < 0.15$. Only statistical errors are shown.}
	\label{fig:E866_triple_7.2_nlo}	 
\end{figure}

For comparison we show our results for a different $M$ bin in Fig.\ \ref{fig:E866_triple_7.2_nlo}. With the parameters determined above our full model reproduces the measured spectrum very well. In contrast to the LO approach of Sec.\ \ref{sec:lo} we do not need a $K$ factor to describe the absolute height of the spectrum. At the same time the width $D$ of the intrinsic (non-perturbative) $k_T$ distributions changes only little ($D=0.5$ GeV
vs. $D=0.45$ GeV) when passing from the LO to the NLO calculation. Note, however, that Fig.\ \ref{fig:E866_triple_4.2_nlo_decompose} indicates, that at least the results for the $q\bar q$ corrections deviate from a Gaussian behavior at $p_T \ge 3$ GeV. Thus the contributions of some of the hard NLO processes are only significant in the high (perturbative) $p_T$ region. Therefore, to describe the low $p_T$ regime at NLO one still requires almost the same non-perturbative input for the intrinsic parton $k_T$, i.e., only little transverse momentum is generated dynamically and the width parameter $D$ does not change considerably.

\subsection{E866 - $M$ spectrum}
\label{subsubsec:nlo-res-E866-M}

In this section we compare our results with the measured $M$ spectrum from E866, which are basically the $p_T$ spectra integrated over $p_T$. The double-differential cross section is given by the E866 collaboration as
\begin{equation}
	M^3\frac{\diffd \sigma}{\diffd M \diffd x_F}\ .         \label{eq:E866_double_xsec}	
\end{equation}	
Again the data are given in several bins of $M$ and $x_F$ and for every datapoint the 
average values $\left< M \right>$ and $\left< x_F \right>$ are provided. 
For the different contributions in our model we calculate the quantity
of Eq.~(\ref{eq:E866_double_xsec}) by integrating over $p_T^2$ for every datapoint using these 
averaged values:
\begin{align}
	& M^3\frac{\diffd \sigma}{\diffd M \diffd x_F}
 	\rightarrow \left<M\right>^3 \int_0^{(p_T)^2_\textrm{max}} \diffd p_T^2
	      \frac{\diffd \sigma}{\diffd M \diffd x_F \diffd p_T^2} \nonumber \\
	&=\left<M\right>^3 \int_0^{(p_T)^2_\textrm{max}} \diffd p_T^2\ 
	 2 \left<M\right> \frac{\diffd \sigma}{\diffd M^2 \diffd x_F \diffd p_T^2} 
           \left(\left< M \right>, \left< x_F \right> \right) \ .
	   \label{eq:exp_double_xsec}
\end{align}
The maximal possible $p_T$ is determined by the kinematics to
\begin{align}
	P_1 + P_2 &= q + X \\
	\Rightarrow (P_1 + P_2 - q)^2 &= X^2 = M_R^2 \\
	\Rightarrow S + M^2 - M_R^2 &= 2\ (P_1 + P_2)\ q \nonumber \\
	                            &= 2 \sqrt{S} E \nonumber \\
				    &= 2 \sqrt{S} \sqrt{M^2 + (p_T)^2_\textrm{max}
				                        + x_F^2 (q_z)_\text{max}^2}
\end{align}							
\begin{align}
	\Rightarrow E^2 &=  M^2 + (p_T)^2_\textrm{max} + x_F^2 (q_z)_\text{max}^2 \\
	              &= \frac{(S+M^2-M_R^2)^2}{4S} \\	    
	\Rightarrow (p_T)^2_\textrm{max} &= \frac{(S+M^2-M_R^2)^2}{4S} - M^2 - x_F^2 (q_z)_\text{max}^2 \ ,
	\label{eq:pt2max}
\end{align}	
where $M_R^2$ is the minimal invariant mass of the undetected remnants. For pp and pn collisions we choose a value
of $M_R=2m_N$ and for $\overline{\text{p}}$p a value of $M_R=1.1$ GeV, which can be interpreted as two times
a diquark mass. Note that at c.m.\ energies of $\sqrt{S} \approx 15.3$ GeV (E537), $\sqrt{S} \approx 27.4$ GeV (E288, E439)
and $\sqrt{S} \approx 38.8$ GeV (E605, E772, E866) we are not really sensitive to these values if they 
stay at or below a few GeV.

\begin{figure}[H]
	\centering
	\includegraphics[angle=-90,keepaspectratio,width=0.45\textwidth]
                        {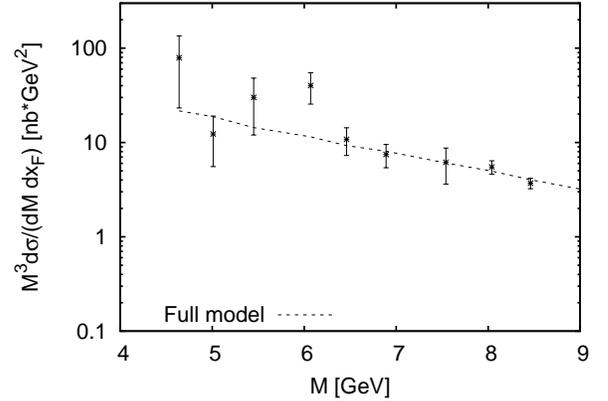}
	\caption{$M$ spectrum obtained from our full model.
		 Everywhere $D=0.45$ GeV, $\Gamma=0.2$ GeV, $\lambda=5$ MeV,
		 $\kappa_q=1$ and $\kappa_g=2$.
		 The PDFs are the MSTW2008LO68cl set.
		 Data are from E866 binned with $-0.05 < x_F < 0.05$.
		 Only statistical errors shown.}
	\label{fig:E866_M}	 
\end{figure}

In Fig.\ \ref{fig:E866_M} we compare our result for the double differential cross section to the data from E866. The slope and absolute height of the curve agree with the data quite well. However, since here the experimental error bars are rather large, we will make comparisons to $M$ spectra from other experiments in Secs.\ \ref{subsub:nlo-res-E605} and \ref{subsub:nlo-res-E439} to test our model further.

\subsection{E772}
\label{subsub:nlo-res-E772}
Experiment E772 \cite{McGaughey:1994dx} measured dimuon production in pd collisions at $S \approx 1500$ GeV$^2$. For the calculation of the triple differential cross section we again use Eq.~(\ref{eq:exp_triple_xsec}) and for the average values of $M$ and $x_F$ we use the center of the $M$ and $x_F$ bins. Since the experiment was done on deuterium we have calculated pp and pn cross sections and then averaged.
\begin{figure}[H]
	\centering
	\includegraphics[angle=-90,keepaspectratio,width=0.45\textwidth]
                        {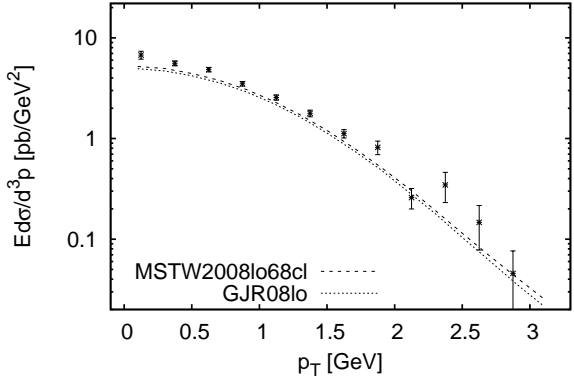}
	\caption{$p_T$ spectrum obtained from our full model with different PDF sets. Everywhere $D=0.45$ GeV, $\Gamma=0.2$ GeV, $\lambda=5$ MeV, $\kappa_q=1$ and $\kappa_g=2$. Data are from E772 binned with $5$ GeV $< M < 6$ GeV, $0.1 < x_F < 0.3$. Only statistical errors are shown.}
	\label{fig:E772_triple_5_nlo_pdfcomp}	 
\end{figure}	

\begin{figure}[H]
	\centering
	\includegraphics[angle=-90,keepaspectratio,width=0.45\textwidth]
                        {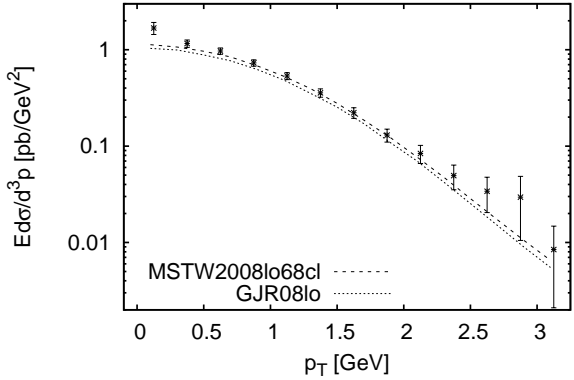}
	\caption{$p_T$ spectrum obtained from our full model with different PDF sets. Everywhere $D=0.45$ GeV, $\Gamma=0.2$ GeV, $\lambda=5$ MeV, $\kappa_q=1$ and $\kappa_g=2$. Data are from E772 binned with $7$ GeV $< M < 8$ GeV, $0.1 < x_F < 0.3$. Only statistical errors are shown.}
	\label{fig:E772_triple_7_nlo_pdfcomp}	 
\end{figure}

In Figs.~\ref{fig:E772_triple_5_nlo_pdfcomp} and \ref{fig:E772_triple_7_nlo_pdfcomp} we compare the results of our full model with different PDF sets to triple differential data from E772 in different $M$ bins. Agreement is again quite well, however, the shape of the spectrum seems to favor a slightly smaller value for $D$, which would enhance the spectrum near $p_T=0$. Nevertheless, we have chosen $D=0.45$ GeV, since this value allows us to describe the data from several different experiments with only minor deviations.

\subsection{E605}
\label{subsub:nlo-res-E605}
Experiment E605 \cite{Moreno:1990sf} measured dimuon production in pCu collisions at $S \approx 1500$ GeV$^2$. For the calculation of the triple differential cross section we again use Eq.~(\ref{eq:exp_triple_xsec}) and for the average value of $M$ we use the center of the $M$ bin. For the $p_T$ spectrum E605 gives $x_F=0.1$. For the double differential cross section we use Eq.~(\ref{eq:exp_double_xsec}). Since the experiment was done on copper we have calculated pp and pn cross sections and then averaged (29 protons and 34 neutrons).

\begin{figure}[H]
	\centering
	\includegraphics[angle=-90,keepaspectratio,width=0.45\textwidth]
                        {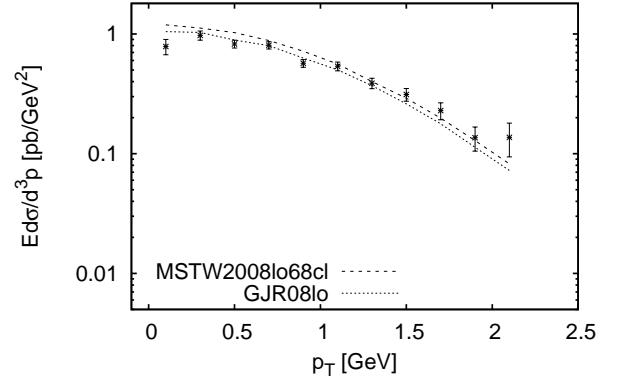}
	\caption{$p_T$ spectrum obtained from our full model with different PDF sets. Everywhere $D=0.45$ GeV, $\Gamma=0.2$ GeV, $\lambda=5$ MeV, $\kappa_q=1$ and $\kappa_g=2$. Data are from E605 binned with $7$ GeV $< M < 8$ GeV, $x_F=0.1$. Only statistical errors are shown.}
	\label{fig:E605_triple_7_nlo_pdfcomp}	 
\end{figure}	
In Fig.~\ref{fig:E605_triple_7_nlo_pdfcomp} we compare the results of our full model with different PDF sets to triple differential data from E605. Here the shape of the spectrum confirms our chosen value for $D$ and the overall agreement is good.

\begin{figure}[H]
	\centering
	\includegraphics[angle=-90,keepaspectratio,width=0.45\textwidth]
                        {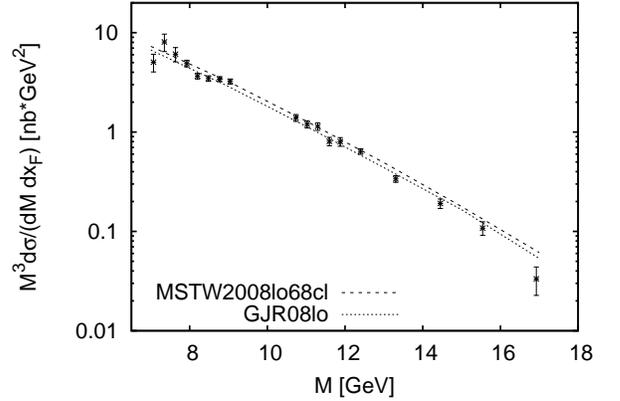}
	\caption{$M$ spectrum obtained from our full model.
		 Everywhere $D=0.45$ GeV, $\Gamma=0.2$ GeV, $\lambda=5$ MeV, $\kappa_q=1$
		 and $\kappa_g=2$. 
		 Data are from E605 with $x_F=0.125$.
		 Only statistical errors shown.}
	\label{fig:E605_M}	 
\end{figure}	
In Fig.\ \ref{fig:E605_M} we plot our result with different PDF sets for the double differential cross section together with the data from E605. Again agreement is quite good over the entire range of $M$.

\subsection{E288}
\label{subsub:nlo-res-E288}

Experiment E288 \cite{Ito:1980ev} measured dimuon production in pA collisions at $S \approx 750$ GeV$^2$. For the calculation of the triple differential cross section we again use Eq.~(\ref{eq:exp_triple_xsec}) and for the average value of $M$ we use the center of the $M$ bin. For the $p_T$ spectrum E288 gives for the rapidity $y=0.03$ and we thus chose $x_F=0$ for our calculation. The experiment was done on different nuclei and only data averaged over the results from these nuclei have been presented. Therefore, we have calculated pp cross sections only.

\begin{figure}[H]
	\centering
	\includegraphics[angle=-90,keepaspectratio,width=0.45\textwidth]
                        {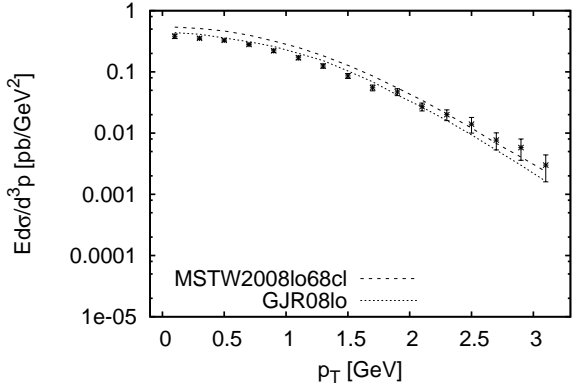}
	\caption{$p_T$ spectrum obtained from our full model with different PDF sets. Everywhere $D=0.45$ GeV, $\Gamma=0.2$ GeV, $\lambda=5$ MeV, $\kappa_q=1$ and $\kappa_g=2$. Data are from E288 binned with $7$ GeV $< M < 8$ GeV, $y=0.03$, we have chosen $x_F=0$ in our calculation. Only statistical errors are shown.}
	\label{fig:E288_triple_7_nlo_pdfcomp}	 
\end{figure}	
In Fig.\ \ref{fig:E288_triple_7_nlo_pdfcomp} we compare the results of our full model with different PDF sets to the triple differential data from E288. Again the agreement with the data is good and confirms our choice of parameters.

\subsection{E439}
\label{subsub:nlo-res-E439}
\begin{figure}[H]
	\centering
	\includegraphics[angle=-90,keepaspectratio,width=0.45\textwidth]
                        {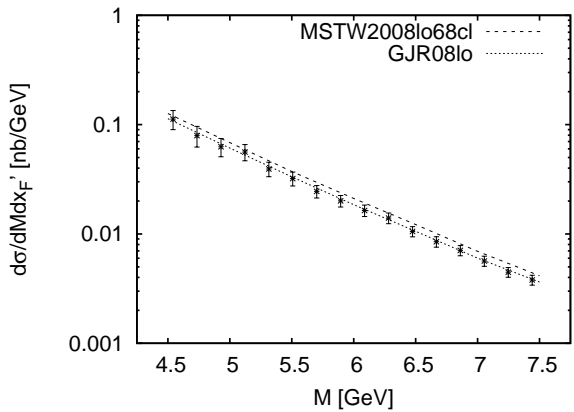}
	\caption{$M$ spectrum obtained from our full model with different PDF sets. Everywhere $D=0.45$ GeV, $\Gamma=0.2$ GeV, $\lambda=5$ MeV, $\kappa_q=1$ and $\kappa_g=2$. Data are from E439 with $x_F' = 0.1$. Only statistical errors are shown.}
	\label{fig:E439_nlo_double}	 
\end{figure}	

The details of experiment E439 and how we calculate the cross section are given in Sec.\ \ref{subsubsec:E439_lo}. In Fig.\ \ref{fig:E439_nlo_double} we compare our results for the double differential cross section with the data from E439. For both PDF sets the absolute height and slope agrees well with the data.

\subsection{E537 (Antiprotons)}
\label{subsub:nlo-res-E537}
Experiment E537 \cite{Anassontzis:1987hk} measured dimuon production in $\overline{\text{p}}$W collisions at $S \approx 236$ GeV$^2$ in an invariant mass range of $4 < M < 9$ GeV. The obtained cross sections are double differential in two of the observables $M$, $x_F$ and $p_T^2$. To calculate the cross sections differential in $p_T^2$ with our model we use
\begin{align}
	\frac{\diffd \sigma}{\diffd x_F \diffd p_T^2}
       &\rightarrow \int_{M^2\textrm{-bin}}
         \frac{\diffd \sigma}{\diffd M^2 \diffd x_F \diffd p_T^2} \diffd M^2 \nonumber \\
       &\approx \sum_i 	\Delta M_i^2
                \frac{\diffd \sigma}{\diffd M^2 \diffd x_F \diffd p_T^2}\left(\left< M_i \right>,
									      \left< x_F \right>,
									      \left< p_T \right>
									\right)
									  \ .
       \label{eq:E537_exp_pT_xsec}
\end{align}
The sum runs over several mass bins, which we choose as $M=4\dots5,5\dots6,6\dots7,7\dots8,8\dots9$ GeV and in each bin
we take the central value for $\left< M_i \right>$. Since the experiment was done on tungsten we have calculated $\overline{\text{p}}$p and $\overline{\text{p}}$n cross sections and then averaged (74 protons and 110 neutrons).

\begin{figure}[H]
	\centering
	\includegraphics[angle=-90,keepaspectratio,width=0.45\textwidth]
                        {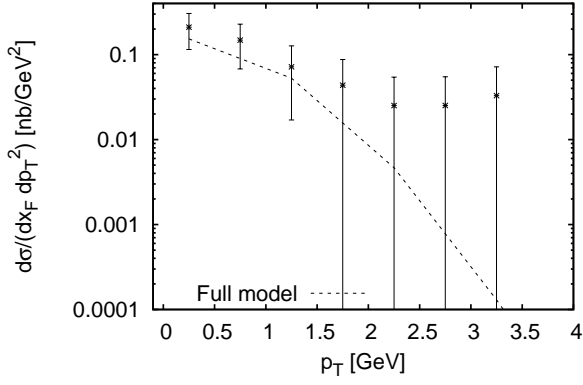}
	\caption{$p_T$ spectrum obtained from our full model. Everywhere $D=0.45$ GeV, $\Gamma=0.2$ GeV, $\lambda=5$ MeV, $\kappa_q=1$ and $\kappa_g=2$. The PDFs are the MSTW2008LO68cl set. Data are from E537 with $4$ GeV $< M < 9$ GeV, $0 < x_F < 0.1$. We have chosen $x_F=0.05$ in our calculation. Only statistical errors are shown.}
	\label{fig:E537_pT_nlo}	 
\end{figure}

We compare the calculated $p_T$ spectrum with the data in Fig.~\ref{fig:E537_pT_nlo}. Our full model is on the lower side of the error bars of the data. However, one should note that the experimental error bars are rather large and thus the possibility to confirm or rule out our model is limited.

\begin{figure}[H]
	\centering
	\includegraphics[angle=-90,keepaspectratio,width=0.45\textwidth]
                        {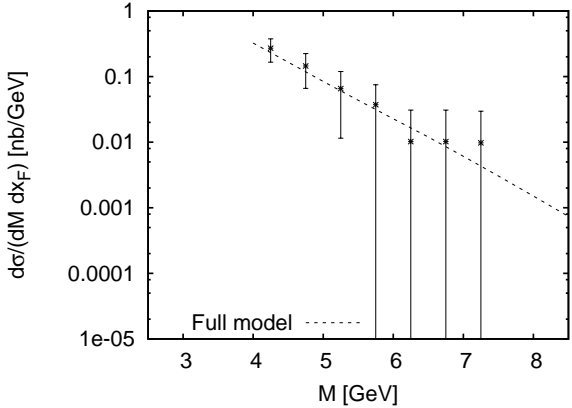}
	\caption{$M$ spectrum obtained from our full model. Everywhere $D=0.45$ GeV, $\Gamma=0.2$ GeV, $\lambda=5$ MeV, $\kappa_q=1$ and $\kappa_g=2$. The PDFs are the MSTW2008LO68cl set. Data are from E537 with $0 < x_F < 0.1$. We have chosen $x_F=0.05$ in our calculation. Only statistical errors are shown.}
	\label{fig:E537_M_nlo}	 
\end{figure}

To calculate the $M$ spectra we use
\begin{align}
	\frac{\diffd \sigma}{\diffd M \diffd x_F}
 	=&\int_0^{(p_T)^2_\textrm{max}} \diffd p_T^2
	      \frac{\diffd \sigma}{\diffd M \diffd x_F \diffd p_T^2} \nonumber \\
	=&\int_0^{(p_T)^2_\textrm{max}} \diffd p_T^2\ 
	 2 M  \frac{\diffd \sigma}{\diffd M^2 \diffd x_F \diffd p_T^2} 
           \left(M, x_F \right) \ ,
\end{align}
with $(p_T)^2_\textrm{max}$ given in Eq.~(\ref{eq:pt2max}). We compare our calculated $M$ spectrum with the data in Fig.~\ref{fig:E537_M_nlo}. Agreement is better than for the $p_T$ spectrum, however, the experimental error bars are again large compared to proton data.

\subsection{Prediction for $\overline{\text{P}}$ANDA}
\label{subsub:nlo-res-panda}
Based on the parameters which we have fixed on the available data above, we here present our predictions for DY pair production at $S=30 \text{ GeV}^2$ in $\overline{\text{p}}$p collisions, where, for example, $\overline{\textrm{P}}$ANDA \cite{Lutz:2009ff} will measure.

\begin{figure}[H]
	\centering
	\includegraphics[angle=-90,keepaspectratio,width=0.4\textwidth]
                        {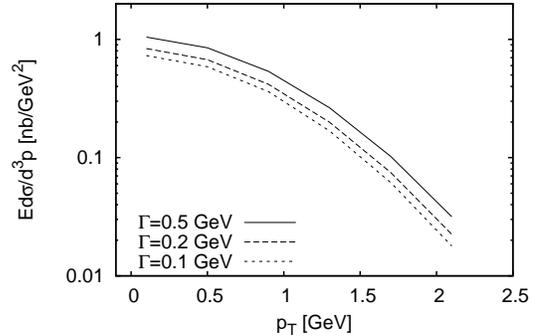}
	\caption{$p_T$ spectrum obtained from our full model for different values of $\Gamma$. Everywhere $D=0.45$ GeV, $\lambda=5$ MeV, $\kappa_q=1$ and $\kappa_g=2$. The PDFs are the MSTW2008LO68cl set. $x_F' = 0$ and $1.5$ GeV $< M < 2.5$ GeV.}
	\label{fig:Panda_pT_M2_varGamma}	 
\end{figure}

\begin{figure}[H]
	\centering
	\includegraphics[angle=-90,keepaspectratio,width=0.4\textwidth]
                        {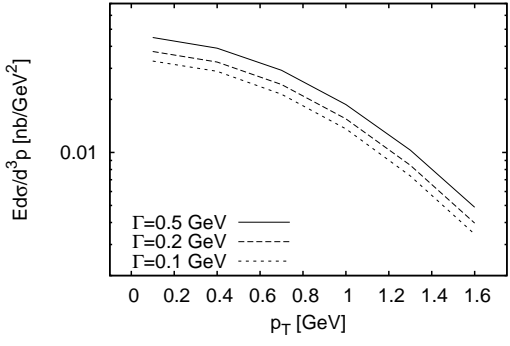}
	\caption{$p_T$ spectrum obtained from our full model for different values of $\Gamma$. Everywhere $D=0.45$ GeV, $\lambda=5$ MeV, $\kappa_q=1$ and $\kappa_g=2$. The PDFs are the MSTW2008LO68cl set. $x_F' = 0$ and $2.5$ GeV $< M < 3.5$ GeV.}
	\label{fig:Panda_pT_M3_varGamma}	 
\end{figure}

For the calculation of the triple differential cross section we use a modified version of Eq.~(\ref{eq:exp_triple_xsec}):
\begin{align}
       &\frac{2\sqrt{S}E}{\pi (S - M^2)} \frac{\diffd \sigma}{\diffd x_F' \diffd p_T^2} \nonumber \\
       &\rightarrow\frac{2\sqrt{S}E}{\pi (S - M^2)}\int_{M^2\textrm{-bin}}
         \frac{\diffd \sigma}{\diffd M^2 \diffd x_F' \diffd p_T^2} \diffd M^2 \nonumber \\
       &\approx 	\frac{2\sqrt{S}E}{\pi (S - \left<M\right>^2)} \cdot \Delta M^2
                \frac{\diffd \sigma}{\diffd M^2 \diffd x_F' \diffd p_T^2}\left(\left< M \right>,
									      x_F',
									       p_T
									\right)
									  \ ,
       \label{eq:PANDA_triple_xsec}
\end{align}
where
\begin{equation}
      E = \sqrt{\left< M \right>^2 + p_T^2 + 
	         (x_F')^2 \left<(q_z')_\textrm{max}\right>^2 }
\end{equation}
and $\Delta M^2 = M_\textrm{max}^2 - M_\textrm{min}^2$ with $M_\textrm{max}$ ($M_\textrm{min}$)
the upper (lower) limit of the bin. For the average value of $M$ we use the center of the $M$ bin and we choose everywhere $x_F'=0$. In Figs.~\ref{fig:Panda_pT_M2_varGamma}-\ref{fig:Panda_pT_M3_varGamma} we show our predictions for different values of $\Gamma$ in different $M$ bins. Note that while at E866 energies we could not discriminate between different $\Gamma$ over a wide range
(cf.\ Fig.\ \ref{fig:E866_triple_4.2_nlo_varGamma}), the results here become more sensitive to this
parameter.

\begin{figure}[H]
	\centering
	\includegraphics[angle=-90,keepaspectratio,width=0.4\textwidth]
                        {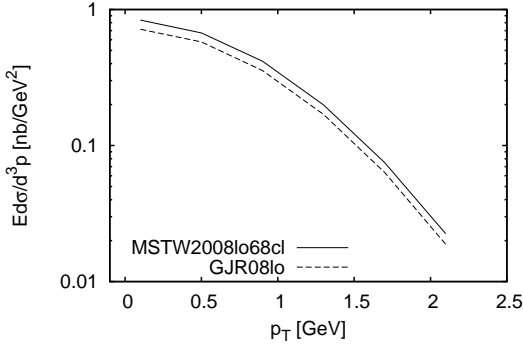}
	\caption{$p_T$ spectrum obtained from our full model for different PDF sets. Everywhere $D=0.45$ GeV, $\Gamma=0.2$ GeV, $\lambda=5$ MeV, $\kappa_q=1$, $\kappa_g=2$. $x_F' = 0$ and $1.5$ GeV $< M < 2.5$ GeV.}
	\label{fig:Panda_pT_M2_PDFcomp}	 
\end{figure}

\begin{figure}[H]
	\centering
	\includegraphics[angle=-90,keepaspectratio,width=0.4\textwidth]
                        {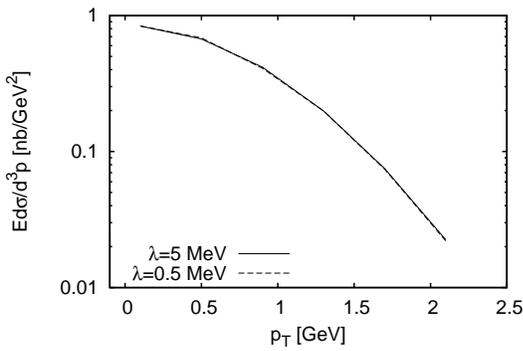}
	\caption{$p_T$ spectrum obtained from our full model for different values of $\lambda$. Everywhere $D=0.45$ GeV, $\Gamma=0.2$ GeV, $\kappa_q=1$ and $\kappa_g=2$. The PDFs are the MSTW2008LO68cl set. $x_F' = 0$ and $1.5$ GeV $< M < 2.5$ GeV. Note that both curves are basically on top of each other.}
	\label{fig:Panda_pT_M2_varomega}	 
\end{figure}

In Fig.\ \ref{fig:Panda_pT_M2_PDFcomp} we compare results for different PDFs and the uncertainty induced
by the choice of different PDFs is comparable to the uncertainties we found for high energies (e.g.\ at E772). 
To ensure that at such low hadronic energies the fictitious gluon mass is still small enough, we study our model at different values of $\lambda$, see Fig.\ \ref{fig:Panda_pT_M2_varomega}. The results coincide, indicating that our standard choice of $\lambda = 5$ MeV is still applicable at these energies.

\begin{figure}[H]
	\centering
	\includegraphics[angle=-90,keepaspectratio,width=0.4\textwidth]
                        {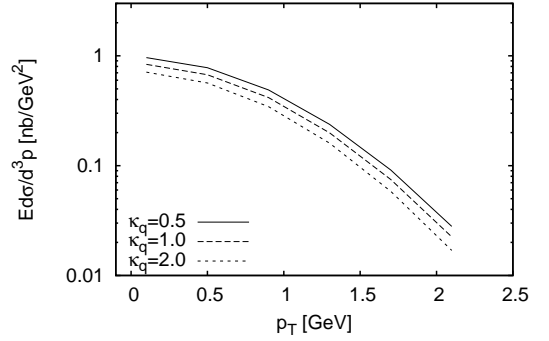}
	\caption{$p_T$ spectrum obtained from our full model with different values of the subtraction
	         parameter $\kappa_q$. Everywhere $D=0.45$ GeV, $\Gamma=0.2$ GeV, $\lambda=5$ MeV
and $\kappa_g=2$. The PDFs are the MSTW2008LO68cl set. $x_F' = 0$ and $1.5$ GeV $< M < 2.5$ GeV.}
	\label{fig:Panda_pT_M2_varkappaq}
\end{figure}

\begin{figure}[H]
	\centering
	\includegraphics[angle=-90,keepaspectratio,width=0.4\textwidth]
                        {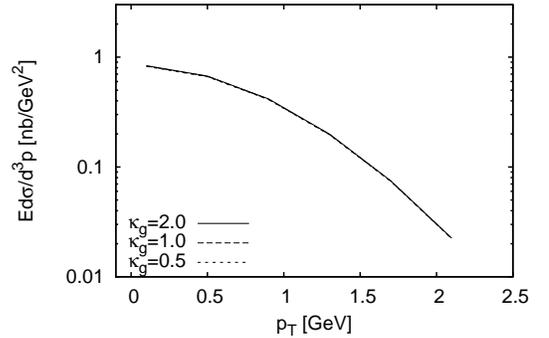}
	\caption{$p_T$ spectrum obtained from our full model with different values of the subtraction
	         parameter $\kappa_g$. Everywhere $D=0.45$ GeV, $\Gamma=0.2$ GeV, $\lambda=5$ MeV
and $\kappa_q=1$. The PDFs are the MSTW2008LO68cl set. $x_F' = 0$ and $1.5$ GeV $< M < 2.5$ GeV.
Note that the curves are practically on top of each other.}
	\label{fig:Panda_pT_M2_varkappag}
\end{figure}

To check the dependence of our results on the choice of the subtraction parameters $\kappa_q$ and $\kappa_g$ (see Sec.\ \ref{subsec:coll_sing} for details) at the low hadronic energies of the $\overline{\textrm{P}}$ANDA kinematics, we again vary one of the two parameters and keep the other one fixed and show our results
in Figs.\ \ref{fig:Panda_pT_M2_varkappaq} and \ref{fig:Panda_pT_M2_varkappag}. The results for
different $\kappa_q$ deviate by about $15\%$, which is comparable to the deviation at E866 energies.
However, the results are practically insensitive to variations in $\kappa_g$.

\begin{figure}[H]
	\centering
	\includegraphics[angle=-90,keepaspectratio,width=0.4\textwidth]
                        {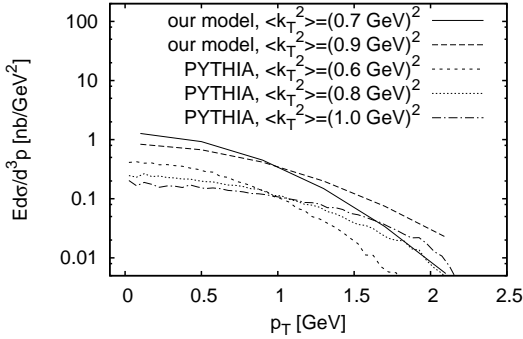}
	\caption{$p_T$ spectrum obtained from our full model and from PYTHIA (see main text for details) 
	         for different values of the
	         average initial $k_T$. Our predictions were calculated with $\Gamma=0.2$ GeV, $\lambda=5$ MeV, $\kappa_q=1$, $\kappa_g=2$ and the MSTW2008LO68cl PDF set. $x_F' = 0$ and $1.5$ GeV $< M < 2.5$ GeV.}
	\label{fig:Panda_pT_M2_PYTHIA}	 
\end{figure}

Finally in Fig.\ \ref{fig:Panda_pT_M2_PYTHIA} we compare our predictions with a PYTHIA calculation 
(PYTHIA version 6.225, CTEQ5L PDFs), each for different values of the average initial $k_T$.
As explained above, 
PYTHIA calculations for E866 conditions seem to prefer a somewhat smaller width for the initial $k_T$ 
distribution, 
$\langle k_T^2 \rangle = (0.8 \text{ GeV})^2$ instead of $(0.9 \text{ GeV})^2$.
Since various calculations (\cite{GallmeisterDiss},\cite{Gallmeister:2009ht}) hint to some monotonic 
dependence of the initial 
$k_T$ with the underlying $\sqrt{S}$, we would expect that at $\overline{\text{P}}$ANDA
energies, a somewhat smaller 
$\langle k_T^2 \rangle$ should be used. Therefore, in Fig.\ \ref{fig:Panda_pT_M2_PYTHIA}, 
we also show calculations with
$\langle k_T^2 \rangle = (0.6 \text{ GeV})^2$ for PYTHIA and $\langle k_T^2 \rangle = (0.7 \text{ GeV})^2$ 
($D=0.35 \text{ GeV}$) for our model.  
For PYTHIA, already with $\langle k_T^2 \rangle = (0.6 \text{ GeV})^2$, the difference in the functional 
behaviour is rather large compared to the PYTHIA calculations with the higher parameter values. 
This may be taken as some 
hint for the theoretical uncertainties. On the other hand, the intrinsic $k_T$ in PYTHIA is some effective 
parameter. It is not clear, whether this parameter should follow the same energy dependence in pp and in 
p$\overline{\text{p}}$ collisions, since multiple effects are encoded. Note, that the PYTHIA results shown in
Fig.\ \ref{fig:Panda_pT_M2_PYTHIA} were not multiplied by a $K$ factor, i.e.\ $K=1$.

\section{Conclusions}
\label{sec:conc}
In this paper we have extended a phenomenological model of DY pair production \cite{Eichstaedt:2009cn}. We have included all relevant processes up to NLO in the strong coupling $\alpha_s$ to account for the missing strength in the LO calculation ($K$ factor). To describe DY pair $p_T$ spectra we introduced initial parton transverse momentum distributions and to regularize the otherwise divergent $p_T$ spectra at NLO we distributed the masses of the quarks with spectral functions. To avoid double-counting we introduced a subtraction
scheme which removes those $O(\alpha_s)$ contributions which are usually absorbed into the renormalized
quark PDFs.

The results show that with our choice for the width of the initial parton transverse momentum distribution, $D\approx 0.45$ GeV ($\langle k_T^2 \rangle \approx (0.9 \text{ GeV})^2$), the shape of the $p_T$ spectra is reproduced very well. We note, however, that this width might be $S$ dependent, which introduces additional
uncertainties. The spectral functions (quark mass distributions) serve as a regulator for the NLO order processes, which for massless quarks are divergent for $p_T\rightarrow 0$. However, at high energies (for example at E866) over a wide range the results depend only weakly on the width $\Gamma$ of the mass distribution. 
In addition we could fix the subtraction parameters $\kappa_q$ and $\kappa_g$ at E866 energies and found that
they are of natural magnitude ($O(1)$).

In summary we found that in our phenomenological model we can reproduce measured $p_T$ and $M$ spectra for DY pair production in pp and pA reactions at different hadronic energies {\it without} the need for a $K$ factor. The comparison to $\overline{\textrm{p}}$A data from E537 is inconclusive, however, our full model is still within the large (compared to the pp and pA data) error bars. The general agreement of our results with the data from different experiments indicates that we effectively have parametrized the soft initial state interactions in the nucleon by fixing our parameters on the available data. Using this framework we have obtained predictions for DY pair production at the low hadronic energy regime, where future experiments, for example $\overline{\textrm{P}}$ANDA, are aiming at. We have found that our predictions become more sensitive to the mass distribution width $\Gamma$, which we could not reliably fix at higher energies. In addition we found some sensitivity
on the subtraction parameter $\kappa_q$, which is comparable to the finding at high energies (E866).
Nevertheless, our model provides a narrow band of estimates for the fully differential DY pair production cross section at low energies.

\section*{Acknowledgements}

The work of H.v.H.\ was supported by the Hessian LOEWE initiative through the Helmholtz
International Center for FAIR.
F.E.\ was supported by HGS-HIRe. This publication
represents a component of my (F.E.) doctoral (Dr. rer. nat.) thesis in the Faculty
of Physics at the Justus-Liebig-University Giessen, Germany.

\begin{appendix}

\begin{widetext}
\section{Gauge invariance}
\label{app:gauge}

\subsection{Electromagnetic sector}
\label{app:gauge-em}
By assigning to the annihilating quark and antiquark different masses $m_1$ and $m_2$,
see Fig.~\ref{fig:vertex-correction}, gauge invariance is broken at the quark-photon vertex. One can easily see this by contracting the quark current with the photon momentum $q_\mu$:
\begin{align}
	&\, \bar v(p_2,m_2)\,\gamma^\mu\,u(p_1,m_1) \cdot q_\mu \nonumber \\
	=&\,	\bar v(p_2,m_2)\,\slashed{q}\, u(p_1,m_1) \nonumber \\
	=&\,	\bar v(p_2,m_2)\,(\slashed{p}_1 +\slashed{p}_2)\,u(p_1,m_1) \nonumber \\
	=&\,	\bar v(p_2,m_2)\,(m_1-m_2)\,u(p_1,m_1)  \neq 0\ .
\end{align}
However, in the full amplitudes for DY pair production of Secs.~\ref{sec:lo} and \ref{sec:nlo}
this is not an issue since gauge invariance
is preserved at the lepton-photon vertex. To realize this one has to 
look at the gauge dependent part of the photon propagator. The propagator has the following Lorentz structure:
\begin{align}
  G(q) \sim \left(g_{\mu\nu} - \xi \frac{q_\mu q_\nu}{q^2}\right) , \
  \label{eq:photon-propagator-structure}
\end{align}
with a gauge parameter $\xi$.
Now we insert the gauge dependent part of Eq.~(\ref{eq:photon-propagator-structure}) between the quark and lepton currents and exploit the Dirac equation:
\begin{align}
	&\,\bar v(p_2,m_2)\,\gamma^\mu\,u(p_1,m_1) \cdot
	 \left(q_\mu q_\nu\right) \cdot
	 \bar u(k_1,m)\,\gamma^\nu\,v(k_2,m) \nonumber \\
        =&\, \bar v(p_2,m_2)\,\slashed{q}\, u(p_1,m_1) \cdot
             \bar u(k_1,m)\,\slashed{q}\, v(k_2,m) \nonumber \\
	=&\, \bar v(p_2,m_2)\,(\slashed{p}_1 +\slashed{p}_2)\,u(p_1,m_1)
             \cdot
     	     \bar u(k_1,m)\,(\slashed{k}_1+\slashed{k}_2)\, v(k_2,m) \nonumber \\
        =&\, \bar v(p_2,m_2)\,(m_1-m_2)\,u(p_1,m_1)
	     \cdot
	     \bar u(k_1,m)\,(m-m)\, v(k_2,m) \nonumber \\
        =&\, 0 \ ,
\end{align}	
and so the amplitude is invariant under gauge transformations of the electromagnetic field.
\subsection{Strong sector}
\label{app:gauge-qcd}
As shown in Figs.~\ref{fig:vertex-correction}, \ref{fig:wave-renorm}, \ref{fig:gluon-brems}
and \ref{fig:gluon-compton}, we always keep the quark masses fixed at any quark-gluon vertex.
This ensures that all our calculations are also gauge invariant in the strong sector.
This is trivial for the gluon bremsstrahlung and gluon Compton scattering processes.
For the vertex correction (VC) and self energy diagrams the proof proceeds as follows:
We denote the LO ($\alpha_s^0$) DY amplitude with $M_\text{LO}$ and the vertex correction amplitude with $M_\text{VC}$. Then 
\begin{align}
	M_\text{LO} &\sim \bar v(p_2,m_2)\, \gamma^\mu \, u(p_1,m_1) \cdot L_\mu(q^2) \ , \\
	M_\text{VC} &\sim \bar v(p_2,m_2)\, \Gamma^\mu(q^2) \, u(p_1,m_1) \cdot 
	    L_\mu(q^2) \ , 
\end{align}
where $L_\mu$ is the leptonic part and $\Gamma^\mu$ is of order $\alpha_s$. Also
of order $\alpha_s$ are the interferences of the LO process with the self energy diagrams in Fig.~\ref{fig:wave-renorm} and we have to 
include them by dressing the external quark legs with field strength
renormalization factors of $\sqrt{Z_2(m_i)}$ with
$Z_2 = 1 + \delta Z_2 + O(\alpha_s^2)$.
Then to order $\alpha_s$ the amplitude for the process $q\bar q \rightarrow l^+l^-$
can be written as
\begin{align}
	M &= \sqrt{Z_2(m_1)}\sqrt{Z_2(m_2)} \cdot 
	        \bar v(p_2,m_2)\, \left(i e_q \gamma^\mu + \Gamma^\mu(q^2)\right) 
                \, u(p_1,m_1) \cdot L_\mu(q^2) \nonumber \\
	  &= \left(1+ \frac{1}{2}\delta Z_2(m_1) + O(\alpha_s^2)\right) 		           \left(1+ \frac{1}{2}\delta Z_2(m_2) + O(\alpha_s^2)\right)
	     \cdot 	
		\bar v(p_2,m_2)\, \left(i e_q \gamma^\mu + \Gamma^\mu(q^2)\right) 
                \, u(p_1,m_1) \cdot L_\mu(q^2) \nonumber \\
	  &= \bar v(p_2,m_2)\, 
	        \left[i e_q \gamma^\mu\left(1+\frac{1}{2}\delta Z_2(m_1)
				  +\frac{1}{2}\delta Z_2(m_2)\right)
				  + \Gamma^\mu(q^2)\right] 
                \, u(p_1,m_1) \cdot L_\mu(q^2)
	     + O(\alpha_s^2) \label{qcd_gi_master}
\end{align}
To prove gauge invariance we now have to calculate the gluon gauge dependent
parts of $\Gamma^\mu$ and $\delta Z_2$, which we denote by the index $\text{g}$. 
We begin with $\Gamma^\mu_\text{g}$. We denote the gluon momentum with $k$, insert only the gauge dependent part of the gluon propagator ($k^\alpha k^\beta$) and again exploit the Dirac equation:
\begin{align}
	 \bar v(p_2,m_2)\, \Gamma^\mu_\text{g}\, u(p_1,m_1)
	            =&\, \int \diffd^4k\, \bar v(p_2,m_2)\, (i g \gamma_\alpha t^a)
	                 \frac{i(-\slashed{p}_2 - \slashed{k}+m_2)}{(p_2+k)^2-m_2^2}
			 i e_q \gamma^\mu
	                 \frac{i(\slashed{p}_1 - \slashed{k}+m_1)}{(p_1-k)^2-m_1^2}
			 (i g \gamma_\beta t^a)\, u(p_1,m_1)
			 \cdot \frac{i k^\alpha k^\beta}{k^2} \frac{1}{k^2} \nonumber \\
		    =&\, - e_q 4\pi \alpha_s (t^a t^a)
		         \int \diffd^4k\, \bar v(p_2,m_2)\, \slashed{k} 
	                 \frac{-\slashed{p}_2 - \slashed{k}+m_2}{(p_2+k)^2-m_2^2}
			 \gamma^\mu
	                 \frac{\slashed{p}_1 - \slashed{k}+m_1}{(p_1-k)^2-m_1^2}
			 \slashed{k}\, u(p_1,m_1)
			 \cdot \frac{1}{k^4} \nonumber \\
		    =&\, - e_q 4\pi \alpha_s (t^a t^a) \int \diffd^4k\, \bar v(p_2,m_2)\, 
			  \left((-\slashed{p}_2 -m_2) - (-\slashed{p}_2-\slashed{k} - m_2)
			  \right)
	                 \frac{-\slashed{p}_2 - \slashed{k}+m_2}{(p_2+k)^2-m_2^2}
			 \gamma^\mu
	                 \frac{\slashed{p}_1 - \slashed{k}+m_1}{(p_1-k)^2-m_1^2}
			 \slashed{k} \nonumber \\
                     &\phantom{- e_q 4\pi \alpha_s (t^a t^a) \int \diffd^4k}
			\times 
			 \, u(p_1,m_1)
			 \cdot \frac{1}{k^4} \nonumber \\
		    =&\, - e_q 4\pi \alpha_s (t^a t^a)\int \diffd^4k\, \bar v(p_2,m_2)\, 
			  (-1)\, \gamma^\mu
	                 \frac{\slashed{p}_1 - \slashed{k}+m_1}{(p_1-k)^2-m_1^2}
			 \left((-\slashed{p}_1 +\slashed{k}+m_1) - (-\slashed{p}_1 + m_1)
			  \right)
                         \, u(p_1,m_1)
			 \cdot \frac{1}{k^4} \nonumber \\
		    =&\, - e_q 4\pi \alpha_s (t^a t^a) \int \diffd^4k\, \bar v(p_2,m_2)\, 
			  (-1)\, \gamma^\mu
			 (-1)
                         \, u(p_1,m_1)
			 \cdot \frac{1}{k^4} \nonumber \\
		    =&\, - e_q 4\pi \alpha_s (t^a t^a) \int \diffd^4k\, \bar v(p_2,m_2)\, 
			  \frac{\gamma^\mu}{k^4} u(p_1,m_1) \ .
			\label{gauge-Gamma-mu}
\end{align}			 
The renormalization factor $\delta Z_2^\text{g}(m)$ is connected to
the QCD quark selfenergy via \cite{Peskin:1995ev}
\begin{align}
  \delta Z_2^\text{g}(m) = \left. \frac{\partial \Sigma^g(p)}{\partial \slashed{p}} \right|_{\slashed{p}=m}
  \ .
  \label{Z2_def}
\end{align}
The gauge dependent part of the quark selfenergy is given by
\begin{align}
  	-i \Sigma^g(p) =&\, \int\, \diffd^4k
			(i g \gamma_\alpha t^a)
			\frac{i(\slashed{p}_1 - \slashed{k} +m_1)}
			     {(p_1 - k)^2 - m_1^2}
			(i g \gamma_\beta t^a)
			\cdot \frac{i k^\alpha k^\beta}{k^2} \frac{1}{k^2} \nonumber \\
\Rightarrow \Sigma^g(p) =&\, i 4\pi \alpha_s (t^a t^a)	
			\int\, \diffd^4k\,
			\slashed{k}
			\frac{\slashed{p}_1 - \slashed{k} +m_1}
			     {(p_1 - k)^2 - m_1^2}
			\slashed{k}
			\cdot \frac{1}{k^4}
  \label{quark-selfen} \ .
\end{align}
Now we find for the renormalization factor:
\begin{align}
  \delta Z_2^\text{g}(m) =&\, i 4\pi \alpha_s (t^a t^a)	
			\int\, \diffd^4k\,
			\slashed{k} \cdot
			\frac{\partial}{\partial \slashed{p}}
			\left. \left( \frac{\slashed{p} - \slashed{k} +m}{(p - k)^2 - m^2}
			\right) \right|_{\slashed{p}=m}
			\cdot \slashed{k}
			\cdot \frac{1}{k^4} \nonumber \\
	=&\,	 	i 4\pi \alpha_s (t^a t^a)	
			\int\, \diffd^4k\,
			\slashed{k} \cdot
			\left. \left( \frac{1}{\slashed{p}^2 - \slashed{p}\slashed{k} - \slashed{k}\slashed{p}
			                       +\slashed{k}^2 - m^2}
					       +\frac{(\slashed{p} - \slashed{k} +m)
					              (-2\slashed{p}+2\slashed{k})}
						{\left(\slashed{p}^2 - \slashed{p}\slashed{k} - \slashed{k}\slashed{p}
			                       +\slashed{k}^2 - m^2 \right)^2}
			\right) \right|_{\slashed{p}=m}
			\cdot \slashed{k}
			\cdot \frac{1}{k^4} \nonumber \\
	=&\,	 	i 4\pi \alpha_s (t^a t^a)	
			\int\, \diffd^4k\,
				\left( \frac{1}{- 2 m \slashed{k} +\slashed{k}^2}
					       +\frac{(- \slashed{k} + 2m)
					              (-2m+2\slashed{k})}
						{\left(- 2 m \slashed{k} +\slashed{k}^2\right)^2}
				\right)
			\cdot \frac{1}{k^2} \nonumber \\
	=&\,	 	i 4\pi \alpha_s (t^a t^a)	
			\int\, \diffd^4k\,
				\left( \frac{- 2 m \slashed{k} +k^2 + (- \slashed{k} + 2m)
					              (-2m+2\slashed{k})}
						      {k^2\left(4 m^2 -4m\slashed{k} +k^2\right)^2}
				\right)
			\cdot \frac{1}{k^2} \nonumber \\
	=&\,	 	i 4\pi \alpha_s (t^a t^a)	
			\int\, \diffd^4k\, \cdot \frac{-1}{k^4} \ .
  \label{Z2g}
\end{align}
Thus
\begin{align}
	&\bar v(p_2,m_2)\, \left[
                i e_q \gamma^\mu \left(\frac{1}{2} \delta Z_2^\text{g}(m_1)
				 +\frac{1}{2} \delta Z_2^\text{g}(m_2)\right)
		+ \Gamma^\mu_g \right] \, u(p_1,m_1) \nonumber \\
	=\, 
	&\bar v(p_2,m_2)\, \left[
                i e_q \gamma^\mu \left(i 4\pi \alpha_s (t^a t^a)\int\, \diffd^4k\, \cdot \frac{-1}{k^4} \right)
		+ - e_q 4\pi \alpha_s (t^a t^a) \int \diffd^4k\, \frac{\gamma^\mu}{k^4} \right] \, u(p_1,m_1) 
		\nonumber \\
	=\,	&0
\end{align}
and so the amplitude in Eq.~(\ref{qcd_gi_master}) does not depend on the gluon gauge.

\section{$F_1$}
\label{app:F1}
As already mentioned in Sec.~\ref{subsubsec:nlo-vc-formf} we unintentionally renormalized
the charge at the quark-photon vertex by assigning different masses $m_1$ and $m_2$ to
the annihilating quark and antiquark, which breaks gauge invariance at the quark-photon vertex,
see Appendix \ref{app:gauge-em}. Thus
\begin{align}
	\underset{q^2 \rightarrow 0}{\lim} F_1(q^2,m_1^2,m_2^2) \neq 1 \ , 
\end{align}	
which we illustrate in Fig.~\ref{fig:F_1_smallq2}. There we plot the real part of
the correction $\delta F_1$ to $F_1$ to order $\alpha_s$ for small $\sqrt{q^2}$. The correction is defined by
\begin{align}
	F_1 = 1 + \frac{\alpha_s}{4\pi}\, \delta F_1 \ .
	\label{F_1-correction}
\end{align}	
As one can see, $\delta F_1$ approaches zero for the case of equal quark masses 
$m_1=m_2=0.1$ GeV, as it should. However, for different quark masses (in our example plot:
$m_1=0.1$ GeV, $m_2=0.5$ GeV) this is clearly not the case.

\begin{figure}[htbp]
	\centering
	\includegraphics[keepaspectratio,angle=-90,width=0.45\textwidth]{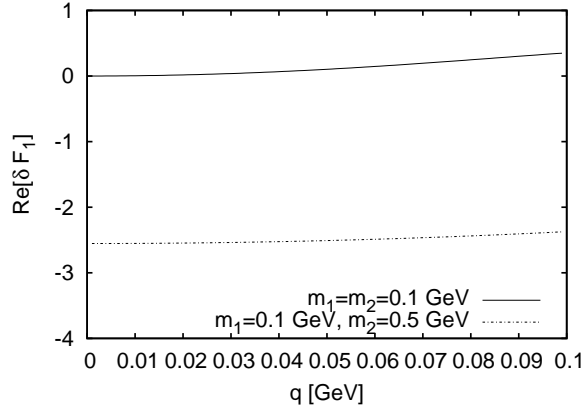}
	\caption{Correction to $F_1$ at order $\alpha_s$ for equal quark masses (solid)
		 and different quark masses (dashed). See main text for details.}
	\label{fig:F_1_smallq2}	 
\end{figure}	
This behavior could potentially spoil our calculations. One should note, however, that
we always stay far away from $q^2=0$, since $q^2=M^2$ sets the hard scale for our
calculations. Therefore, reasonable physical values of $q^2$ are larger than 1 GeV.
To study the influence of different quark masses in the physically interesting
range of $q^2$ we devise the following scheme: \\
to calculate the hadronic cross section
we weight the partonic subprocess cross sections by quark mass distributions (spectral
functions), see Eqs.~(\ref{spec-func-pminus}-\ref{spec-func-mod-width}). Thus also the form factor $F_1(q^2,m_1^2,m_2^2)$ is
weighted by these distributions in our calculation. Therefore, it is worthwhile to compare
the weighted form factors for different masses, $F_1(q^2,m_1^2,m_2^2)$, and for equal masses, $F_1(q^2,m^2,m^2)$, for physically interesting $q^2$. Because of Eq.~(\ref{F_1-correction}) it suffices to compare only the corrections $\delta F_1$. We define
\begin{align}
	\delta \hat F_1(q^2) = \int_0^{m_N^2} \diffd m^2 A(p)\, \delta F_1(q^2,m^2,m^2) \ ,
\end{align}	
with the spectral function $A(p)$ defined in Eq.~(\ref{spec-func}). Now we know that
$\delta F_1(q^2,m^2,m^2)$ shows the correct low $q^2$ behavior,
\begin{align}
	\underset{q^2\rightarrow 0}{\lim}\, \delta F_1(q^2,m^2,m^2) = 0 \ ,
\end{align}	
and we know that the spectral function is normalized to 1, see Eq.~(\ref{spec-func-norm}).
Therefore, also the weighted correction shows the right behavior for $q^2 \rightarrow 0$:
\begin{align}
	\underset{q^2\rightarrow 0}{\lim}\, \delta \hat F_1(q^2) = 
        \int_0^{m_N^2} \diffd m^2 A(p) \cdot \underset{q^2\rightarrow 0}{\lim}\, \delta F_1(q^2,m^2,m^2) = 1 \cdot 0 = 0 \ .
\end{align}
Next we define the weighted correction for different masses:
\begin{align}
	\delta \tilde F_1(q^2) = \int_0^{m_N^2} \diffd m_1^2 \int_0^{m_N^2} \diffd m_1^2 
	                         A(p_1)\, A(p_2)\,
	                         \delta F_1(q^2,m_1^2,m_2^2) \ .
\end{align}
In Fig.~\ref{fig:dF_1_comp} we compare the real parts of $\delta \hat F_1(q^2)$ and 
$\delta \tilde F_1(q^2)$ for $\sqrt{q^2} = 1\ \dots\ 20 \text{ GeV}$. As one can see they agree very well over the entire range. Thus we conclude that the wrong behavior of 
$\delta F_1(q^2,m_1^2,m_2^2)$ near $q^2=0$ ultimately does not affect our calculations.

\begin{figure}[htbp]
	\centering
	\includegraphics[keepaspectratio,angle=-90,width=0.45\textwidth]{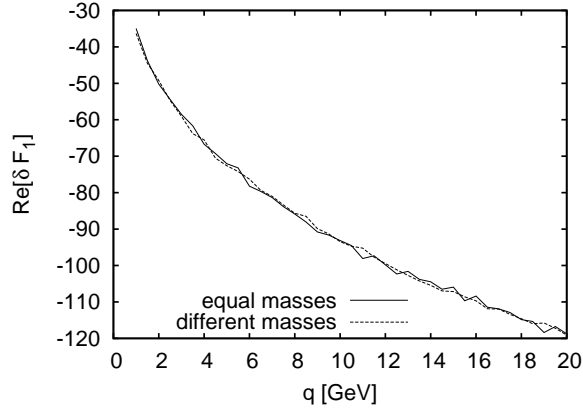}
	\caption{Comparison of the real parts of the weighted corrections to $F_1$ at order $\alpha_s$ for equal quark masses (solid) and different quark masses (dashed). For the spectral functions a width $\Gamma=0.2$ GeV and a large quark momentum component $p^+ = \frac{q}{2}$ was chosen, cf.~Eqs.~(\ref{spec-func-pminus}-\ref{spec-func-mod-width}). The gluon mass was set to $\lambda = 0.05$ GeV, cf.~Eq.~(\ref{vc-kappa}). Both curves agree well over the entire range.}
	\label{fig:dF_1_comp}	 
\end{figure}

\section{Kinematics}
\label{app:nlo-kin}

The calculation of the hadronic cross sections for the LO process, see Sec.~\ref{sec:lo}, gluon bremsstrahlung, see Sec.~\ref{subsec:nlo-brems}, and gluon Compton scattering, see Sec.~\ref{subsec:nlo-compton}, requires to
remove unphysical solutions for the momentum fractions $x_i$.
We closely follow the arguments presented in \cite{Eichstaedt:2009cn} and refer to this publication for more details. 

\subsection{Bremsstrahlung}
\label{appsub:nlo-kin-brems}
We begin with the symmetric case of gluon bremsstrahlung. The hadronic cross section
reads, cf.~Eqs.~(\ref{genkin-master},\ref{hadr_full_xsec_offshell}),
\begin{align}
	\frac{\diffd \sigma_\text{B}}{\diffd M^2 \diffd p_T^2 \diffd x_F} =& 
			   \fint_0^1 \textrm{d}x_1 \fint_0^1 \textrm{d}x_2
       		           \int \textrm{d}{\vec p}_{1_\perp} \int \textrm{d}{\vec p}_{2_\perp}
			   \int \textrm{d}{m_1^2} \int \textrm{d}{m_2^2} \nonumber \\
		          &\times \sum_i q_i^2 {\hat f}_i(x_1,{\vec p}_{1_\perp},m_1^2,q^2)
				{\hat f}_{\, \bar i}(x_2,{\vec p}_{2_\perp},m_2^2,q^2) \cdot
			   \frac{2\sqrt{s} p_\text{cm} (q_z)_\textrm{max}}{E_q}
	\cdot \frac{\diffd \hat\sigma_\text{B}}{\diffd M^2 \diffd t}\,
	\delta\left((p_1+p_2-q)^2 - \lambda^2 \right)
			   \ .        \label{hadr_xsec_nlo_brems}
\end{align}
Again, the $\tilde f_i$ are our unintegrated parton distributions, see Sec.~\ref{subsubsec:spectralfunc}, the partonic cross section $\frac{\diffd \hat\sigma_\text{B}}{\diffd M^2 \diffd t}$ is given in Eq.~(\ref{gluon-brems-part-xsec})
and $\lambda$ is the fictitious gluon mass, introduced to regulate the soft divergence. Now we collect everything except
$\delta$- and $\Theta$-functions in $F$ and rewrite the cross section as
\begin{align}
	\frac{\textrm{d} \sigma_\text{B}}{\diffd M^2\ \diffd x_F \diffd p_T^2} 
	                   =& \fint_0^1 \textrm{d}x_1 \fint_0^1 \textrm{d}x_2
	                   \int \textrm{d}{\vec p}_{1_\perp} 
			   \int \textrm{d}{\vec p}_{2_\perp}
			   \int \textrm{d}{m_1^2} \int \textrm{d}{m_2^2}\,
		           F(x_1,{\vec p}_{1_\perp},m_1^2,x_2,{\vec p}_{2_\perp},m_2^2,M^2)\, \nonumber   \\ 
			   &\times\, \delta\left((p_1+p_2-q)^2 - \lambda^2 \right)\,
			   \Theta(E_g)
			   \ .
			\label{hadr_xsec_nlo_brems_phase_space_only}
\end{align}
\end{widetext}
The $\delta$-functions in Eq.\ (\ref{hadr_xsec_nlo_brems_phase_space_only}) must be worked out in a way 
that allows to discern physical and unphysical solutions for the momentum fractions $x_i$ in order
to perform the $\fint$-integrations. For this aim it is useful to rewrite the parton 
momenta in terms of different variables:
\begin{align}
	\hat q &=  p_1 + p_2 \ , \\
	k &= \frac{1}{2} \left( p_2 - p_1 \right) \ .
	\label{eq:qhat}
\end{align}
Inverting the last two equations, we can use the on-shell conditions for the partons to obtain
\begin{equation}
	m_1^2 = p_1^2 
	  = \left(\frac{1}{2} \hat q - k \right)^2      
	  = \frac{1}{4} \hat q^2 - k \cdot \hat q + k^2 \label{onshell_cond1_full}
\end{equation}	  
and
\begin{equation}
	m_2^2 = p_2^2 
	  = \left(\frac{1}{2} \hat q + k \right)^2       
	  = \frac{1}{4} \hat q^2 + k \cdot \hat q + k^2  \label{onshell_cond2_full} \ .
\end{equation}	  
Adding and subtracting Eqs.\ (\ref{onshell_cond1_full}) and (\ref{onshell_cond2_full})
yields
\begin{align}
	k^2      &= -\frac{1}{4} \hat q^2 + \frac{m_1^2+m_2^2}{2}\ , \label{ksquare} \\
	k\cdot q  &= \frac{m_2^2-m_1^2}{2}		     \label{kq}	\ .
\end{align}	
Solving Eq.\ (\ref{ksquare}) for $k^+$ yields
\begin{equation}
	k^+ = \frac{\vec k_{\perp}^2 - \frac{1}{4}\hat q^2 + \frac{m_1^2+m_2^2}{2}}{k^-} \ .
\end{equation}
Inserting this result into Eq.\ (\ref{kq}) leads to an equation quadratic in $k^-$:
\begin{align}
	m_2^2-m_1^2 &= k^+ \hat q^- + k^- \hat q^+ - 2 \vec k_{\perp} \cdot \overrightarrow{(\hat q_{\perp})}  \nonumber    \\
	  &= \frac{\vec k_{\perp}^2 - \frac{1}{4}\hat q^2 + \frac{m_1^2+m_2^2}{2}}{k^-}
	     \hat q^- + k^- \hat q^+ - 2 \vec k_{\perp} \cdot \overrightarrow{(\hat q_{\perp})} 
						     \label{kq_comp}
\end{align}
\begin{align}
        \Rightarrow 0 =& \left (k^-\right)^2 \hat q^+ + k^- 
	                         (-2\vec k_{\perp} \cdot \overrightarrow{(\hat q_{\perp})}
				  -m_2^2+m_1^2)  \nonumber \\
                   &+ \left( \vec k_{\perp}^2 - \frac{1}{4}\hat q^2 
				              +\frac{m_1^2+m_2^2}{2}\right)\hat q^- \ .
\end{align}	
The solutions are
\begin{align}
	&(k^-)_\pm = \frac{\vec k_{\perp} \cdot \overrightarrow{(\hat q_{\perp})}}{\hat q^+}
	            +\frac{m_2^2-m_1^2}{2\hat q^+} \nonumber \\
             	&\pm \sqrt{ \left( \frac{\vec k_{\perp} \cdot \overrightarrow{(\hat q_{\perp})}}{\hat q^+} +\frac{m_2^2-m_1^2}{2\hat q^+} \right)^2 
	       + \frac{\hat q^-}{\hat q^+} 
	       \left(\frac{1}{4} \hat q^2 - \vec k_{\perp}^2 -\frac{m_1^2+m_2^2}{2}\right) } \ .
	       \label{kminus_sol}
\end{align}	
Inserting (\ref{kminus_sol}) into (\ref{kq_comp}) gives the solutions for $k^+$:
\begin{align}
	&(k^+)_\mp = \frac{\hat q^+}{\hat q^-} \left( \frac{\vec k_{\perp} \cdot \overrightarrow{(\hat q_{\perp})}}{\hat q^+}
	            +\frac{m_2^2-m_1^2}{2\hat q^+} \right. \nonumber \\
             	&\mp \left. \sqrt{ \left( \frac{\vec k_{\perp} \cdot \overrightarrow{(\hat q_{\perp})}}{\hat q^+} +\frac{m_2^2-m_1^2}{2\hat q^+} \right)^2 
	       + \frac{\hat q^-}{\hat q^+} 
	       \left(\frac{1}{4} \hat q^2 - \vec k_{\perp}^2 -\frac{m_1^2+m_2^2}{2}\right) }
\  \right)
		\ .
  		\label{kplus_sol}
\end{align}
Rewriting Eqs.\ (\ref{x1}) and (\ref{x2}) in terms of $\hat q$ and $k$ we obtain the solutions
for the parton momentum fractions ($P_2^+=P_1^-$):
\begin{align}
	(x_1)_\pm = \frac{p_1^-}{\sqrt{S}} 
	    = \frac{1}{P_1^-} \left( \frac{1}{2} \hat q^- - (k^-)_\pm \right) \label{x1_full}
\end{align}
and
\begin{align}
	(x_2)_\mp = \frac{p_2^+}{\sqrt{S}} 
	    = \frac{1}{P_1^-} \left( \frac{1}{2} \hat q^+ + (k^+)_\mp \right) \label{x2_full} \ .
\end{align}
Since there are two solutions for $k^-$ and $k^+$, respectively, we also get two solutions
for $x_1$, $x_2$. To determine which set of $x_1,x_2$ and thus $k^+,k^-$ has to be chosen we take
the limit of zero parton transverse momentum and vanishing masses,
\begin{align}
	(k^-)_\pm &\rightarrow \pm \sqrt{\frac{\hat q^-}{\hat q^+} \frac{1}{4} \hat q^2} = \pm \frac{\hat q^-}{2} 
									\label{kminus_coll} \ ,\\
	(k^+)_\mp &\rightarrow \mp \sqrt{\frac{\hat q^+}{\hat q^-} \frac{1}{4} \hat q^2} = \mp \frac{\hat q^+}{2} 
									\label{kplus_coll} \ .
\end{align}	
Inserting expressions (\ref{kminus_coll}) and (\ref{kplus_coll}) into (\ref{x1_full}) and (\ref{x2_full}) 
yields two solutions for the momentum fractions,
\begin{equation}
	(x_1)_\pm \rightarrow \frac{1}{P_1^-} \begin{cases} 
					0 \\
					\hat q^-
				  \end{cases} 
\end{equation}				 
and
\begin{equation}
	(x_2)_\mp \rightarrow \frac{1}{P_1^-} \begin{cases}
					0 \\
					\hat q^+
				  \end{cases} \ .
	\label{eq:x2_cases}
\end{equation}			  	  
The upper solutions correspond to the unphysical case $x_1 = x_2 = 0$. Thus we only keep the lower solutions when evaluating the phase space integrals. This requires the integrals in Eq. (\ref{hadr_xsec_nlo_brems_phase_space_only}) to be 
evaluated in the correct order, otherwise one cannot disentangle the two different 
solutions for $x_1$ and $x_2$.

We begin by introducing several integrals over $\delta$-functions in Eq.\  
(\ref{hadr_xsec_nlo_brems_phase_space_only}).
In this way we will transform the integration variables to the above chosen $\hat q$ and $\vec k_\perp$:
\begin{widetext}
\begin{align}
  \frac{\textrm{d} \sigma_\text{B}}{\diffd M^2\ \diffd x_F \diffd p_T^2} 
	                   = &\fint_0^1 \textrm{d}x_1 \fint_0^1 \textrm{d}x_2
	                   \int \textrm{d}{\vec p}_{1_\perp} 
			   \int \textrm{d}{\vec p}_{2_\perp}
			   \int \textrm{d}\overrightarrow{(\hat q_{\perp})}
			   \int \textrm{d}{\vec k}_{\perp}
			   \int \textrm{d}{m_1^2} 
			   \int \textrm{d}{m_2^2}
			   \int\diffd \hat q^+ \int \diffd \hat q^- 
			   F(x_1,{\vec p}_{1_\perp},m_1^2,x_2,{\vec p}_{2_\perp},m_2^2,M^2) \nonumber \\ 
		           &\times \delta\left(\hat q^+ - (p_1^+ + p_2^+)\right)\, 
			           \delta\left(\hat q^- - (p_1^- + p_2^-)\right)\,
			   \delta^{(2)}\left(\overrightarrow{(\hat q_{\perp})} - \left({\vec p}_{1_\perp} + {\vec p}_{2_\perp} \right)\right)\,
			   \delta^{(2)}\left({\vec k}_{\perp} - \frac{1}{2}\left({\vec p}_{1_\perp} - {\vec p}_{2_\perp} \right)\right)
			   \nonumber \\
			  &\times
			  \delta\left((p_1+p_2-q)^2 - \lambda^2 \right)\,
			   \Theta \left(E_g \right)
			\label{full_integral_correct} \ .
\end{align}
First we perform
\begin{equation}
	\int \textrm{d}{\vec p}_{1_\perp} \int \textrm{d}{\vec p}_{2_\perp}
	\delta^{(2)}\left(\overrightarrow{(\hat q_{\perp})}  - \left({\vec p}_{1_\perp} + {\vec p}_{2_\perp} \right)
		    \right)\,
	\delta^{(2)}\left({\vec k}_{\perp} - \frac{1}{2}\left({\vec p}_{2_\perp} - {\vec p}_{1_\perp} 
				\right)\right)
				= 1 \label{eq:nlo-kin-brems-pperps}\ .
\end{equation}	
\end{widetext}
Now we calculate the integral
\begin{align}
	\fint_0^1 \textrm{d}x_1 \fint_0^1 \textrm{d}x_2\  
	 \delta\left(\hat q^+ - (p_1^+ + p_2^+)\right)\, \delta\left(\hat q^- - (p_1^- + p_2^-)\right)\ . 
\end{align}       
According to Eqs.\ (\ref{kminus_sol})-(\ref{x2_full}) the $\delta$-functions in this expression
have two possible solutions for each $p_1^-$ and $p_2^+$. However, as explained above,
we now have to explicitly remove the unphysical solutions $(x_1)_+$ and $(x_2)_-$ , which are the ones
corresponding to the upper sign in Eqs.\ (\ref{kminus_sol}) and (\ref{kplus_sol}):
\begin{widetext}
\begin{align}
	&\fint_0^1 \textrm{d}x_1 \fint_0^1 \textrm{d}x_2\  
	 \delta(\hat q^+ - (p_1^+ + p_2^+))\, \delta(\hat q^- - (p_1^- + p_2^-)) \nonumber \\
       =&\fint_0^1 \textrm{d}x_1 \fint_0^1 \textrm{d}x_2\  
	 \delta\left(\hat q^+ 
	       - \frac{\left(\frac{1}{2}\overrightarrow{(\hat q_{\perp})} - {\vec k}_{\perp}\right)^2 + m_1^2}{x_1P_1^-}
	       - x_2 P_1^- \right)\,
	 \delta\left(\hat q^- - x_1 P_1^-
               - \frac{\left(\frac{1}{2}\overrightarrow{(\hat q_{\perp})}  + {\vec k}_{\perp}\right)^2 + m_2^2}{x_2 P_1^-}
	       \right)
		 \nonumber \\
       =&\int_0^1 \textrm{d}x_1 \int_0^1 \textrm{d}x_2\  
	 \delta\left(\hat q^+ 
	       - \frac{\left(\frac{1}{2}\overrightarrow{(\hat q_{\perp})} - {\vec k}_{\perp}\right)^2 + m_1^2}{(x_1)_-P_1^-}
	       - (x_2)_+ P_1^- \right)\,
	 \delta\left(\hat q^- - (x_1)_- P_1^-
               - \frac{\left(\frac{1}{2}\overrightarrow{(\hat q_{\perp})}  + {\vec k}_{\perp}\right)^2 + m_2^2}{(x_2)_+ P_1^-}
	       \right)
		 \nonumber \\
       =&\left| \left(P_1^-\right)^2 - \frac{ \left[
	           \left(\frac{1}{2}\overrightarrow{(\hat q_{\perp})} - {\vec k}_{\perp}\right)^2 + m_1^2 \right]
	           \left[\left(\frac{1}{2}\overrightarrow{(\hat q_{\perp})} + {\vec k}_{\perp}\right)^2 + m_2^2\right]}{(x_1)_-^2 (x_2)_+^2 \left( P_1^- \right)^2}\right|^{-1} \cdot
		   \Theta\left(1-(x_1)_-\right)\, \Theta\left((x_1)_-\right)\,
                   \Theta\left(1-(x_2)_+\right)\, \Theta\left((x_2)_+\right) \ .	
	 \label{eq:nlo-kin-brems-x1x2}
\end{align}       
Using $\diffd \hat q^+ \diffd \hat q^- = 2 \diffd \hat q_0 \diffd \hat q_z$ we can evaluate one of the remaining 
integrals of Eq.\ (\ref{full_integral_correct}) with the help of the $\delta$--function:
\begin{align}
	2&\int \diffd \hat q_0\, \delta\left((p_1+p_2-q)^2 - \lambda^2 \right) = 
	2\int \diffd \hat q_0\, \delta\left((\hat q-q)^2 -\lambda^2 \right)
	= \frac{1}{E_g}
\end{align}	
with $E_g = \sqrt{(\vec r)^2 + \lambda^2} = \sqrt{\left(\overrightarrow{(\hat q)}- \vec q\right)^2 + \lambda^2} $.
Collecting the pieces, Eq.\ (\ref{full_integral_correct}) simplifies to
\begin{align}
  \frac{\textrm{d} \sigma_\text{B}}{\diffd M^2\ \diffd x_F \diffd p_T^2} 
	                   = &\int_{(q_z)_\textrm{min}}^{(q_z)_\textrm{max}}\diffd \hat q_z
			   \int_0^{\left|\overrightarrow{(\hat q_{\perp})}\right|_\textrm{max}} \textrm{d}\overrightarrow{(\hat q_{\perp})}
			   \int_0^{|{\vec k}_{\perp}|_\textrm{max}} \textrm{d}{\vec k}_{\perp}
			   \int_0^{(m_1^2)_\text{max}} \textrm{d}{m_1^2} 
			   \int_0^{(m_2^2)_\text{max}} \textrm{d}{m_2^2}
			   \,F( (x_1)_-,\hat {\vec p}_{1_\perp},m_1^2,(x_2)_+,\hat {\vec p}_{2_\perp},m_2^2,M^2) \nonumber \\
	&\times \Theta \left(E_g \right) \frac{1}{E_g} \cdot
       	\left| \left(P_1^-\right)^2 - \frac{ \left[
	           \left(\frac{1}{2}\overrightarrow{(\hat q_{\perp})} - {\vec k}_{\perp}\right)^2 + m_1^2 \right]
	           \left[\left(\frac{1}{2}\overrightarrow{(\hat q_{\perp})} + {\vec k}_{\perp}\right)^2 + m_2^2\right]}{(x_1)_-^2 (x_2)_+^2 \left( P_1^- \right)^2}\right|^{-1} 
	           \nonumber \\
	           &\times
		   \Theta\left(1-(x_1)_-\right)\,\Theta\left((x_1)_-\right)\,
                   \Theta\left(1-(x_2)_+\right)\,\Theta\left((x_2)_+\right)
			\label{hadr_xsec_final} \ .
\end{align}
Now $(x_1)_-,\hat {\vec p}_{1_\perp},(x_2)_+$ and $\hat {\vec p}_{2_\perp}$ are fixed:

\begin{align}
	(x_1)_- &= \frac{1}{P_1^-}\left(\frac{\hat q^-}{2} 
			 - \frac{\vec k_{\perp} \cdot \overrightarrow{(\hat q_{\perp})}}{\hat q^+} -\frac{m_2^2-m_1^2}{2\hat q^+}
             	+ \sqrt{ \left( \frac{\vec k_{\perp} \cdot \overrightarrow{(\hat q_{\perp})}}{\hat q^+} +\frac{m_2^2-m_1^2}{2\hat q^+} \right)^2 
	       + \frac{\hat q^-}{\hat q^+} 
	       \left(\frac{1}{4} \hat q^2 - \vec k_{\perp}^2 -\frac{m_1^2+m_2^2}{2}\right) } \right)\ , \\
	(x_2)_+ &= \frac{1}{P_1^-} \left(\frac{\hat q^+}{2} 
			 + \frac{\vec k_{\perp} \cdot \overrightarrow{(\hat q_{\perp})}}{\hat q^-} +\frac{m_2^2-m_1^2}{2\hat q^-}
             	+ \sqrt{ \left( \frac{\vec k_{\perp} \cdot \overrightarrow{(\hat q_{\perp})}}{\hat q^-} +\frac{m_2^2-m_1^2}{2\hat q^-} \right)^2 
	       + \frac{\hat q^+}{\hat q^-} 
	       \left(\frac{1}{4} \hat q^2 - \vec k_{\perp}^2 -\frac{m_1^2+m_2^2}{2}\right) } \right)\ , \\
	\hat {\vec p}_{1_\perp}	&= \frac{1}{2}\overrightarrow{(\hat q_{\perp})} - {\vec k}_{\perp} \ , \\
	\hat {\vec p}_{2_\perp}	&= \frac{1}{2}\overrightarrow{(\hat q_{\perp})} + {\vec k}_{\perp} \ , \\
	\vec k_{\perp} \cdot \overrightarrow{(\hat q_{\perp})} &= k_\perp \hat q_\perp \cos \phi_{k_\perp}
\end{align}
with
\begin{align}
	\hat q^+ &= E_q + E_g + \hat q_z \ , \\
	\hat q^- &= E_q + E_g - \hat q_z \ , \\
	E_q &= \sqrt{M^2 + p_T^2 + q_z^2 } \ , \\
	E_g &= \sqrt{\overrightarrow{(\hat q_{\perp})}^2 + \hat q_z^2 
		     -2\overrightarrow{(\hat q_{\perp})}\cdot \vec q_\perp
		     -2\hat q_z \cdot q_z
		     + p_T^2 + q_z^2 + \lambda^2} \ , \\
        \overrightarrow{(\hat q_{\perp})}\cdot \vec q_\perp &= \hat q_\perp p_T \cos \phi_{\hat q_\perp} \ , \\
        q_z &= x_F (q_z)_\textrm{max} \ .
\end{align}
The integration limits can now be found from general considerations:
$(m_2^2)_\text{max}$ is fixed by the condition that $(x_1)_-$ and $(x_2)_+$ must be real
numbers,
\begin{align}
	(m_2^2)_\text{max} = -2\vec k_{\perp} \cdot \overrightarrow{(\hat q_{\perp})}
			       + m_1^2 + \hat q^+ \hat q^-
			     - \sqrt{4\hat q^+ q^- m_1^2 
				     + 4\hat q^+ \hat q^- \left(\frac{1}{2}\overrightarrow{(\hat q_{\perp})} - {\vec k}_{\perp}\right)^2} \ .
    \label{eq:nlo_kin_brems_m2max}
\end{align}
From $(m_2^2)_\text{max} \overset{!}{>} 0$ we find
\begin{align}
	(m_1^2)_\text{max} = 2\vec k_{\perp} \cdot \overrightarrow{(\hat q_{\perp})}
			     + \hat q^+ \hat q^-
			     - 2\sqrt{\hat q^+ \hat q^- \left(\frac{1}{2}\overrightarrow{(\hat q_{\perp})} + {\vec k}_{\perp}\right)^2} \ ,
\end{align}
and from $(m_1^2)_\text{max} \overset{!}{>} 0$ follows
\begin{align}
	|{\vec k}_{\perp}|^2_\textrm{max} = \frac{\hat q^+ \hat q^- 
                                \left(\hat q^+ \hat q^- - \hat q^2_\perp \right)}
				{4\left(\hat q^+ \hat q^- - \hat q^2_\perp \cos^2(\phi_{k_\perp}) \right)} \ . \label{eq:nlo_kin_brems_kperpmax}
\end{align}
The energy of the incoming partons must be less than the energy of the hadronic system,
$\hat q_0 < \sqrt{S}$, and so
\begin{align}
	|\overrightarrow{(\hat q_{\perp})}|_\textrm{max} =
	p_T \cos \phi_{\hat q} + \sqrt{p_T^2 \left(\cos^2 \phi_{\hat q}- 1\right) - 
	                               \left(q_z -\hat q_z\right)^2 + 
	                               \left(\sqrt{S}-E_q\right)^2 - \lambda^2}\ .
\end{align}
Finally, $\hat q_{\perp}$ is a real number and thus
\begin{align}
  \left(\hat q_z\right)^\textrm{max}_\textrm{min} = 
	  q_z \pm \sqrt{p_T^2\left(\cos^2 \phi_{\hat q_\perp} -1 \right) + \left(\sqrt{S}-E_q\right)^2 -\lambda^2} \ .
\end{align}

\subsection{Leading order process}
\label{appsub:lo-kin}
The kinematics of the LO cross section is a special case of the bremsstrahlung kinematics of Appendix \ref{appsub:nlo-kin-brems}. Namely at LO the four-momentum of the incoming partons is equal to the four-momentum of the virtual photon: $\hat q = q$. From Eq.~(\ref{part_offshell_xsec}) we note for the partonic cross section
\begin{align}
  \frac{\diffd \hat \sigma_\text{LO}}{\diffd M^2 \diffd x_F \diffd p_T^2}
  \sim \delta\left(M^2 - (p_1+p_2)^2 \right) \delta\left(p_T^2 - (\vec p_{1_\perp} + \vec p_{2_\perp})^2\right)
  \delta\left(x_F - \frac{(p_1)_z + (p_2)_z}{(q_z)_\text{max}}\right) \ .
\end{align}
Now employing Eqs.~(\ref{eq:qhat}-\ref{eq:x2_cases}) and Eqs.~(\ref{eq:nlo-kin-brems-pperps}-
\ref{eq:nlo-kin-brems-x1x2}) and everywhere replacing $\hat q$ by $q$ we find for the LO hadronic cross section
\begin{align}
  \frac{\diffd \sigma_\text{LO}}{\diffd M^2 \diffd x_F \diffd p_T^2} =&
  	\int 2 \diffd q_0
	\int \diffd q_z
	\int \textrm{d}\vec q_{\perp}
        \int \textrm{d}{\vec k}_{\perp}
	\int \textrm{d}{m_1^2} 
	\int \textrm{d}{m_2^2}
	\,F_\text{LO}( (x_1)_-,\hat {\vec p}_{1_\perp},m_1^2,(x_2)_+,\hat {\vec p}_{2_\perp},m_2^2,M^2) \nonumber \\
	&\times \left| \left(P_1^-\right)^2 - \frac{ \left[
	\left(\frac{1}{2}{\vec q}_{\perp} - {\vec k}_{\perp}\right)^2 + m_1^2 \right]
	\left[\left(\frac{1}{2}{\vec q}_{\perp} + {\vec k}_{\perp}\right)^2 + m_2^2\right]}{(x_1)_-^2 (x_2)_+^2 \left( P_1^- \right)^2}\right|^{-1} 
	           \nonumber \\
	           &\times
		   \Theta\left(1-(x_1)_-\right)\,\Theta\left((x_1)_-\right)\,
                   \Theta\left(1-(x_2)_+\right)\,\Theta\left((x_2)_+\right) \nonumber \\
		   &\times
		   \delta\left(M^2 - q^2 \right) \delta\left(p_T^2 - (\vec q_{\perp})^2\right)
		   \delta\left(x_F - \frac{q_z}{(q_z)_\text{max}}\right) \ .
\end{align}
With help of the three remaining $\delta$-functions we can now easily perform the four integrations over the components of $q$:
\begin{align}
  \frac{\textrm{d} \sigma_\text{LO}}{\diffd M^2\ \diffd x_F \diffd p_T^2} 
	                   =& \int_0^{2\pi} \diffd \phi_\perp 
			     \int_0^{({\vec k}_{\perp})^2_\textrm{max}} \frac{1}{2}
			     \diffd ({\vec k}_{\perp})^2
			     \int_0^{(m_1)^2_\textrm{max}} \diffd m_1^2
			     \int_0^{(m_2)^2_\textrm{max}} \diffd m_2^2 \
			     \frac{\pi\ (q_z)_\textrm{max}}{E_q}  \nonumber \\
			&\times   \left| \left(P_1^-\right)^2 - \frac{ \left[
			\left(\frac{1}{2}{\vec q}_{\perp} - {\vec k}_{\perp}\right)^2 + m_1^2 \right]
			\left[\left(\frac{1}{2}{\vec q}_{\perp} + {\vec k}_{\perp}\right)^2 + m_2^2\right]}{(x_1)_-^2 (x_2)_+^2 \left( P_1^- \right)^2}\right|^{-1} 
		   F_\text{LO}((x_1)_-,\hat {\vec p}_{1_\perp},m_1^2,
		          (x_2)_+,\hat {\vec p}_{2_\perp},m_2^2,M^2) \nonumber \\
	           &\times
		   \Theta\left(1-(x_1)_-\right)\,\Theta\left((x_1)_-\right)\,
                   \Theta\left(1-(x_2)_+\right)\,\Theta\left((x_2)_+\right) \ .
\end{align}
The integration limits are now recovered from Eqs.~(\ref{eq:nlo_kin_brems_m2max}-\ref{eq:nlo_kin_brems_kperpmax}), again by replacing $\hat q$ with $q$.
\subsection{Compton scattering}
\label{appsub:nlo_kin_compton}
At this point we like to stress the kinematical differences between bremsstrahlung and Compton scattering.
For bremsstrahlung we have a quark from nucleon 1 annihilating with an antiquark from nucleon 2 or vice
versa. However, we treat quarks and antiquarks on equal footing and distribute their masses with
the same spectral function, cf.~Sec.~\ref{subsubsec:spectralfunc}. Thus we can easily take care of both
cases by simply summing over all quark- {\it and} antiquark-flavors in Eq.~(\ref{hadr_xsec_nlo_brems}).
Gluon Compton scattering is different since we keep the gluons massless and the simplification from
above does not apply anymore. However, we can calculate one of the two cases, for example quark/antiquark
from nucleon 1 annihilates with gluon from nucleon 2, and then simply find the other case by symmetry considerations: nucleon 1 and 2 are defined by their direction of motion along the $z$-axis. Thus by changing
$z$ to $-z$ and so $x_F$ to $-x_F$ we find that the second case corresponds to the first case with $x_F 
\rightarrow -x_F$. The hadronic cross section therefore reads, compare with Eq.\ (\ref{hadr_xsec_nlo_brems}),
\begin{align}
	\frac{\diffd \sigma_\text{C}}{\diffd M^2 \diffd p_T^2 \diffd x_F} =& 
			   \fint_0^1 \textrm{d}x_1 \fint_0^1 \textrm{d}x_2
       		           \int \textrm{d}{\vec p}_{1_\perp} \int \textrm{d}{\vec p}_{2_\perp}
			   \int \textrm{d}{m_1^2} \nonumber \\
		          &\times \sum_i q_i^2 ({\hat f}_i)_1(x_1,{\vec p}_{1_\perp},m_1^2,q^2)
			  {\tilde g}_2(x_2,{\vec p}_{2_\perp},q^2) \cdot
			   \frac{2\sqrt{s} p_\text{cm} (q_z)_\textrm{max}}{E_q}
	\cdot \frac{\diffd \hat\sigma_\text{C}}{\diffd M^2 \diffd t}\,
	\delta\left((p_1+p_2-q)^2 \right)  \nonumber \\
	+& \fint_0^1 \textrm{d}x_1 \fint_0^1 \textrm{d}x_2
       		           \int \textrm{d}{\vec p}_{1_\perp} \int \textrm{d}{\vec p}_{2_\perp}
			   \int \textrm{d}{m_1^2} \nonumber \\
		          &\times \left. \sum_i q_i^2 ({\hat f}_i)_2(x_1,{\vec p}_{1_\perp},m_1^2,q^2)
			  {\tilde g}_1(x_2,{\vec p}_{2_\perp},q^2) \cdot
			   \frac{2\sqrt{s} p_\text{cm} (q_z)_\textrm{max}}{E_q}
	\cdot \frac{\diffd \hat\sigma_\text{C}}{\diffd M^2 \diffd t}\,
	\delta\left((p_1+p_2-q)^2 \right) \right|_{x_F\rightarrow -x_F} \nonumber \\
	=& \frac{(\diffd \sigma_\text{C})_{12}}{\diffd M^2 \diffd p_T^2 \diffd x_F} 
	   +
	   \left. \frac{(\diffd \sigma_\text{C})_{21}}{\diffd M^2 \diffd p_T^2 \diffd x_F} \right|_{x_F \rightarrow -x_F}
	\ .        \label{hadr_xsec_nlo_compton}
\end{align}
The indices $1$ and $2$ for the parton distributions denote the parent nucleons (p,n,$\overline{\text{p}}$). $\tilde g$ is the transverse momentum dependent gluon distribution function and we choose it in analogy
with the transverse momentum dependent quark distribution function of Eq.\ (\ref{ftilde}),
\begin{equation}
   {\tilde g}(x,{\vec p}_{\perp},q^2) = g(x_i,q^2) \cdot f_{\perp}({\vec p}_{i_\perp})
   \label{eq:gtilde}
\end{equation}	
with $f_\perp$ defined in Eq.~(\ref{fperp}) and with the usual gluon PDF $g$. Now can we proceed similarly to Sec.~\ref{appsub:nlo-kin-brems}:
\begin{align}
  \frac{(\diffd \sigma_\text{C})_{12}}{\diffd M^2 \diffd p_T^2 \diffd x_F}
	=& \fint_0^1 \textrm{d}x_1 \fint_0^1 \textrm{d}x_2
	                   \int \textrm{d}{\vec p}_{1_\perp} 
			   \int \textrm{d}{\vec p}_{2_\perp}
			   \int \textrm{d}\overrightarrow{(\hat q_{\perp})}
			   \int \textrm{d}{\vec k}_{\perp}
			   \int \textrm{d}{m_1^2} 
			   \int\diffd \hat q^+ \int \diffd \hat q^- 
			   F(x_1,{\vec p}_{1_\perp},m_1^2,x_2,{\vec p}_{2_\perp},M^2) \nonumber \\ 
		           &\times \delta\left(\hat q^+ - (p_1^+ + p_2^+)\right)\, 
			           \delta\left(\hat q^- - (p_1^- + p_2^-)\right)\,
			   \delta^{(2)}\left(\overrightarrow{(\hat q_{\perp})} - \left({\vec p}_{1_\perp} + {\vec p}_{2_\perp} \right)\right)\,
			   \delta^{(2)}\left({\vec k}_{\perp} - \frac{1}{2}\left({\vec p}_{1_\perp} - {\vec p}_{2_\perp} \right)\right)
			   \nonumber \\
			  &\times
			  \delta\left((p_1+p_2-q)^2 - m_1^2 \right)\ .
	                  \label{eq:nlo-kin-compton-full-integral-correct} 
\end{align}
We use Eqs.~(\ref{eq:nlo-kin-brems-pperps}) and (\ref{eq:nlo-kin-brems-x1x2}) and
\begin{align}
     \int \textrm{d}{m_1^2}\,  \delta\left((p_1+p_2-q)^2 - m_1^2 \right) = 1
\end{align}
to find 
\begin{align}
  \frac{(\diffd \sigma_\text{C})_{12}}{\diffd M^2 \diffd p_T^2 \diffd x_F}
	=&\int_{(q_z)_\textrm{min}}^{(q_z)_\textrm{max}}\diffd \hat q_z
			   \int_0^{\left|\overrightarrow{(\hat q_{\perp})}\right|_\textrm{max}} \textrm{d}\overrightarrow{(\hat q_{\perp})}
			   \int_{(\hat q_0)_\text{min}}^{(\hat q_0)_\text{max}} 2 \diffd \hat q_0
			   \int_0^{|{\vec k}_{\perp}|_\textrm{max}} \textrm{d}{\vec k}_{\perp}
			   \,F(x_1,{\vec p}_{1_\perp},m_1^2,x_2,{\vec p}_{2_\perp},M^2) \nonumber \\
	&\times \left| \left(P_1^-\right)^2 - \frac{ \left[
	           \left(\frac{1}{2}\overrightarrow{(\hat q_{\perp})} - {\vec k}_{\perp}\right)^2 + m_1^2 \right]
	           \left[\left(\frac{1}{2}\overrightarrow{(\hat q_{\perp})} + {\vec k}_{\perp}\right)^2 \right]}{(x_1)_-^2 (x_2)_+^2 \left( P_1^- \right)^2}\right|^{-1} 
	           \nonumber \\
	           &\times
		   \Theta\left(1-(x_1)_-\right)\,\Theta\left((x_1)_-\right)\,
                   \Theta\left(1-(x_2)_+\right)\,\Theta\left((x_2)_+\right)
		   \label{eq:nlo-kin-compton-final-integral} \ .
\end{align}
Now $(x_1)_-,\hat {\vec p}_{1_\perp},(x_2)_+$ and $\hat {\vec p}_{2_\perp}$ are fixed by
\begin{align}
	(x_1)_- &= \frac{1}{P_1^-}\left(\frac{\hat q^-}{2} 
			 - \frac{\vec k_{\perp} \cdot \overrightarrow{(\hat q_{\perp})}}{\hat q^+} +\frac{m_1^2}{2\hat q^+}
             	+ \sqrt{ \left( \frac{\vec k_{\perp} \cdot \overrightarrow{(\hat q_{\perp})}}{\hat q^+} -\frac{m_1^2}{2\hat q^+} \right)^2 
	       + \frac{\hat q^-}{\hat q^+} 
	       \left(\frac{1}{4} \hat q^2 - \vec k_{\perp}^2 -\frac{m_1^2}{2}\right) } \right)\ , \\
	(x_2)_+ &= \frac{1}{P_1^-} \left(\frac{\hat q^+}{2} 
			 + \frac{\vec k_{\perp} \cdot \overrightarrow{(\hat q_{\perp})}}{\hat q^-} -\frac{m_1^2}{2\hat q^-}
             	+ \sqrt{ \left( \frac{\vec k_{\perp} \cdot \overrightarrow{(\hat q_{\perp})}}{\hat q^-} -\frac{m_1^2}{2\hat q^-} \right)^2 
	       + \frac{\hat q^+}{\hat q^-} 
	       \left(\frac{1}{4} \hat q^2 - \vec k_{\perp}^2 -\frac{m_1^2}{2}\right) } \right)\ , \\
	\hat {\vec p}_{1_\perp}	&= \frac{1}{2}\overrightarrow{(\hat q_{\perp})} - {\vec k}_{\perp} \ , \\
	\hat {\vec p}_{2_\perp}	&= \frac{1}{2}\overrightarrow{(\hat q_{\perp})} + {\vec k}_{\perp} \ , \\
	\vec k_{\perp} \cdot \overrightarrow{(\hat q_{\perp})} &= k_\perp \hat q_\perp \cos \phi_{k_\perp} \ , \\
	m_1^2 &= \left( \hat q_0 - E_q\right)^2 - \left(\overrightarrow{(\hat q_{\perp})} - \vec q_\perp\right)^2		     - \left(q_z - \hat q_z\right)^2
\end{align}
with
\begin{align}
	\hat q^+ &= \hat q_0 + \hat q_z \ , \\
	\hat q^- &= \hat q_0 - \hat q_z \ , \\
	E_q &= \sqrt{M^2 + p_T^2 + q_z^2 } \ , \\
        \overrightarrow{(\hat q_{\perp})}\cdot \vec q_\perp &= \hat q_\perp p_T \cos \phi_{\hat q_\perp} \ , \\
        q_z &= x_F (q_z)_\textrm{max} \ .
\end{align}
The integration limits can now be found from general considerations.
$|{\vec k}_{\perp}|_\textrm{max}$  is fixed by the condition that $(x_1)_-$ and $(x_2)_+$ must be real
numbers:
\begin{align}
  |{\vec k}_{\perp}|_\textrm{max} = -\frac{\left|\overrightarrow{(\hat q_{\perp})}\right| \cos \phi_{k_\perp} m_1^2}{2(\hat q^+ \hat q^- - \hat q_\perp^2 \cos^2 \phi_{k_\perp})}
  + \sqrt{ \left(\frac{\left|\overrightarrow{(\hat q_{\perp})}\right| \cos \phi_{k_\perp} m_1^2}{2(\hat q^+ \hat q^- - \hat q_\perp^2 \cos^2 \phi_{k_\perp}}\right)^2 + 
           \left(\frac{-\frac{m_1^4}{4} - \frac{1}{4}\hat q^+ \hat q^-(\hat q^+ \hat q^- - \hat q_\perp^2)
	   +\hat q^+ \hat q^- \frac{m_1^2}{2}}{\hat q_\perp^2 \cos^2 \phi_{k_\perp} - \hat q^+ \hat q^-} 
	   \right)} 
   \ .
\end{align}
From $0 < m_1^2 < m_N^2$ one finds
\begin{align}
  (\hat q_0)_\text{min} &= E_q + \sqrt{\overrightarrow{(\hat q_{\perp})}^2 + \hat q_z^2 
		     -2\overrightarrow{(\hat q_{\perp})}\cdot \vec q_\perp
		     -2\hat q_z \cdot q_z
		     + p_T^2 + q_z^2} \ , \\
  (\hat q_0)_\text{max} &= E_q + \sqrt{\overrightarrow{(\hat q_{\perp})}^2 + \hat q_z^2 
		     -2\overrightarrow{(\hat q_{\perp})}\cdot \vec q_\perp
		     -2\hat q_z \cdot q_z
		     + p_T^2 + q_z^2 + m_N^2}  \ . 
\end{align}
Since the energy of the incoming partons cannot be larger than the hadronic energy, we have
$\hat q_0 < \sqrt{S}$ and thus
\begin{align}
  \left|\overrightarrow{(\hat q_{\perp})}\right|_\text{max} = p_T \cos \phi_{\hat q_\perp}
	+ \sqrt{p_T^2 \left(\cos^2 \phi_{\hat q_\perp} -1\right) - \left(q_z - \hat q_z\right)^2 - m_N^2 + \left( \sqrt{S} - E_q \right)^2} \ .
\end{align}
Finally, $\hat q_{\perp}$ is a real number and thus
\begin{align}
  \left(\hat q_z\right)^\textrm{max}_\textrm{min} = 
	  q_z \pm \sqrt{p_T^2\left(\cos^2 \phi_{\hat q_\perp} -1 \right) + \left(\sqrt{S}-E_q\right)^2 - m_N^2} \ .
\end{align}

\end{widetext}

\end{appendix}

\raggedright
\bibliographystyle{apsrev4-1}
\bibliography{literature}

\end{document}